\documentclass[aps,prl,twocolumn,superscriptaddress,groupedaddress]{revtex4-1}
\usepackage{hyperref}
\usepackage{amsfonts}
\usepackage{bbm}
\usepackage{graphicx}
\usepackage{dcolumn}
\usepackage{bm}
\usepackage{amssymb, amsmath}
\usepackage{enumerate}
\usepackage{slashed}
\usepackage{simplewick}
\usepackage{comment}
\usepackage{color}
\usepackage{soul}
\usepackage{braket}

\newcommand{\ba}{\begin{align}}
\newcommand{\ee}{\end{equation}}
\newcommand{\be}{\begin{equation}}

\def\12{\frac{1}{2}}

\newcommand{\en}{\end{align}}

\usepackage{color}

\def\_{{\:\!}}

\def\beq{\begin{equation}}
\def\eeq{\end{equation}}
\def\bea{\begin{eqnarray}}
\def\eea{\end{eqnarray}}

\begin{document}

\setcounter{secnumdepth}{6}

\title{Dynamical transitions from slow to fast relaxation in random open quantum systems}
\author{Dror Orgad$^1$, Vadim Oganesyan$^{2,3}$, and Sarang Gopalakrishnan$^{4}$}
\affiliation{$^1$Racah Institute of Physics, The Hebrew University, Jerusalem 91904, Israel \\
$^2$Department of Physics and Astronomy, College of Staten Island, CUNY, Staten Island, NY 10314, USA \\
$^3$Center for Computational Quantum Physics, Flatiron Institute, 162 5th Avenue, New York, NY 10010, USA \\
%$^4$Department of Physics, The Pennsylvania State University, University Park, PA 16802, USA \\
\mbox{$^4$Department of Electrical and Computer Engineering, Princeton University, Princeton, NJ 08540, USA}}

%\date{\today}% It is always \today, today,
             %  but any date may be explicitly specified

\begin{abstract}

We explore the effects of spatial locality on the dynamics of random quantum systems subject to a Markovian noise.
To this end, we study a model in which the system Hamiltonian and its couplings to the noise are
random matrices whose entries decay as power laws of distance, with distinct exponents $\alpha_H, \alpha_L$.
The steady state is always featureless, but the rate at which it is approached exhibits three
phases depending on $\alpha_H$ and $\alpha_L$: a phase where the approach is asymptotically exponential as a
result of a gap in the spectrum of the Lindblad superoperator that generates the dynamics, and two gapless phases with
subexponential relaxation, distinguished by the manner in which the gap decreases with system size.
Within perturbation theory, the phase boundaries in the $(\alpha_H, \alpha_L)$ plane differ for weak and strong decoherence,
suggesting phase transitions as a function of noise strength. We identify nonperturbative effects that prevent such phase transitions in
the thermodynamic limit.

\end{abstract}

%\pacs{Valid PACS appear here}% PACS, the Physics and Astronomy
                             % Classification Scheme.
%\keywords{Suggested keywords}%Use showkeys class option if keyword
                              %display desired
\maketitle
%\tableofcontents

The dynamics of generic quantum systems has been a central theme in contemporary many-body physics,
spanning disciplines from quantum information to condensed matter and high-energy physics.
A key conceptual tool in this context is random matrix theory (RMT), which
%similar to the maximum-entropy principle in statistical mechanics,
prescribes studying systems governed by dynamics that is as random
as is allowed by the symmetries and other constraints of the underlying problem of interest.
%RMT was first applied to quantum chaos in a series of works in the 1980s**.
RMT has been used over the past four decades to study quantum chaos in closed systems that lack spatial structure~\cite{Beenakker97, d2016quantum}.
Recently, various extensions of RMT that include forms of spatial structure were considered. These range
%A key recent focus has been on extending it by including various forms of spatial structures. to systems with spatial structure: many different %forms of spatial structure have been incorporated,
from banded random matrices~\cite{Mirlin-PRBM} (which represent generic local \emph{single-body} problems), to random circuits~\cite{fisher2022random}
(which represent random many-body problems with no structure beyond the spatial locality of interactions),
 and the SYK model~\cite{rosenhaus2019introduction} (in which interactions are \emph{few-body} but not otherwise local).
 Such explorations have led to a deeper understanding of quantum chaos, entanglement dynamics, and related questions.

Despite some early applications of RMT to \emph{open} quantum systems \cite{Haake-PRL89,Haake-PRA90,Chalker-JMP00,nonhermitian}, studies
of systems whose Hamiltonian and couplings to a Markovian bath are drawn from RMT ensembles have only recently appeared
\cite{xu2019extreme,Poles2018,randlind,tc,Prosen-randlind,Prosen-PRX,Luitz-hierarchy,Timm-RMT,Poles21,Presilla,Luitz-metastable,Prosen-SYK,Ryu-SYK}.
A notable conclusion that has emerged is that such fully nonlocal open systems are rapidly equilibrating,
i.e., the spectrum of their Lindblad superoperator is generically \emph{gapped} in the thermodynamic limit.
This conclusion is supported by numerical evidence, exact solutions, and general bounds \cite{randlind,tc}.
In contrast, one does not expect a gap in the opposite limit of local dissipative dynamics where the slowest-relaxing modes are long-wavelength spatial probability fluctuations, which decay through diffusion.
For many-body systems with few-body interactions, the connectivity graph is more complicated but is still local in Fock space,
hence suggesting a gapless Lindbladian, consistent with numerics \cite{Luitz-hierarchy}.
The discrepancy between the local and nonlocal regimes indicates that there must be a phase transition between them.

\begin{figure}[t!!!]
\begin{center}
\begin{tabular}{ll}
\includegraphics[width = 0.235\textwidth]{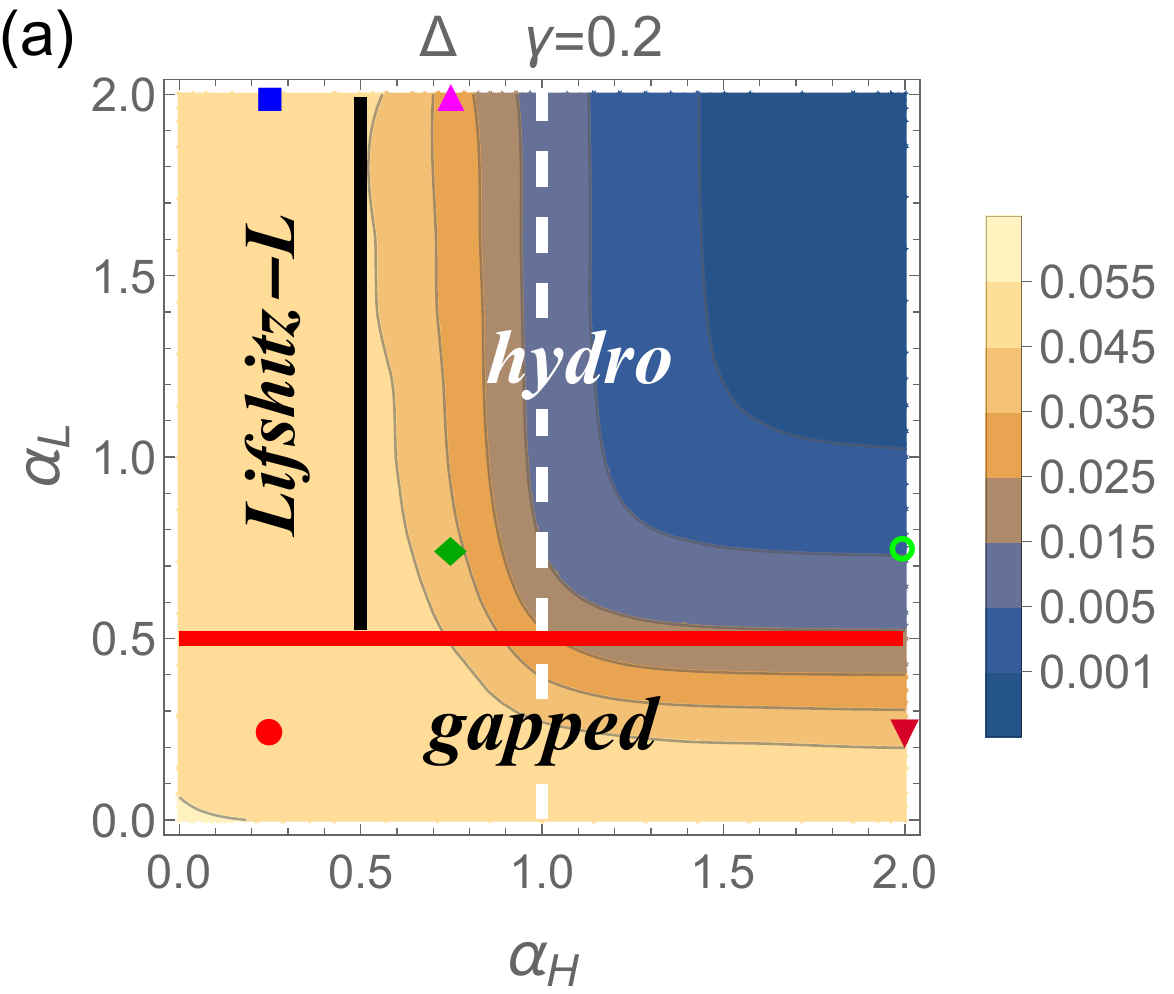} & \!\!\!
\includegraphics[width = 0.235\textwidth]{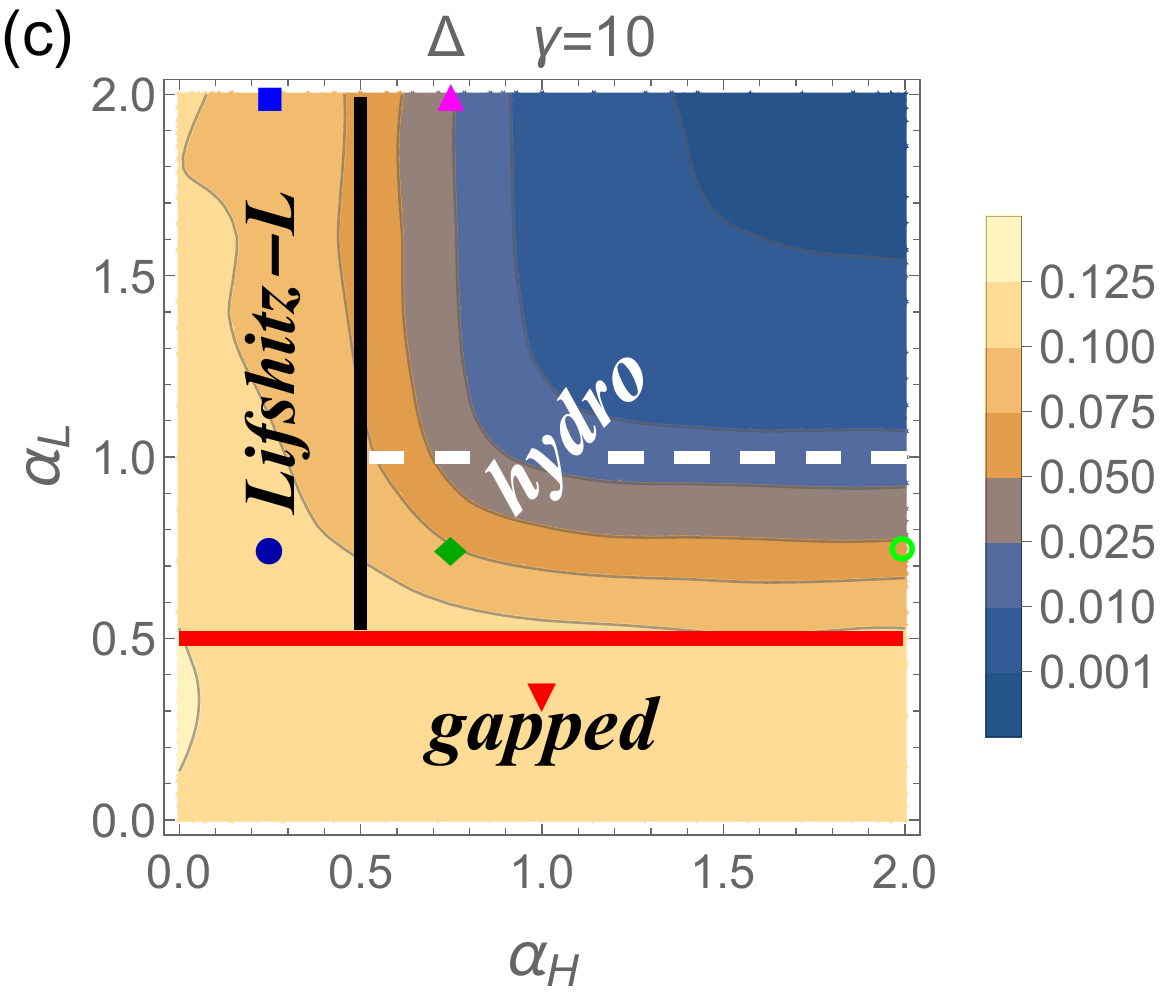} \\
\includegraphics[width = 0.23\textwidth]{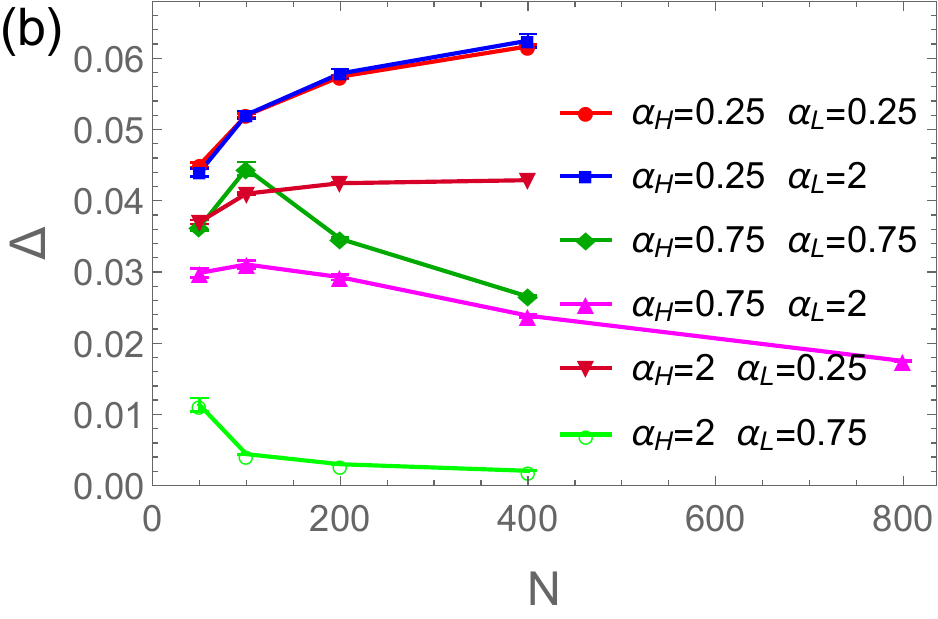} & \!\!
\includegraphics[width = 0.23\textwidth]{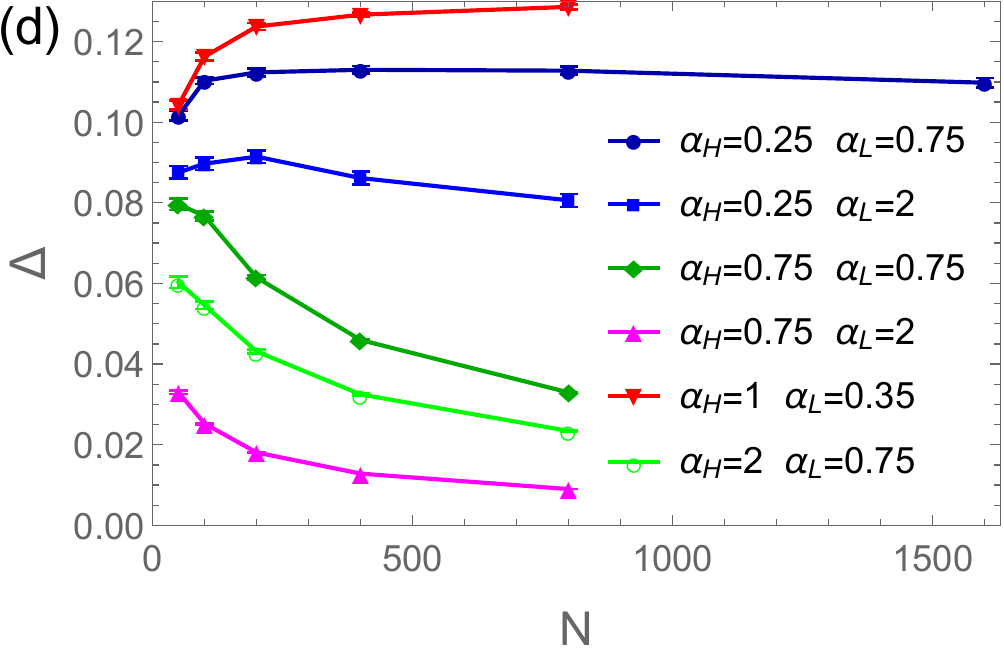}
\end{tabular}
\caption{(a) The Lindbladian spectral gap as function of the exponents $\alpha_H, \alpha_L$ at weak decoherence $\gamma=0.2$ and $N=100$.
The solid lines mark the $N\rightarrow\infty$ phase transitions between gapped, hydrodynamic, and Lifshitz phases.
The dashed line marks a change in the populations content of the slowest decaying eigenvector.
%The dashed line marks a crossover within the
%hydrodynamic and (for small $\gamma$) the gapped phases, as discussed in the main text.
(b) $N$-dependence of the gap for selected values of $(\alpha_H,\alpha_L)$ indicated
by colored symbols in (a). (c)-(d) Similar data at strong decoherence $\gamma=10$.
%For large $N$ the slowest decaying nontrivial eigenstate resides largely outside the population subspace in the region
%to the left of the dashed line (small $\gamma$) and in the region containing the green symbols (large $\gamma$).}
}
\label{fig:phasediag}
\end{center}
\end{figure}

In this work we identify such phase transitions by exploring an ensemble of master equations constructed from
power-law random banded matrices (PRBMs). PRBMs can be regarded as random hopping models in one dimension, with hopping that
falls off as a power $\alpha$ of the distance between two sites~\cite{Mirlin-PRBM, PhysRevB.62.7920, PhysRevB.61.R11859, Kravtsov-DOS}.
They interpolate between conventional random matrices in the $\alpha \rightarrow 0$ limit and short-range hopping systems with power-law
localized eigenvectors for large $\alpha$.
These models have been studied extensively in the Hamiltonian case \cite{Mirlin-PRBM}, where a localization transition occurs
at $\alpha = 1$. Here, we analyze related ensembles for \emph{open} systems, whose Hamiltonian and couplings to a Markovian noise
of strength $\gamma$ are given by $N\times N$  PRBMs with two distinct powers $\alpha_H$ and $\alpha_L$.
%(** Say something about possible realizations? **)
We find a rich phase diagram, shown in Fig. \ref{fig:phasediag}, containing three dynamical phases:
(i)~a gapped phase in which the relaxation rate remains independent of $N$,
(ii)~a ``hydrodynamic'' phase where the relaxation rate falls off as a power law of $N$ and the slowest-relaxing modes are long-wavelength fluctuations,
and (iii)~a ``Lifshitz'' phase where the relaxation rate falls off logarithmically in $N$, and the slowest-relaxing modes are localized perturbations in real space.
Notably, we find that the limits $N\rightarrow \infty$ and $\gamma\rightarrow 0$ (or $\gamma\rightarrow \infty$) do not always commute,
and finite-$N$ systems with given $(\alpha_H,\alpha_L)$ may exhibit quite different behaviors for small and large $\gamma$.
However, we show that in the $N\rightarrow \infty$ limit the weak- and strong-decoherence regimes connect smoothly and any phase transitions
as a function of $\gamma$ (apart from the appearance of mid-gap states reported for
the pure RMT case \cite{randlind}) are avoided due to nonperturbative effects.
%no phase transitions vs. $\gamma$ (beyond the one we previously reported in Ref. **); the weak- and strong-dissipation limits are smoothly connected, due to nonperturbative effects that we discuss.

%\begin{figure*}[!t]
%\begin{center}
%\includegraphics[width = 0.92\textwidth]{fig-phase-diag}
%\caption{(a)-(c): Data at weak dissipation $\gamma = 0.1$. (a)~Spectral gap of the Liouvillian as a function of the exponents $\alpha_H, \alpha_L$. The solid lines mark the phase transitions between gapped, hydrodynamic, and Lifshitz phases; the dashed line marks a crossover within the hydrodynamic phase, discussed at length in the main text. (b)~Inverse participation ratio (IPR) of the lowest nontrivial eigenvector of $\mathcal{L}$ in the basis of energy eigenstates of the Hamiltonian. (c)~IPR of the same eigenvector, but in the position basis ***need to remind me what this means***. (d)-(f)~Data at strong dissipation $\gamma = 10$: (d)~spectral gap, (e)~IPR of the lowest nontrivial eigenvector in the eigenbasis of the jump operator $L$, and (f)~IPR in the position basis. (g)-(i): System-size dependence at $\gamma = 10$ for selected values of $(\alpha_H, \alpha_L)$ marked as colored dots in panel~(d). (g)~Spectral gap, (h)~IPR in $L$-eigenbasis, (i)~IPR in the position basis.}
%\label{phasediag}
%\end{center}
%\end{figure*}

\emph{Model}.---We consider systems described by noisy dynamics of the form $H(t) = H + \xi(t) L$, where $\xi$ is a Gaussian Markovian noise with variance $\gamma$.
The Hamiltonian $H$ and the jump operator $L$ are $N \times N$ random matrices \cite{HL-note}, whose elements in the position basis are $G_{ij}f_{ij}$. Here, $G$ is a
matrix from the Gaussian orthogonal ensemble and $f_{ij}=1/(\delta_{ij}+|i-j|^{\alpha})$, where the exponent $\alpha$ generally takes different values, $\alpha_H$ and $\alpha_L$,
for $H$ and $L$. We normalize $G$ such that the variance of the spectrum of both $H$ and $L$ is $1/2$ for all $\alpha_H$,$\alpha_L$.
% (in any given realization) acting on the system. In the computational basis, $H$ and $L$ have matrix elements falling off with distance like $1/|i - j|^{\alpha_k}$ where $k = H, L$: i.e., $\alpha_H$ and $\alpha_L$ are generally different.
%We fix the normalization of $H$ and $L$ so that both have unit bandwidth (i.e., the standard deviation of the set of eigenvalues is unity) regardless of the exponents.
%
The noise-averaged dynamics is described by the Lindblad master equation
\beq
\partial_t \rho = \mathcal{L} \rho \equiv -i [H, \rho] + \gamma(L \rho L - L^2 \rho/2 - \rho L^2/2).
\eeq
%
%We will sometimes decompose $\mathcal{L} = \mathcal{L}_H + \gamma \mathcal{L}_d$, where the two terms refer to the Hamiltonian and dissipative parts of the dynamics.
%
The eigenvalues of the Lindbladian superoperator $\mathcal{L}$ occupy the complex half-plane $\mathrm{Re}(\lambda)\leq 0$ and are either real or form complex conjugated pairs \cite{randlind}.
The steady state ($\lambda=0$) of the specified model is always the maximally mixed state $\rho_0 =  \mathbb{I}/N$. The remaining right eigenvectors of $\mathcal{L}$
%i.e., $\mathcal{L}$ right eigenvectors $\rho_i$, with $i=1,\cdots ,N^2-1$,
are traceless matrices, $\rho_i$, $i=1,\cdots ,N^2-1$, that are either Hermitian or form Hermitian conjugated pairs.
A general density matrix can be expanded as $\rho(t)=\rho_0+\sum_{i=1}^{N^2-1}(a_i e^{\lambda_i t} \rho_i +{\rm H.c.})$ and its late-time approach
to $\rho_0$ is governed by the eigenvalue with the smallest negative real part, $-\Delta$, and its corresponding eigenvector $\rho_1$.
(This is always true in finite systems, but important exceptions exist in the thermodynamic limit~\cite{PhysRevLett.121.086803, PhysRevX.11.031019, Mori-PRL23}.)
As $N \rightarrow \infty$, $\Delta$ may tend to a positive value (i.e., is ``gapped'') or approach zero (``gapless''), and we compute its dependence on
$\alpha_H$, $\alpha_L$, and $\gamma$.
%For the most part we treat the case of a single noise channel, but comment on the case of multiple channels at the end.

%The Lindbladian superoperator $\mathcal{L}$ has a spectrum consisting of a steady state (with zero eigenvalue), which for the specified model is always the maximally mixed density matrix $\rho_{\mathrm{s.s.}} =  \mathbb{I}/N$.
%The remaining $N^2-1$ eigenstates are traceless matrices with eigenvalues in the complex half-plane $\mathrm{Re}( z) < 0$.
%%and $N^2 - 1$ eigenvalues in the complex half-plane $\mathrm{Re}( z) < 0$. For the class of problems we consider the steady state is always the maximally mixed density matrix, $\rho_{\mathrm{s.s.}} = N^{-1} \mathbb{I}$. The remaining eigenvectors are traceless (generally non-Hermitian) matrices.
%The late-time approach to the steady state is governed by the eigenvalue of $\mathcal{L}$ with the smallest negative real part.
%(This statement is always true at very late times in finite systems, but has important exceptions in the thermodynamic limit~\cite{PhysRevLett.121.086803, PhysRevX.11.031019}.)
% When $N \rightarrow \infty$ this eigenvalue either tends to a negative constant (i.e., is ``gapped'') or approaches zero (``gapless''), and we compute its behaviour as function of the parameters
%$\alpha_H$, $\alpha_L$, and $\gamma$.
%%
%%We compute the phase diagram as a function of the three parameters $\alpha_H, \alpha_L, \gamma$.
%For the most part we treat the case of a single noise channel, but comment on the case of multiple channels at the end.

\emph{Overview of PRBMs}.---We will invoke the spectral properties of PRBMs and thus briefly review
their properties \cite{Mirlin-PRBM,suppmat}. (i)~For $\alpha < 1/2$, PRBMs are akin to
structureless random matrices: their eigenstates are random vectors
and their eigenvalues follow a Wigner semicircle distribution. (ii)~For $1/2 < \alpha < 1$, almost all eigenstates $\ket{v}$ are extended,
as revealed by their inverse participation ratio (IPR) $I=\sum_{i=1}^N |v_i|^4$ that vanishes in the large-$N$ limit.
However, they typically exhibit sparse spatial structure spanning only a fraction of the sites. Concomitantly, the eigenvalue
distribution becomes unbound due to Gaussian tails \cite{Kravtsov-DOS} consisting of states that are localized around
potential extremes and are unable to find any resonances within the system. These tail states are subextensive in number but,
as we will show, may dominate the late-time dynamics. (iii)~For $\alpha > 1$, all eigenstates are localized with power-law decay $|v_i|\sim 1/i^\alpha$.
%\begin{figure}[t!!!]
%\begin{center}
%\includegraphics[width=0.46\textwidth]{Fig2}
%%\includegraphics[width=0.4\textwidth]{smallg-IPRX}
%\caption{(a) PRBM eigenvalue distribution for various exponents $\alpha$. We use a normalization that keeps
%the spectrum variance at $1/2$ for all $\alpha$. Gaussian tails appear for $\alpha>1/2$. (b) The average IPR
%of the eigenstates. The bands half width corresponds to the IPR standard deviation. Localization transition occurs
%at $\alpha=1$ but localized tail states exist for $\alpha>1/2$.}
%\label{fig:PRBM}
%\end{center}
%\end{figure}

\emph{Rate equations: small $\gamma$}.---We begin by discussing the limit of small or large $\gamma$ at finite $N$, where
the analysis is facilitated by the ability to perturbatively eliminate all but $N$ of the eigenvectors of $\mathcal{L}$.
As noted above, the limits $\gamma \to 0, \infty$ and $N \to \infty$ do not always commute and we will address this issue
later on.
%the analysis is facilitated by the ability to perturbatively eliminate all but $N$ of the eigenstates of $\mathcal{L}$.
%As we will find below, the limits $\gamma \to 0, \infty$ and $N \to \infty$ do not always commute,
Consider first the case $\gamma=0$. Here, the eigenvectors of $\mathcal{L}$ are
$|ij) \equiv |i\rangle \langle j|$ with eigenvalues $i(E_i - E_j)$, where $H|i\rangle=E_i|i\rangle$.
The $N$ eigenvectors of the form $|ii)$ have zero eigenvalue, i.e., are steady states.
%in the absence of decoherence any energy eigenvector is a steady state.
Following the convention in the NMR literature we dub them ``$H$-populations'' and the other
$N(N-1)$ states ``$H$-coherences''. At first order in $\gamma$ the noise does not couple populations and coherences,
and one can write down classical rate equations for the populations \cite{randlind}, $\partial_t|ii)=\sum_j A_{ij}|jj)$,
where %\cite{randlind}
\begin{equation}
\label{eq:Asmall}
A_{ij}=\gamma(|\langle i|L|j\rangle|^2-\delta_{ij}\langle i|L^2|j\rangle).
\end{equation}

When $\alpha_H < 1/2$, the eigenbasis of $H$ is effectively random, leading to rates $A_{ij}$ that are approximately
chi-squared distributed with a mean and a standard deviation that scale as $\gamma/N$. We have previously shown that such
conditions lead to a gap $\Delta=\gamma/2$ \cite{randlind}. Conversely, when $\alpha_H>1$ the $H$-eigenvectors
are localized. Analytical progress can be made by modeling them as a set of power-law envelopes centered on each of the $N$
sites (ignoring their mutual orthogonality) and by averaging $A_{ij}$ over the statistics of $L$.
Within this "mean-field" approximation $A$ is similar to a Hamiltonian whose hopping
amplitudes between sites $i, j$ vary as $|i - j|^{-2\alpha'}$, where
$\alpha' \equiv \min(\alpha_H, \alpha_L)$, and whose rows sum up to zero \cite{suppmat}.
%and the average hopping rate between sites $i, j$ scales as $|i - j|^{-2\alpha'}$, where
%$\alpha' \equiv \min(\alpha_H, \alpha_L)$ \cite{suppmat}. One can make analytical progress by employing a "mean-field"
%approximation where $A_{ij}$ is replaced by its average with respect to the $L$-randomness. The resulting rate equations
%are translationally invariant and can be diagonalized using Fourier modes \cite{suppmat}.
The mean-field analysis predicts a gap when $\alpha'<1/2$,
a superdiffusive relaxation for $1/2<\alpha'<3/2$ with a lowest eigenvalue that vanishes as $N^{1-2\alpha'}$,
and diffusive dynamics where this eigenvalue vanishes as $N^{-2}$ for $\alpha'>3/2$. Solving
the rate equations numerically yields a qualitatively similar behavior with a gapped phase for $\alpha_L<1/2$ and a gapless phase for $\alpha_L>1/2$,
albeit with a slower decay of the lowest eigenvalue with $N$ as compared to the mean-field prediction \cite{suppmat}.
For $1/2 \leq \alpha_H \leq 1$ the typical eigenstates of $H$ do not have a simple description. Our numerical results indicate
a gapped phase for $\alpha_L<1/2$ and a "weakly gapless" behavior for $\alpha_L>1/2$, where $\rho_1$ is a localized population
in the Lifshitz tail of $H$ whose eigenvalue slowly decreases with $N$ \cite{suppmat}.

\emph{Rate equations: large $\gamma$}.---A similar analysis can be carried out at large $\gamma$ \cite{randlind}. Here, one begins by diagonalizing the dissipative part of $\mathcal{L}$, finding eigenvectors
of the form $|\mu\nu) = |\mu\rangle \langle \nu|$, with eigenvalues $-(\gamma/2)(\kappa_\mu - \kappa_\nu)^2$, where $L |\mu\rangle = \kappa_\mu|\mu\rangle$. Again, there are $N$ eigenvectors with zero eigenvalue corresponding
to ``$L$-populations''. Eliminating their coupling to the remaining $N(N-1)$ ``$L$-coherences" to second-order in $H$ leads to rate equations $\partial_t|\mu\mu)=\sum_\nu A_{\mu\nu}|\nu\nu)$
with transition rates
\begin{equation}
\label{amatrix}
A_{\mu\nu}= \frac{4}{\gamma} \frac{|\langle\mu |H| \nu\rangle |^2(\kappa_\mu-\kappa_\nu)^2}{(\kappa_\mu - \kappa_\nu)^4+(2/\gamma)^2(\langle\mu|H|\mu\rangle - \langle\nu|H|\nu\rangle)^2}.
\end{equation}
Probability conservation enforces $A_{\mu\mu}=-\sum_{\nu\neq\mu}A_{\mu\nu}$. In the strict large-$\gamma$ limit at finite $N$, one would ignore the $\gamma$-dependent part of the denominator. %in Eq.~\eqref{amatrix}.
However, this term regularizes the effective dynamics for all finite $\gamma$.  % and expedites the analysis of its expected behavior.
Hence, we discuss Eq.~\eqref{amatrix} below and contrast it with the unregularized form in the supplemental material \cite{suppmat}.

%We now discuss the $(\alpha_H, \alpha_L)$ phase diagram at large dissipation.
When $\alpha_L < 1/2$, the spectrum of $L$ is \emph{bounded} with extended states,
causing $H$ to act as a featureless random perturbation between  $L$-populations. Consequently,
one can coarse-grain Eq. (\ref{amatrix}) in $\kappa$-space and replace $|\langle\mu |H| \nu\rangle |^2$ by its average
to find a gap $\Delta\simeq 2/\gamma$ \cite{randlind,suppmat}.
For $1/2 < \alpha_L < 1$, most of the $L$-eigenvectors are still delocalized. However, typical realizations of $L$ also have
spatially-localized tail states whose eigenvalues are far from the rest of the spectrum of $L$.
The matrix elements out of these tail states are suppressed according to Eq.~\eqref{amatrix}.
As a rough estimate, in a sample of size $N$ the extremal eigenvalue resides approximately $\sqrt{\log N}$
away from the bulk of the spectrum \cite{suppmat}. $\rho_1$ is localized on this extremal state,
and the gap closes logarithmically in system size.
When $\alpha_L>1$ the eigenvectors of $L$ are localized and its spectrum is unbounded. Consider the case $\alpha_L=\infty$, where
they are roughly localized on sites and the dominant dependence of $A_{\mu\nu}$ comes from $|\langle\mu |H| \nu\rangle |^2$, %the Hamiltonian matrix elements
scaling as $|\mu - \nu|^{-2\alpha_H}$.
For $\alpha_H<1/2$, these elements fluctuate sufficiently weakly that one can still coarse grain \cite{suppmat}. Since the $L$-spectrum is unbounded,
tail states set a logarithmically decaying gap. For $\alpha_H>1/2$ the effective hopping between $L$-populations is local,
leading to hydrodynamic behavior with extended eigenvectors and a gap that decays as a power-law with $N$. Numerically,
we find that this behaviour persists down to $\alpha_L=3/2$, where the gap is again set by tail states \cite{suppmat}.

\emph{Comparison of small and large $\gamma$}.---We briefly summarize our findings using the rate equations.
(a)~When $\alpha_L < 1/2$, a gapped phase is predicted for all $\gamma$.
(b)~When $\alpha_H, \alpha_L$ are both sufficiently large ($\alpha_H > 1, \alpha_L > 3/2$), a gapless phase is predicted for all $\gamma$.
(c)~Elsewhere, the rate equations for small and large $\gamma$ yield incompatible results. For $\alpha_H<1/2$, $\alpha_L>1/2$ they
suggest a gap-closing transition at finite $\gamma$, and in the remaining part of the $(\alpha_H,\alpha_L)$ plane they disagree
on the way the gap closes with increasing $N$. As we will argue, these discrepancies are absent for sufficiently large $N$.
%
%
%(c)~For other values of $(\alpha_H, \alpha_L)$,
%the rate equations for small and large $\gamma$ yield incompatible predictions, suggesting the existence of gap-closing
%phase transitions at finite $\gamma$. As we will argue, these apparent phase transitions are absent for sufficiently large $N$ where in fact the system is always weakly gapless.

\emph{Numerical investigation of $\cal L$}.---We have contrasted the above predictions against the spectrum of the full Lindbladian
(which is an $N^2 \times N^2$ matrix) for a relatively small system size $N=100$, where a fine sweep across
parameter space is feasible. We then examined larger systems of up to $N=1600$ at selected points in the $(\alpha_H,\alpha_L)$ plane.
At these sizes, we do not have access to the full spectrum of $\mathcal{L}$ but we can find the leading two eigenvalues and their
corresponding eigenvectors by the power method.
The resulting phase diagrams (Fig.~\ref{fig:phasediag}) match our expectations from the rate equations in regimes~(a) and~(b) specified above.
In regime~(c), we find behavior that lies beyond the rate equations.

%from the rate equations in the following regimes: (i)~$\alpha_L < 1/2$ and any $\alpha_H$ (gapped for all $\gamma$),
%(ii)~$\alpha_H, \alpha_L > 1/2$ (gapless for all $\gamma$).
%
%In these regimes, the small-$\gamma$, large-$\gamma$, and numerical treatments match.
%
%Elsewhere in the phase diagram,

%Broadly speaking, the resulting phase diagrams shown in Fig. \ref{fig:phasediag}
%match the expectations based on the rate equations. In particular, the small and large $\gamma$ results agree on the presence or
%absence of a gap in the following regimes: (i)~$\alpha_L < 1/2$ and any $\alpha_H$ (always gapped),
%(ii)~$\alpha_H, \alpha_L > 1/2$ (always gapless). However, for $\alpha_H < 1/2, \alpha_L > 1/2$ the small-$\gamma$ spectra indicate
%a gapped phase while the large-$\gamma$ spectra show a weakly gapless phase, thereby suggesting the existence of gap-closing
%phase transitions at finite $\gamma$. As we will argue, these apparent phase transitions, which are also predicted by the rate equations,
%are absent for sufficiently large $N$ where in fact the system is always weakly gapless.

A more sensitive probe than the gap is the nature of $\rho_1$. In the gapless regime we find
that for small $\gamma$ and $N$ it follows the prediction of the rate equations and is extended both in the position and
$H$ eigenbases as long as $\alpha_H>1$, while it is localized in both bases for $1>\alpha_H>1/2$~\cite{suppmat}.
However, as $N$ increases $\rho_1$ becomes delocalized in the entire $\alpha_H>1/2$ gapless regime.
One can characterize the failure of the rate equations by the fraction of the operator norm of $\rho_1$ that lies
in the populations subspace. This is representing how well a population-only approximation (i.e., classical rate equation)
can capture $\rho_1$. As shown by supplemental Fig.~4 \cite{suppmat} the overlap with the populations is large for $\alpha_H>1$,
but diminishes with $N$ for $\alpha_H<1$.
Intuitively, one expects such behavior if $\rho_1$ is hydrodynamic at large $N$, with a population that is modulated \emph{in real space}.
Since the eigenstates of $H$ are delocalized when $\alpha_H<1$, the projectors onto them miss the real-space structure.
By contrast, for $\alpha_H>1$ the eigenstates are localized, so local populations in energy space are a good proxy for local populations
in real space.

\begin{figure}[t!!!]
\begin{center}
\begin{tabular}{ll}
\includegraphics[width = 0.235\textwidth]{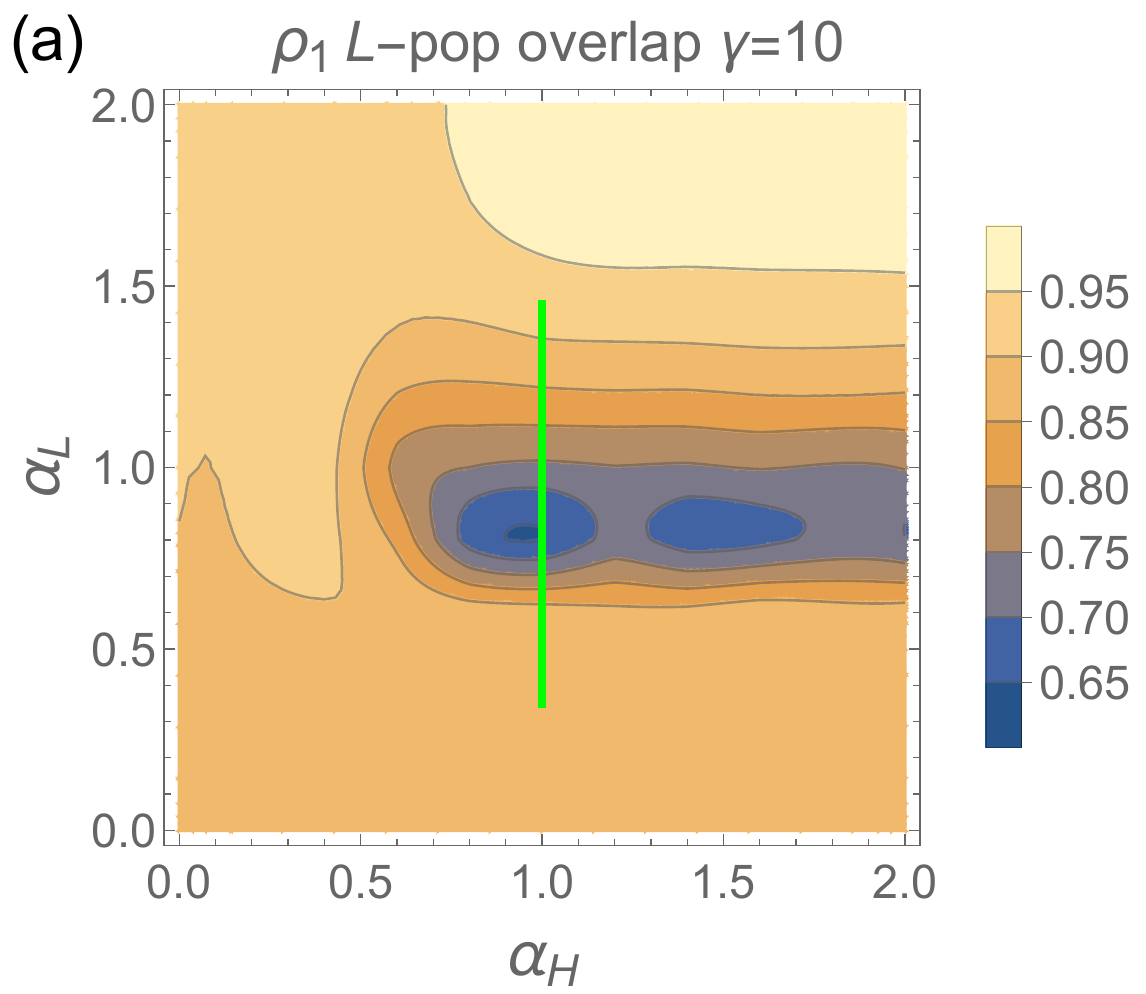} &
\includegraphics[width = 0.2361\textwidth]{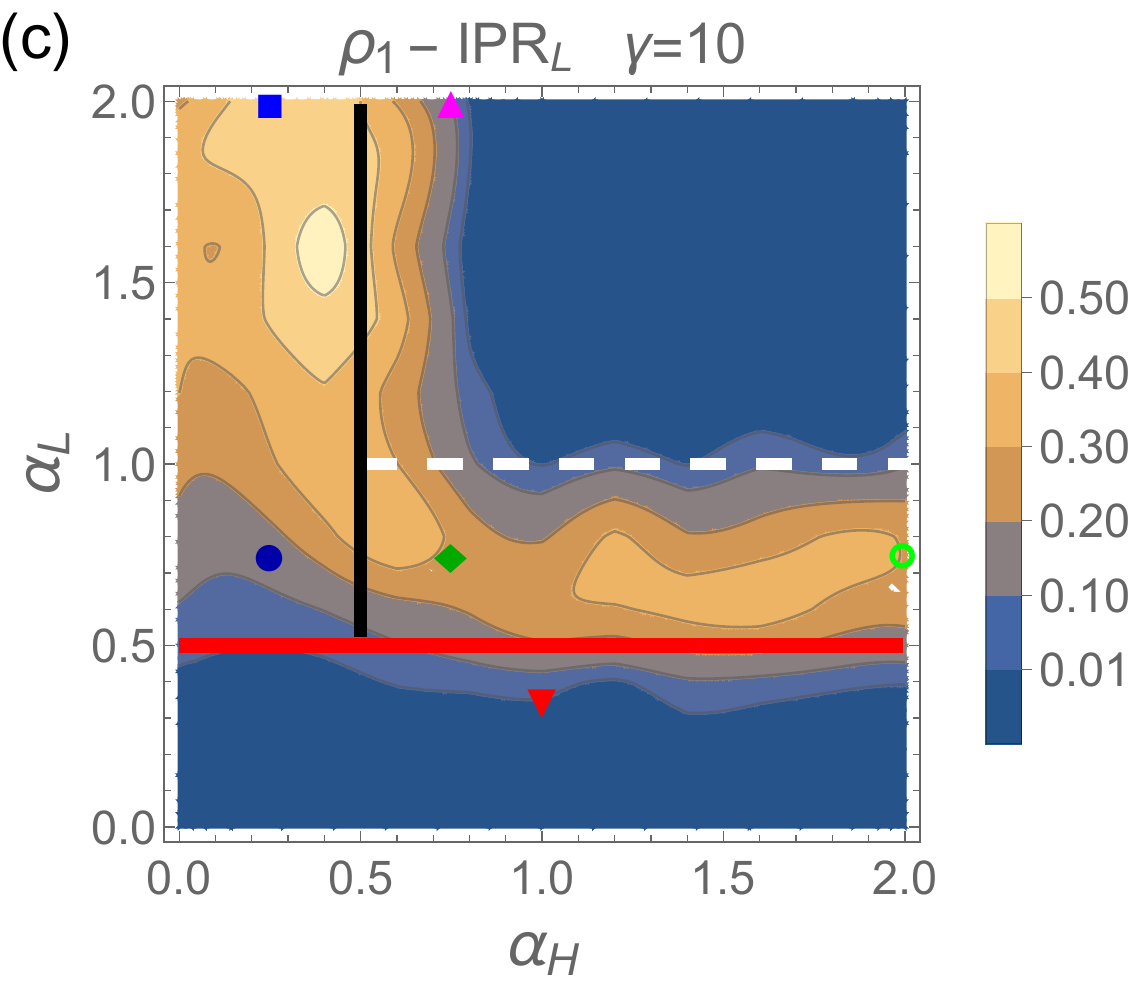} \\
\includegraphics[height = 0.228\textwidth]{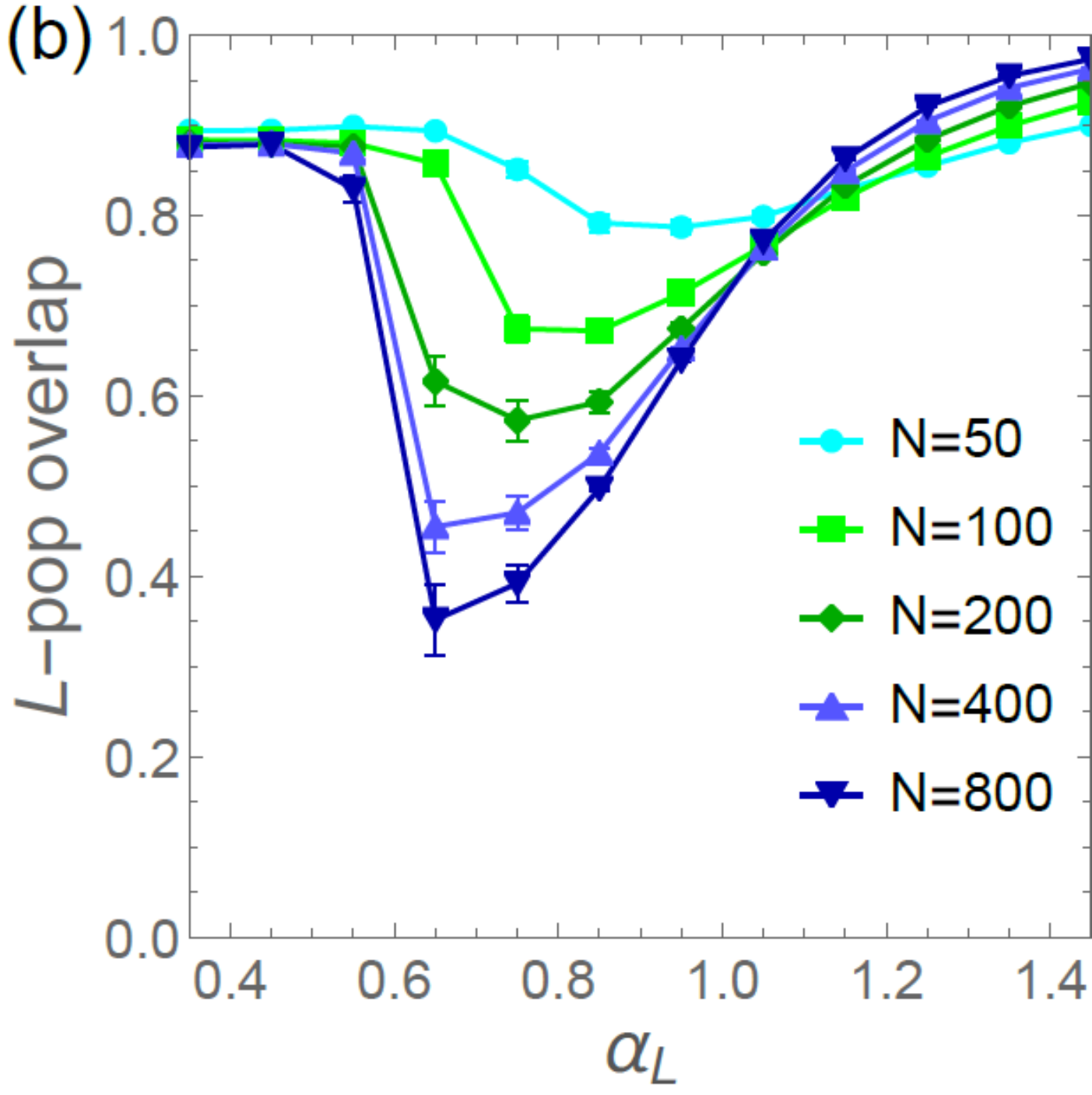} &
\includegraphics[height = 0.22\textwidth]{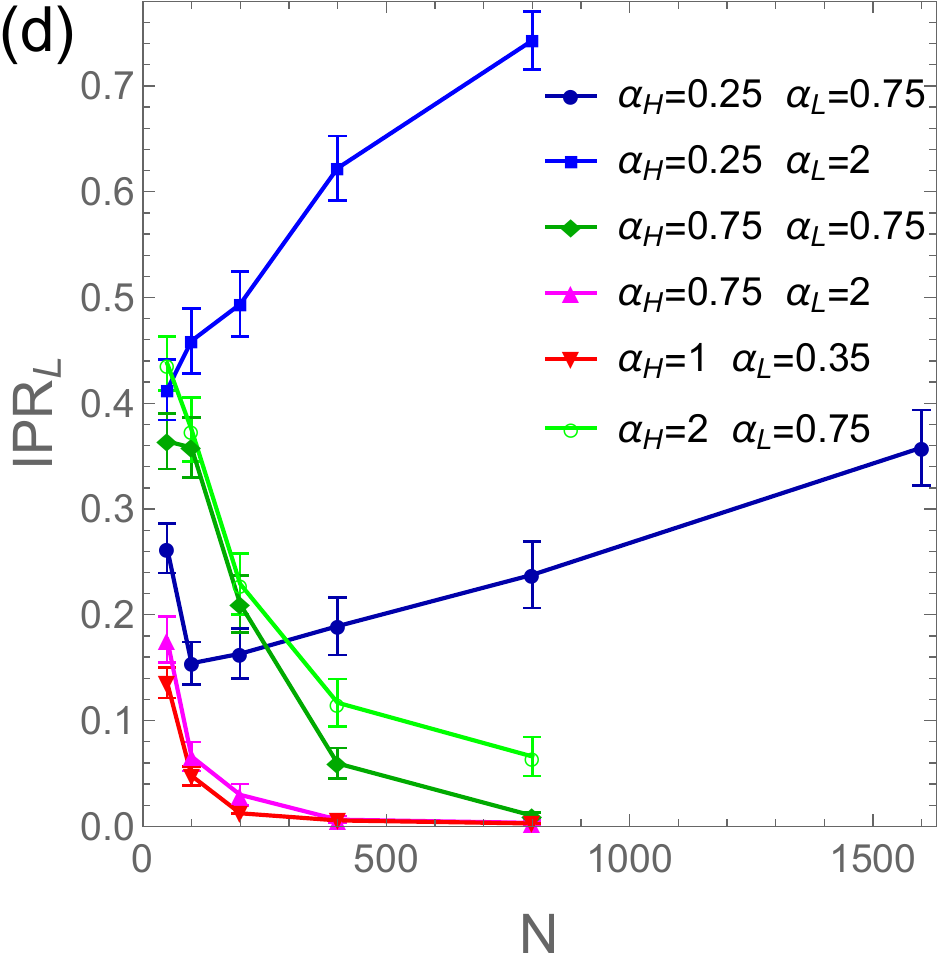}
\end{tabular}
\caption{(a) Overlap between $\rho_1$ and the $L$-populations
at strong decoherence $\gamma=10$ and $N=100$. (b) $N$-dependence of the overlap along the cut shown in (a).
$L$-coherences are essential to describe the state when $\alpha_H>1/2$ and $1>\alpha_L>1/2$.
(c) The IPR of $\rho_1$ in the $L$-eigenbasis.
(d) $N$-dependence of the IPR for the $(\alpha_H,\alpha_L)$ values indicated in (c).
$\rho_1$ is dominated by a tail $L$-population when $\alpha_H<1/2$, $\alpha_L>1/2$.
%The overlap of the state with the $L$-populations
%at strong decoherence $\gamma=10$ and $N=100$. (d) $N$-dependence of the overlap along the cut in (c).
%For large $N$ the state is dominated by coherences when $\alpha_H>1/2$ and $1>\alpha_L>1/2$.}
}
\label{fig:overlap}
\end{center}
\end{figure}
We now support this intuition by analyzing the case $\gamma \ll 1, \alpha_L = \infty, 1/2 < \alpha_H < 1$, corresponding to a system
subject to local noise and a Hamiltonian with power-law hopping and random on-site potentials. Consider a wavepacket initially localized
in real space. In the clean system, it hybridizes via coherent tunneling with states at all distance scales $R$,
with a Rabi frequency $\sim R^{-\alpha_H}$. However, local noise of strength $\gamma$ sets a timescale $\gamma^{-1}$ and a
length-scale $R_\gamma \sim \gamma^{-1/\alpha_H}$ beyond which coherent tunneling is disrupted. For $R > R_\gamma$ transport
is governed by incoherent hopping processes with a rate that is set by Fermi's Golden Rule and scales as $1/R^{2\alpha_H}$.
Since $2\alpha_H > 1$, incoherent hopping is \emph{local} in this regime and the slow modes are accordingly hydrodynamic in real space.
The eigenstates of $H$ are the wrong basis because they are formed by delicate tunneling resonances that any amount of decoherence can disrupt.
Evidently this argument extends to general $\alpha_L > 1$, and an exactly parallel argument can be made for
large $\gamma$ and $1/2 < \alpha_L < 1$.

$\rho_1$ remains delocalized for $\alpha_H>1/2$ also in the strong-decoherence thermodynamic limit.
This is apparent from Fig. \ref{fig:overlap}, showing its IPR in the $L$-population subspace
IPR$_L$=$\sum_\kappa \rho_{\kappa\kappa}^4/(\sum_\kappa \rho_{\kappa\kappa}^2)^2$, where $\rho_{\kappa\kappa}$ are its components
within this subspace. Conforming to the prediction of the rate equations, the crossover regime
$\alpha_H>1/2$, $1>\alpha_L>1/2$ exhibits an eigenvector that is still largely concentrated on a
population of a spatially-localized $L$-tail state at small $N$. However, the IPR$_L$ diminishes with $N$,
and $\rho_1$ becomes modulated in real space. Hence, for similar reasons to those outlined above
its projection onto the $L$-populations also vanishes, see Fig. \ref{fig:overlap}. In contrast,
the IPR$_L$ increases with $N$ when $\alpha_H<1/2$, $\alpha_L>1/2$.
We have confirmed that this is a result of $\rho_1$ becoming more concentrated on a population
of a localized $L$-tail state. Thus, we conclude that the range $\alpha_H<1/2$, $\alpha_L>1/2$ hosts a thermodynamic
Lifshitz phase whose gap vanishes very slowly, as shown by Fig. \ref{fig:phasediag}.
\begin{figure}[t!!!]
\begin{center}
\begin{tabular}{ll}
\includegraphics[width = 0.234\textwidth]{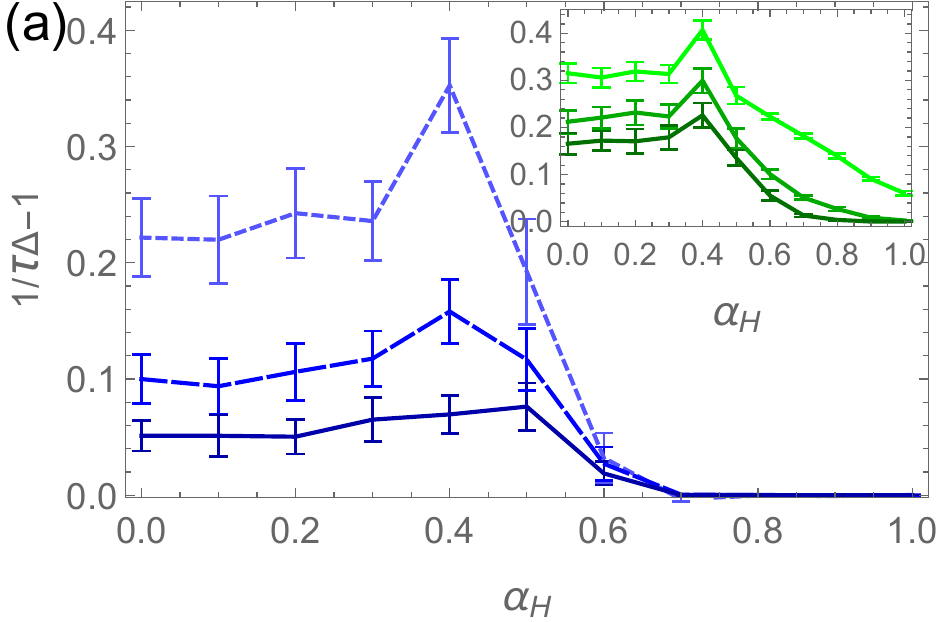} &
\includegraphics[width = 0.234\textwidth]{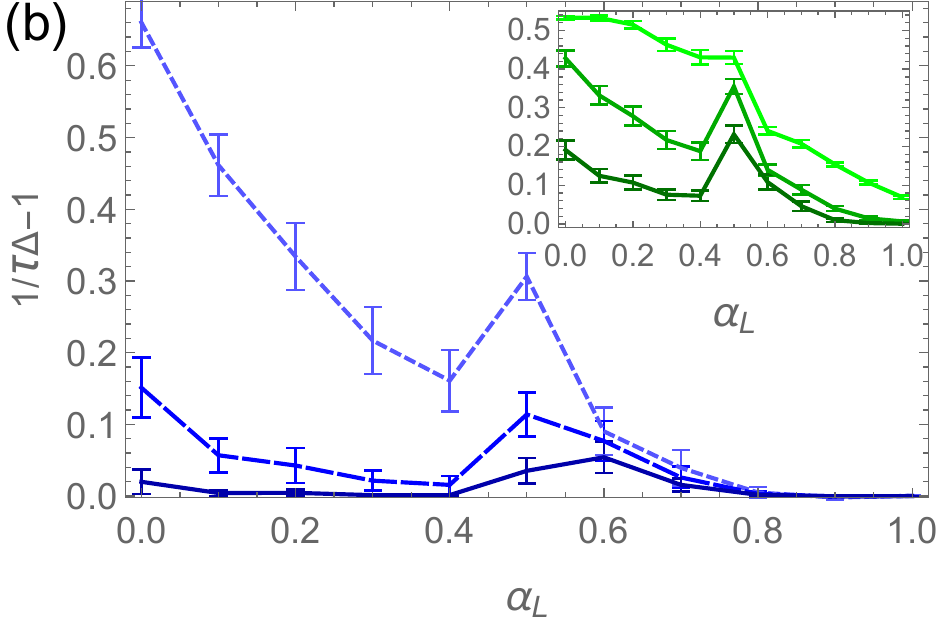}
\end{tabular}
\caption{(a) The relative difference between the spatial average of the relaxation rate $\tau^{-1}$ of local observables and $\Delta$ for
$\gamma=10$, $\alpha_L=1.5$ and $N=400$. The dotted, dashed and solid lines are based on $\tau^{-1}$ extracted by fitting the
relaxation over the range $t=3-6$, $6-9$ and $9-12\Delta^{-1}$, respectively. The inset shows the standard deviation of the relative
difference. (b) The same quantities as a function of $\alpha_L$ for $\alpha_H=1.5$.}
\label{fig:relaxation}
\end{center}
\end{figure}

%\emph{Apparent $\gamma$-dependent transitions}.---
Both the perturbative rate-equation analysis and the available numerical data
point at a transition from a small-$\gamma$ gapped phase to a large-$\gamma$ weakly gapless Lifshitz phase when $\alpha_H<1/2$
and $\alpha_L>1/2$. Nevertheless, we argue that the $N\rightarrow\infty$ spectrum in this range is weakly gapless for all $\gamma$.
The key observation is that the spectrum of $H$ is bounded whereas that of $L$ is unbounded. Hence, in the large-$N$ limit, the largest
energy scale is associated with the Lifshitz tail states of $L$ and grows as $\sqrt{\log N}$. Consequently, as $N\rightarrow\infty$
the noise cannot be treated perturbatively. Rather, the tail states must be diagonalized out first, and only then can one apply the
large-$\gamma$ perturbation theory to treat their mixing with other states via $H$. The resulting gap diminishes
as $1/(\gamma\log N)$ but is challenging to detect: since the fixed-$N$, $\gamma\rightarrow 0$ perturbation theory yields a gap
of order $\gamma$ the tail-state eigenvector extends below it only when $N>\exp(1/\gamma^2)$. For small $\gamma$ this regime is numerically inaccessible.
Instead, the supplemental material demonstrates small-$\gamma$ Lifshitz behavior using a model whose density of
$L$-eigenvalues decays only as $\kappa^{-4}$.

\emph{Discussion}.---Our work focused on the spectral gap $\Delta$. To make contact with the dynamics of local observables
we have followed the evolution of an initial state with $\rho_{ij}=(\delta_{ij}-\delta_{i1}\delta_{j1})/(N-1)$. We observe an asymptotic
exponential approach of every $\rho_{ii}$ to the steady state value $1/N$. The relaxation time is $\Delta^{-1}$ at all sites $i$,
but the onset time of the asymptotic approach varies with $i$ and depends on the overlap $(\rho_1)_{ii}$ with the slowest mode \cite{suppmat}.
At shorter times, the relaxation is faster, due to more rapidly decaying eigenstates. These points are demonstrated by Fig. \ref{fig:relaxation}
and the supplemental material \cite{suppmat}. In terms of the natural scale $\Delta^{-1}$ the asymptotic approach begins earliest in the
hydrodynamic phase, then in the gapped phase and finally in the Lifshitz phase, where most sites have only algebraically small overlap with $\rho_1$.

Often, when classical noise controls the experiment, it couples to a single collective variable, e.g.,
the dipole moment of a chaotic quantum dot. Although we focused on this case, a more general setting
involves multiple decoherence channels with their associated jump operators. %corresponding to many different jump operators.
In the supplemental material we extend our treatment to systems with several PRBM jump operators with exponents $\alpha_{L_k}$ \cite{suppmat}.
Let us briefly quote the results. When $\tilde\alpha_L=\min(\alpha_{L_k}) < 1/2$, the spectrum is gapped, otherwise it is gapless.
A weakly gapless Lifshitz phase occurs when $\tilde\alpha_L>1/2$ and $\alpha_H < 1/2$.
Finally, when all exponents exceed $1/2$ we predict a hydrodynamic regime.
%---In this paper we explored the spectral gaps of random Lindblad equations with varying levels of locality, tuned via the power-law falloff of the Hamiltonian and the decoherence couplings.
%Although we focused on the case of a single decoherence channel the extension to multiple decoherence channels with exponents $\alpha_L^{(i)}$ is straightforward.
%Let us briefly quote the results. When $\alpha_L^{(i)} < 1/2$ for some $i$, the spectrum is gapped; otherwise it is gapless. A weakly gapless phase occurs when all the $\alpha_L^{(i)} > 1/2$ but $\alpha_H < 1/2$.
%Finally, when all exponents exceed $1/2$ we predict a hydrodynamic regime.

Our analysis found three distinct phases as a function of the decay exponents $(\alpha_H, \alpha_L)$,
but no phase transitions as a function of the decoherence strength $\gamma$. Our analysis is consistent
with the possibility of transitions between gapped phases, as in Ref.~\cite{randlind};
indeed, we expect such transitions everywhere in the gapped phase $\alpha_L < 1/2$.

A natural question is whether the transitions we find exhibit nontrivial critical phenomena.
While we have not addressed these in detail, our results shed some light on the matter.
The transition from gapped to hydrodynamic relaxation as one tunes $\alpha_L$ at fixed $\alpha_H \gg 1$ and small $\gamma$
is a transition purely in the decay rates of the hydrodynamic modes: the low-lying eigenvectors themselves evolve smoothly with $\alpha_L$, and show no signs of a diverging length scale. The extended modes do change across the same transition at large $\gamma$,
and further study is required. The $\alpha_H$-tuned transition from Lifshitz to hydrodynamic relaxation at fixed $\alpha_L \gg 1$
appears rather simple: it is a level crossing between the localized Lifshitz tail state and the hydrodynamic mode,
and as such shares some similarities with other spectral ``first-order'' transitions~\cite{PhysRevA.83.051603}.
Finally, the transition between gapped and Lifshitz relaxation at $\alpha_H < 1/2$ as one tunes $\alpha_L$
through $1/2$ is a nontrivial critical point, associated with the emergence of tails in the density of states of PRBMs~\cite{Mirlin-PRBM}.
This transition is a particularly promising candidate for experimental studies in ion traps,
which allow to realize power-law couplings with tunable exponents~\cite{richerme2014non}.

%(Whether transitions between two gapped phases exist, as in Ref.~\cite{randlind}, our analysis does not address.) Experiments in ion traps allow one to tune exponents continuously, so it is realistic to tune across such dynamical phase transitions and probe the associated critical exponents.

% Within perturbation theory such phase transitions appear to exist, but we have identified nonperturbative effects that lead the weak and strong dissipation phases to be smoothly connected for any $(\alpha_H, \alpha_L)$, \emph{except} in the gapped phase where $\alpha_L < 0.5$. (There a phase transition was previously found in Ref. ** between two gapped phases.)  We leave this to future work. It would also be interesting to support the numerical and intuitive arguments of the present work with exact calculations for these ensembles, e.g., using free probability theory.

\begin{acknowledgments}
We acknowledge support by the United States-Israel Binational Science Foundation (Grant No. 2018159).
The Flatiron Institute is a division of the Simons Foundation.
\end{acknowledgments}

\bibliography{PRBMlind}

\begin{thebibliography}{32}
\bibitem [{HL-()}]{HL-note}%
  \BibitemOpen
  \href@noop {} {}\bibinfo {note} {We draw the matrices representing $H$ and
  $L$ in an uncorrelated way. Hence, they do not commute with probability 1. In
  case they commute, the Lindbladian ${\cal L}$ possesses $N$ steady states
  given by the populations $|i\rangle\langle i|$, where $|i\rangle$ are the
  simultaneous eigenstates of $H$ and $L$}\BibitemShut {NoStop}%
\end{thebibliography}

\end{document}

\title{Supplemental Material for: "Dynamical transition from slow to fast relaxation in random open quantum systems"}
\author{Dror Orgad$^1$, Vadim Oganesyan$^{2,3}$, and Sarang Gopalakrishnan$^{4}$}
\affiliation{$^1$Racah Institute of Physics, The Hebrew University, Jerusalem 91904, Israel \\
$^2$Department of Physics and Astronomy, College of Staten Island, CUNY, Staten Island, NY 10314, USA \\
$^3$Center for Computational Quantum Physics, Flatiron Institute, 162 5th Avenue, New York, NY 10010, USA \\
%$^4$Department of Physics, The Pennsylvania State University, University Park, PA 16802, USA \\
\mbox{$^4$Department of Electrical and Computer Engineering, Princeton University, Princeton, NJ 08540, USA}}

\maketitle
\vspace{-1.2cm}

\tableofcontents
%In the following, we provide details about the effective description of the populations subspace at weak and strong decoherence.
%Specifically, we compare the spectral gap, the population DOS and the nature of the slowest decaying mode as calculated from the
%full Lindbladian and from the effective populations Lindbladian. This allows us to assess the range in which the approximation holds. We
%then present numerical results for large systems based on the effective population dynamics, with emphasis on the finite-size
%scaling of the gap together with analytical analysis of the limit of weak decoherence and strongly localized energy eigenstates.

%approximation-based numerical results for large systems, with emphasis on the finite-size scaling of the gap.
%Finally, we analyze analytically the limit of weak decoherence and strongly localized energy eigenstates.
%present numerical results on the finite size scaling of the spectral gap deduced from the effective description
%at the two limits, and compare it to the gap calculated from the full Liouvillian to assess the range at which the approximation holds.
%We also compare the population DOS and the nature of the slowest decaying mode as obtained from the exact and approximate calculations.
%\vspace{-0.1cm}
\section{Power-Law Random Banded Matrices}

%\vspace{-0.2cm}
Consider an $N\times N$ random matrix, $G$, belonging to one of the Gaussian ensembles. From it, one can derive
a power-law random banded matrix (PRBM), $H$, whose elements are defined by
\begin{equation}
\label{eq:PRBM-def}
H_{ij}=f_{ij}G_{ij}, \;\;\;\;\; f_{ij}=f_{|i-j|}=\frac{1}{\delta_{ij}+|i-j|^\alpha}.
\end{equation}
We dub the basis for which Eq. (\ref{eq:PRBM-def}) holds the $x$-basis.
For a system with periodic boundary conditions the appropriate variant is the circulant power-law random banded matrix (CPRBM),
for which
\begin{equation}
\label{eq:PPRBM-def}
f_{ij}=\frac{1}{\delta_{ij}+(N/2-|N/2-|i-j||)^\alpha}.
\end{equation}

%\smallskip
Varying the exponent $\alpha$ changes the scaling of the spread of the spectrum with $N$. In order to avoid this effect we
consider in the following PRBM (and CPRBM) ensembles which are derived from matrices $G$ drawn from the GOE
\begin{equation}
\label{eq:GOE}
P(G)\propto \exp\left[-\frac{a(N,\alpha)}{2}\Tr\left(G^2\right)\right],
\end{equation}
with
\begin{equation}
\label{eq:adef}
a(N,\alpha)=1+\sum_{j=1}^{N-1} f_{N/2,j}^2=2+2h\left(\frac{N}{2}-1,2\alpha\right)\underset{N\rightarrow\infty}{\longrightarrow}
2+2\left[\frac{(N/2)^{1-2\alpha}}{1-2\alpha} + \zeta(2\alpha)\right].
\end{equation}
Here $h(N,\alpha)=\sum_{n=1}^N n^{-\alpha}$ is the harmonic number of order $\alpha$ and $\zeta(\alpha)$ is the zeta function.
Consequently,
%\begin{equation}
%\label{eq:hlargeN}
%h(N,\alpha)\underset{N\rightarrow\infty}{\longrightarrow}\frac{N^{1-\alpha}}{1-\alpha} + \zeta(\alpha),
%\end{equation}
%with $\zeta(\alpha)$ the zeta function.
%Consequently,
\begin{equation}
\label{eq:Hvar}
\langle H_{ij} H_{kl} \rangle = \frac{f_{ij}^2}{2a(N,\alpha)} (\delta_{ik}\delta_{jl} + \delta_{il}\delta_{jk}),
\end{equation}
thus implying that the variance of the sum of elements in a (middle) row of $H$ is 1/2 (up to $1/N$ corrections), irrespective of $\alpha$.
This in turn leads to a spectrum whose standard deviation is $1/\sqrt{2}$.

%\medskip

\begin{figure}[!!!h]
\begin{center}
\includegraphics[width=0.64\textwidth]{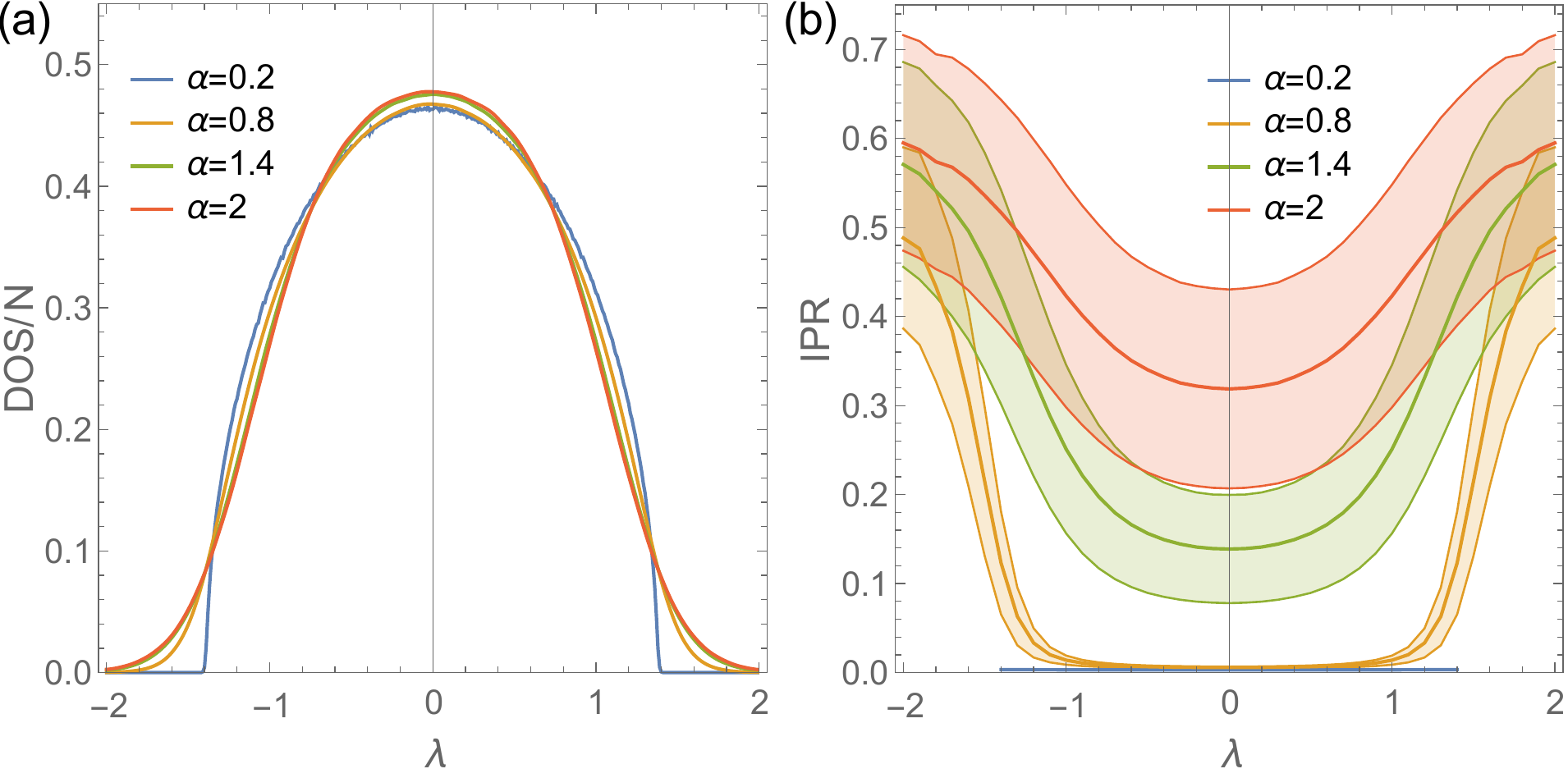}
\vspace{-0.25cm} \caption{(a) PRBM eigenvalue distribution for various exponents $\alpha$. We use a normalization that keeps
the spectrum variance at $1/2$ for all $\alpha$. Gaussian tails appear for $\alpha>1/2$. (b) The average IPR
of the eigenstates. The bands half width corresponds to the IPR standard deviation. Localization transition occurs
at $\alpha=1$ but localized tail states exist for $\alpha>1/2$.}
\label{fig-PRBM}
\end{center}
\end{figure}

\begin{figure}[!!!h]
\begin{center}
\includegraphics[width = 0.49\textwidth]{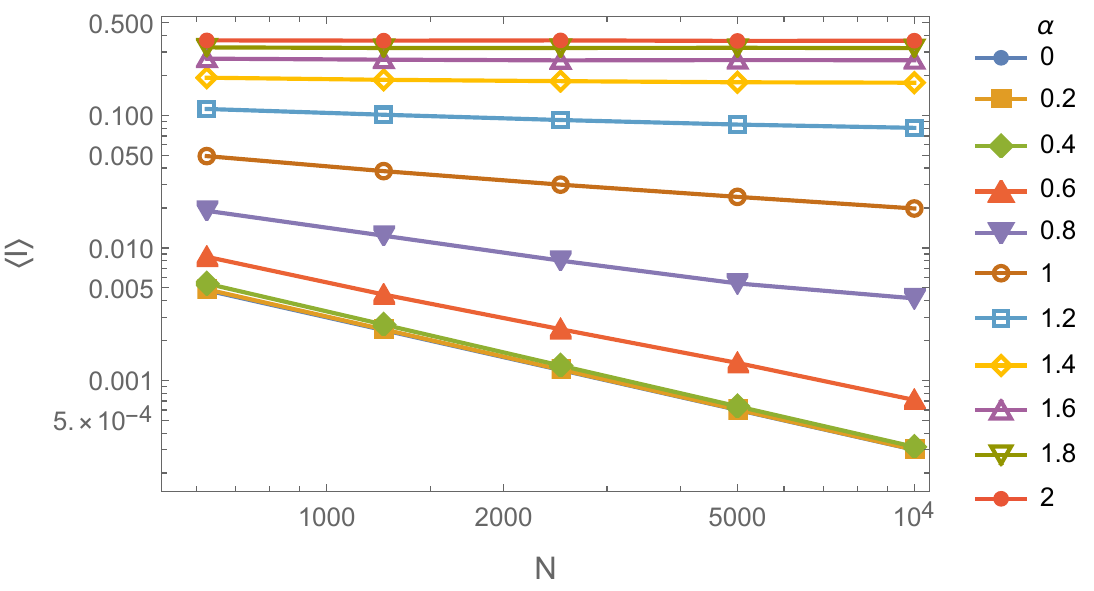}
\hspace{3mm}
\includegraphics[width = 0.47\textwidth]{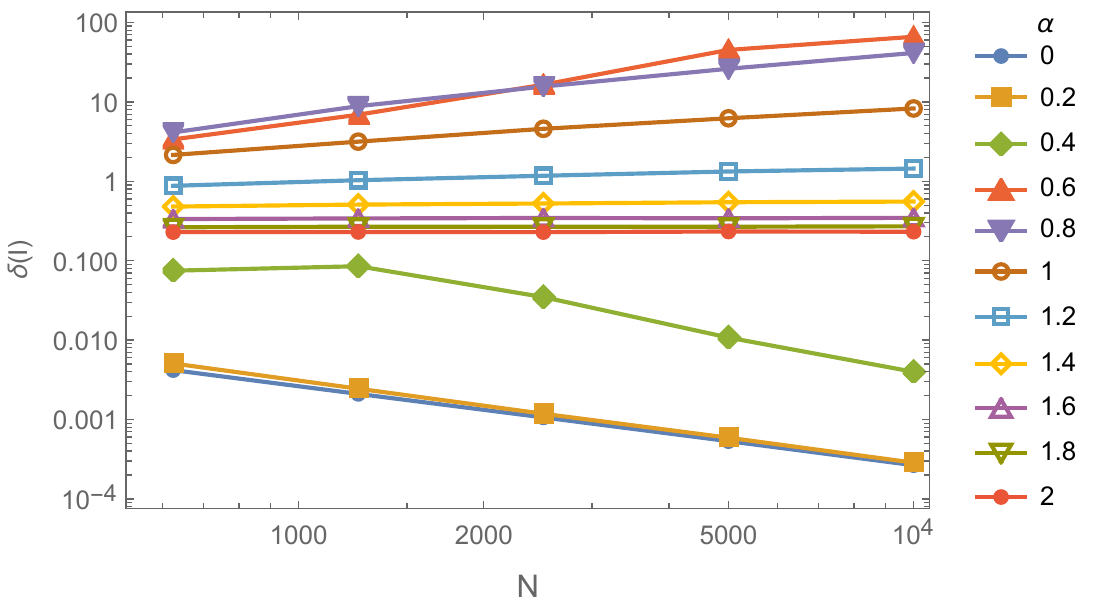}
\end{center}
\vspace{-0.5cm} \caption[]{\small $N$-dependence of the average inverse participation ratio and its relative variance.
For $\alpha<1/2$ both quantities scale as $1/N$. The results shown are for CPRBM but change little for PRBM.}
\label{fig-IPR}
\end{figure}

The spectral properties of the GOE-derived PRBM ensemble were studied by Mirlin \etal~using a non-linear sigma
model (NLSM).\cite{Mirlin-PRBM} For $\alpha<1/2$ all statistical properties are predicted to coincide with those of the GOE.
In particular, the density of states (DOS) is given by Wigner's semicircle law and the eigenstates, $v$, are delocalized, as revealed
by the behavior of their inverse participation ratio (IPR) $I=\sum_{i=1}^N |v_i|^4$, whose mean scales as $\langle I\rangle\propto 1/N$
and whose relative variance follows $\delta(I)=(\langle I^2\rangle- \langle I\rangle^2)/\langle I\rangle^2\propto 1/N$.
Such behavior is confirmed numerically, see Figs. \ref{fig-PRBM}, \ref{fig-IPR}.
\vspace{0.3cm}
%\begin{figure}[!!!h]
%\begin{center}
%\includegraphics[width = 0.9\textwidth]{CPRBM-DOS-IPR}
%\end{center}
%\vspace{-0.5cm} \caption[]{\small The density of states and inverse participation ratio of GOE CPRBMs with $N=1000$ and representative
%values of $\alpha$.}
%\label{fig-CPRBM}
%\end{figure}
%\medskip

\begin{figure}[!!!h]
\begin{center}
\includegraphics[width = 0.43\textwidth]{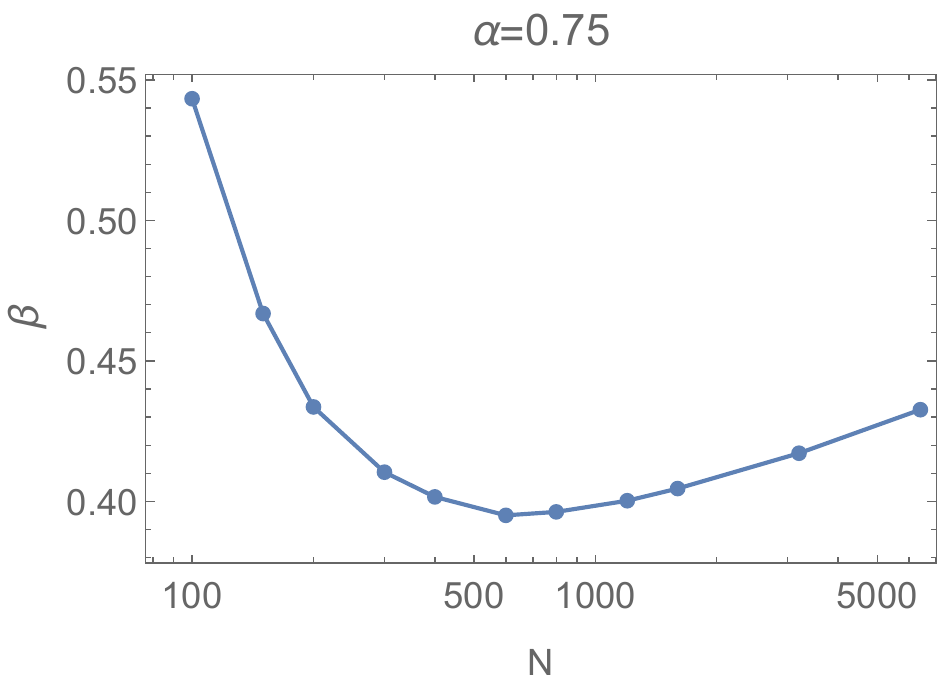}
\hspace{3mm}
\includegraphics[width = 0.444\textwidth]{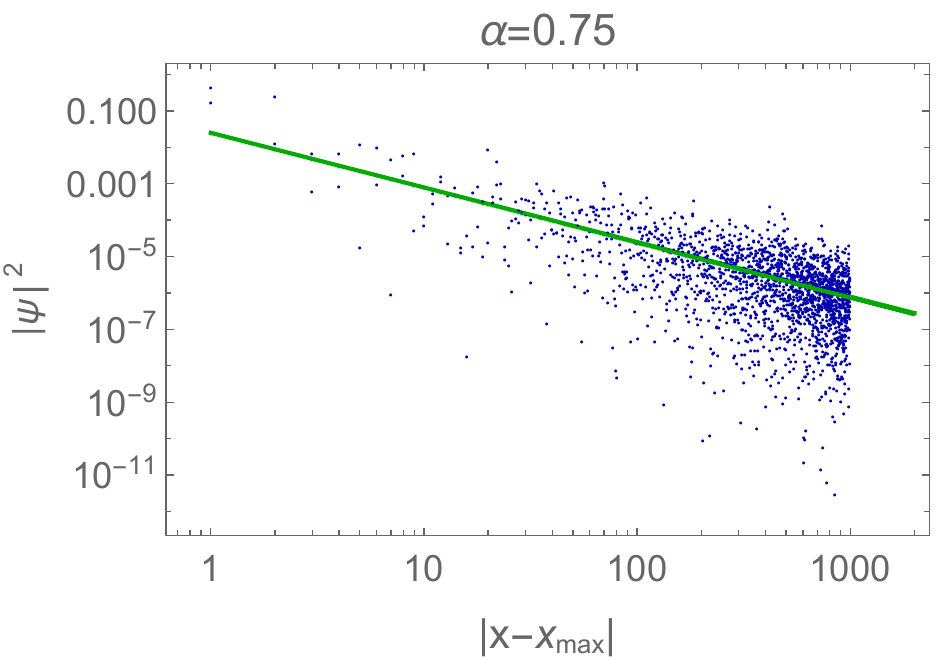}
\end{center}
\vspace{-0.5cm} \caption[]{\small Left: Fitting $n_{\rm loc}$, defined as the average number of states with IPR$>0.1$, to a
power law $n_{\rm loc}=N^\beta$. Right: The decay of a localized tail state as a function of the distance from its peak.
The solid line depicts a $|x-x_{\rm max}|^{-1.5}$ decay.}
\label{fig-loc}
\end{figure}

On the other hand, for $\alpha>1/2$ the DOS deviates from the semicircle and develops Gaussian tails \cite{Kravtsov-DOS}
(we observe similar tails also when the entries of $G$ are box-distributed instead of normal).
In the range $1>\alpha>1/2$ the mean IPR continues to decay with $N$. However, it does so more
slowly than $1/N$, which is the NLSM prediction for $\alpha<1$.
The IPR fluctuations in this range are also stronger than the NLSM predictions, and numerically we find that
$\delta(I)$ increases in the available range of $N$.
We attribute this behaviour to the appearance of localized states
at the tails of the spectrum, as depicted in Fig. \ref{fig-PRBM}. These states decay algebraically $|\psi(x)|\sim |x|^{-\alpha}$
and their number, $n_{\rm loc}$, increases in a sublinear manner with $N$. Results of fitting $n_{\rm loc}$ to a power law
in $N$ together with a typical decay of a localized state are shown in Fig. \ref{fig-loc}. Finally, for $\alpha>1$, $O(N)$
eigenstates are power-law ($|x|^{-\alpha}$) localized leading to $\langle I \rangle>0$.

\section{Effective dynamics of populations}

\subsection{Effective dynamics in the limits of weak and strong decoherence}

\begin{figure}[!!!h]
\begin{center}
\hspace{-4mm}
\includegraphics[height = 0.343\textwidth]{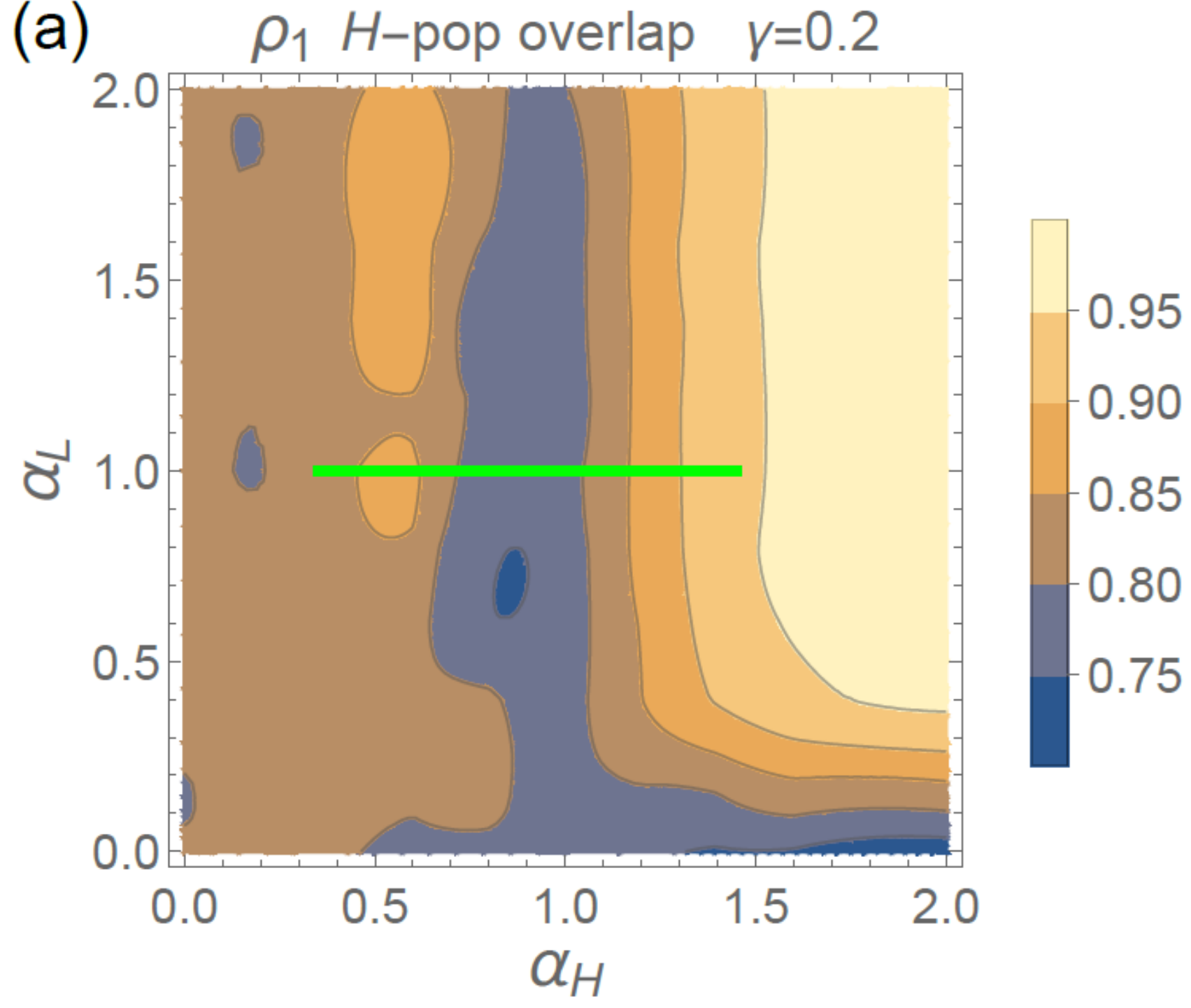}
\hspace{3mm}
\includegraphics[height = 0.348\textwidth]{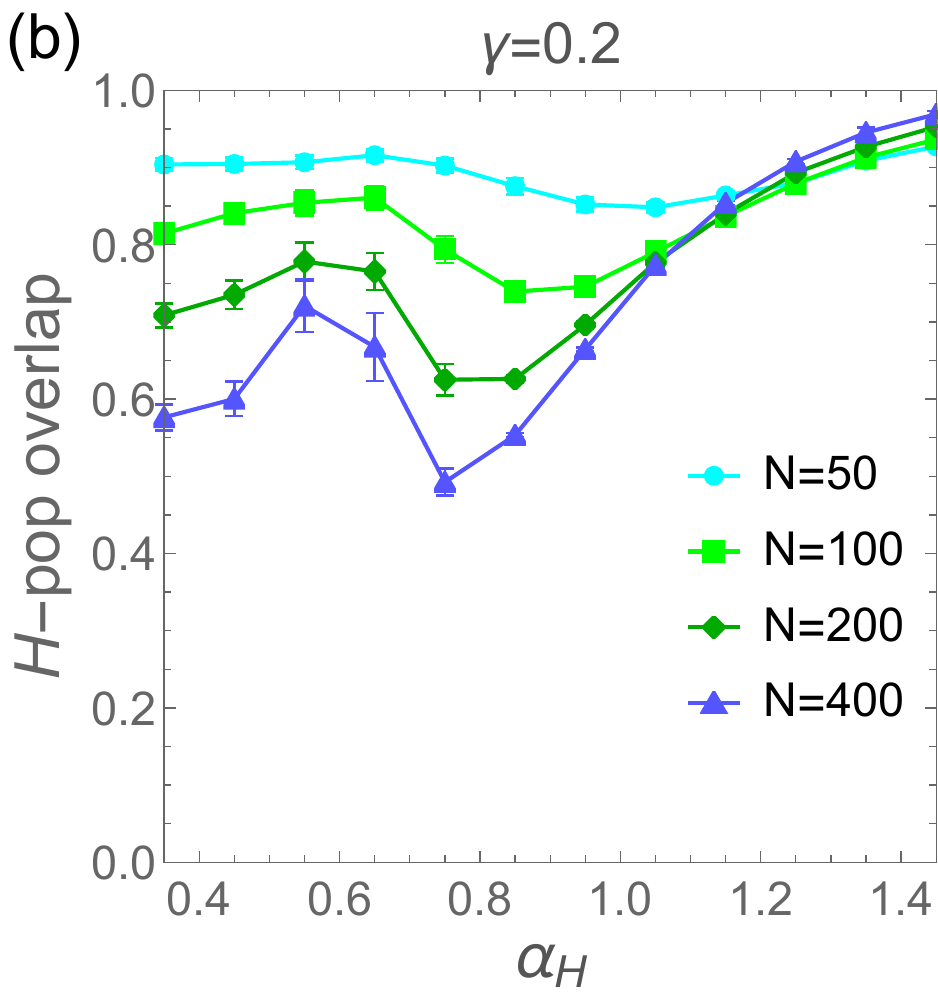}
\end{center}
\vspace{-0.5cm} \caption[]{\small (a) The average overlap (the norm of the projected vector) between the slowest decaying mode $\rho_1$
and the subspace of $H$-populations at weak decoherence $\gamma=0.2$ and $N=100$. (b) $N$-dependence of the overlap along the cut shown in (a).
%$H$-coherences are essential to describe the state when $\alpha_H<1$.
%Left: The average overlap (the norm of the projected vector) between the slowest decaying mode
%of the Lindbladian and the $H$-populations subspace for a system with $\gamma=0.1$ and $N=100$. Right: The overlap with the
%$L$-populations subspace for $\gamma=10$ and $N=100$.}
}
\label{fig-overlap}
\end{figure}

In the limits of vanishing and infinitely strong decoherence the Lindbladian spectrum contains $N$ zero modes given by the populations
of Hamiltonian eigenstates ($H$-populations) and of $L$-eigenstates ($L$-populations), respectively. This degeneracy is lifted away from the strict
$\gamma=0,\infty$ limits with only the infinite temperature thermal state $\rho_{ss}={\mathbbm{1}}/N$ remaining
as a steady state for arbitrary $\gamma$. Nevertheless, one expects that at weak decoherence
the slowest decaying mode consists primarily of $H$-populations, while for strong decoherence it is largely composed of $L$-populations.
This expectation is confirmed numerically, as shown by Fig. 2 of the main text and by supplemental Fig. \ref{fig-overlap}. The figures also
show that the strongest mixing of coherences into the state occurs for $\alpha_H<1$ at weak decoherence and for $\alpha_H>1/2$ and $1>\alpha_L>1/2$
at strong decoherence. The mixing increases with $N$.

More specifically, by considering the limits $\gamma\rightarrow 0,\infty$  for fixed $N$ one may obtain an asymptotically exact description of
the low-lying $\L$-eigenstates in terms of the populations alone \cite{ours}. We begin by considering weak decoherence and express $\L$ in the
eigenbasis of $H$, where $H_{ij}=\epsilon_i\delta_{ij}$. For the case of interest, where $H$ and $L$ are real and symmetric, the effective description
is given by \cite{ours}
\begin{equation}
\label{eq:Asmall}
A_{ij}=-\gamma\left[{L^2}_{ij}\delta_{ij}-(L_{ij})^2\right].
\end{equation}

Conversely, in the limit of strong decoherence we express $\L$ in the eigenbasis of $L$, where $L_{ij}=\kappa_i\delta_{ij}$.
The effective description within the population subspace is governed by the matrix \cite{ours}
\begin{eqnarray}
\label{eq:Alarge}
\nonumber
A_{ij}&=&\frac{4}{\gamma}\frac{H_{ij}^2(\kappa_i-\kappa_j)^2}{(\kappa_i-\kappa_j)^4+(2/\gamma)^2(H_{ii}-H_{jj})^2}
\;\;\;{\rm for}\;\; i\neq j, \\
A_{ii}&=&-\sum_{j\neq i} A_{ij}.
\end{eqnarray}
Physically, Eq. (\ref{eq:Alarge}) describes Hamiltonian-induced hopping between $L$-population via $L$-coherences.
Despite being higher order in $1/\gamma$ we have kept (in the second term of the denominator) the $H$-induced shift
in the eigenvalues of the coherences. As we show below, it may induce delocalization of the slowest decaying mode in the large
$\alpha_H$,$\alpha_L$ regime when $\gamma$ is fixed and $N\rightarrow\infty$.

\subsection{Slowest decaying mode of $A$ for $\gamma\rightarrow 0,\infty$}

\begin{figure}[!!!h]
\begin{center}
\includegraphics[width = \textwidth]{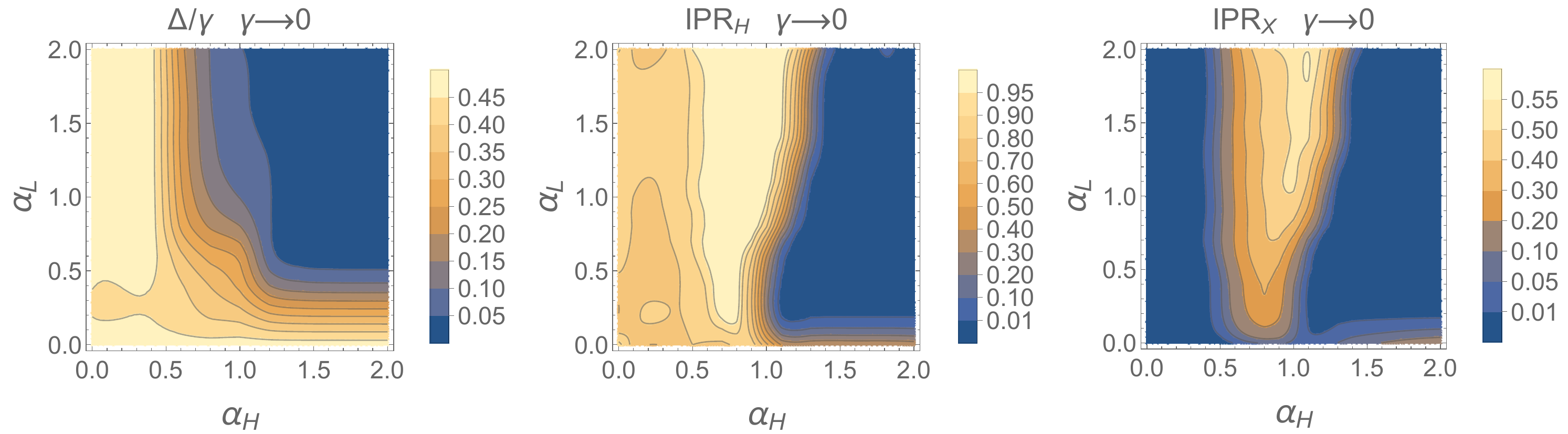}
\end{center}
\vspace{-0.5cm} \caption[]{\small Left: The spectral gap of $A$ in the limit of vanishing decoherence for an $N=12800$ system
whose Hamiltonian and jump operator are PRBMs with exponents $\alpha_H$ and $\alpha_L$, respectively.
Center and right: The inverse participation ratios in energy space and in real space of the corresponding slowest decaying mode
for a system with $N=6400$. The data was averaged over 100 realizations for each $(\alpha_H,\alpha_L)$ point.}
\label{fig-Asmallg}
\end{figure}

Here we present data obtained by diagonalizing $A$ in the limits $\gamma\rightarrow 0,\infty$ for the largest systems
that we considered. The left panel of Fig. \ref{fig-Asmallg} shows the spectral gap for vanishing $\gamma$
while the other two panels concern the corresponding slowest decaying mode and depict its inverse participation ratio
in energy space, IPR$_H$, and in position space, IPR$_X$. The latter are defined via the expansion of the mode in the energy basis
$\rho^{(1)}=\sum_{i=1}^N \rho_i |i\rangle\langle i|$ as
\begin{eqnarray}
\label{eq:IPRHdef}
&&{\rm IPR}_H=\sum_{i=1}^N \rho_i^4, \\
\label{eq:IPRXdef}
&&{\rm IPR}_X=\sum_{i=1}^N \rho_i^4\sum_{x=1}^N |\langle x|i\rangle|^4,
\end{eqnarray}
where $|x\rangle$ is the $x$-basis.

\begin{figure}[!!!t]
\begin{center}
\includegraphics[width = \textwidth]{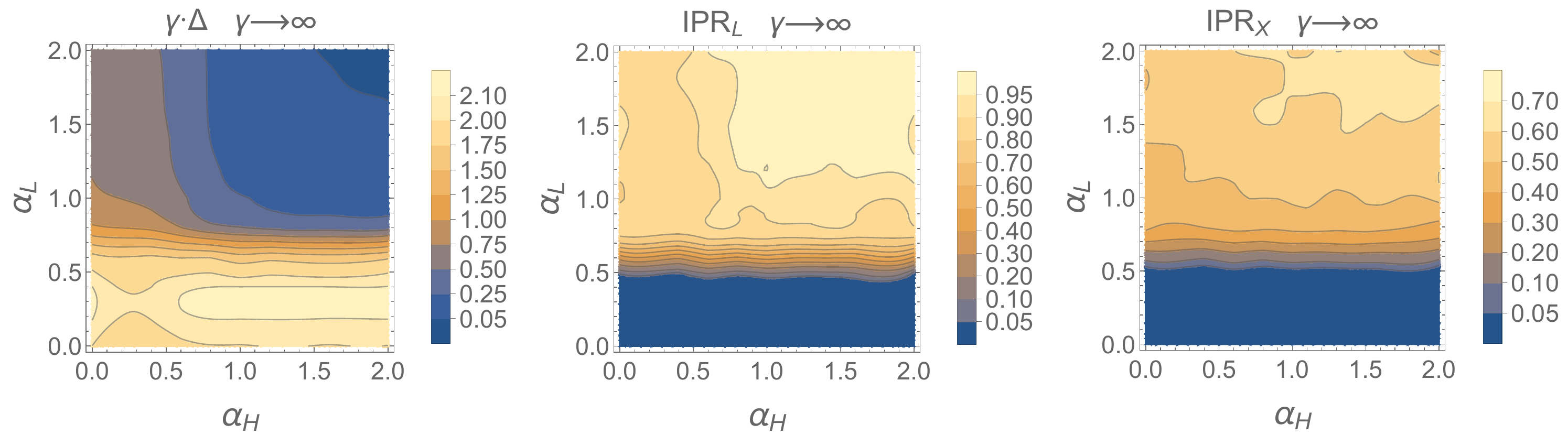}
\end{center}
\vspace{-0.5cm} \caption[]{\small Left: The spectral gap of $A$ in the limit of $\gamma\rightarrow\infty$ for an $N=12800$ system
whose Hamiltonian and jump operator are PRBMs with exponents $\alpha_H$ and $\alpha_L$, respectively.
Center and right: The inverse participation ratios in the $L$-basis and in real space of the corresponding slowest decaying mode
for a system with $N=6400$. We note that an additional region of extended states exists outside the depicted parameters range
for $\alpha_L>3$ and $\alpha_H>3/2$, see bottom row of Fig. \ref{fig-largeg-Agap}.}
\label{fig-Alargeg}
\end{figure}

Fig. \ref{fig-Alargeg} shows similar data obtained by diagonalizing Eq. (\ref{eq:Alarge}) in the limit $\gamma\rightarrow \infty$.
For this case we expand the slowest decaying mode in the $L$-populations and use the expansion coefficients to define the inverse
participation ratio in the $L$-eigenbasis, IPR$_L$, and in the position basis using Eqs. (\ref{eq:IPRHdef},\ref{eq:IPRXdef}).

%\begin{figure}[!!!h]
%\begin{center}
%\includegraphics[width = 0.485\textwidth]{figAaL3p5}
%\hspace{3mm}
%\includegraphics[width = 0.48\textwidth]{figAal3p5-IPRL}
%\end{center}
%\vspace{-0.5cm} \caption[]{\small Left: The spectral gap of $A$ in the limit of $\gamma\rightarrow\infty$ for an $N=12800$ system
%whose Hamiltonian and jump operator are PRBMs with exponents $\alpha_H$ and $\alpha_L$, respectively.
%Center and right: The inverse participation ratios in the $L$-basis and in real space of the corresponding slowest decaying mode
%for a system with $N=6400$. The data was averaged over 100 realizations for each $(\alpha_H,\alpha_L)$ point. }
%\label{fig-AaL3p5}
%\end{figure}

\subsection{Finite size scaling of $\Delta(A)$}

\begin{figure}[!!!h]
\begin{center}
\includegraphics[width = 0.485\textwidth]{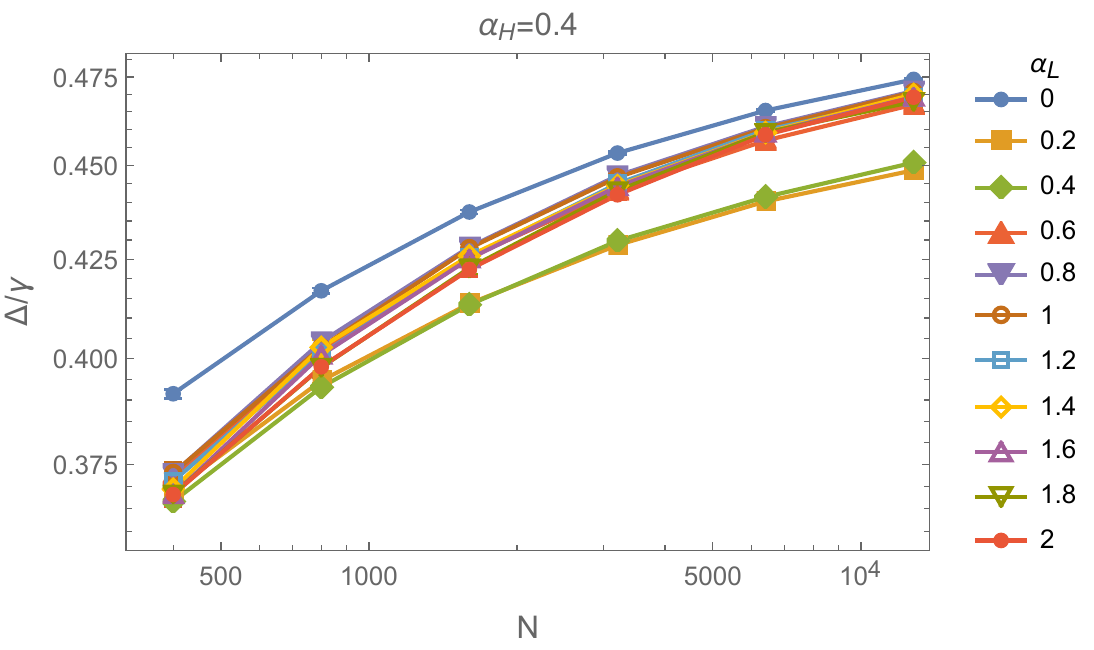}
\hspace{3mm}
\includegraphics[width = 0.471\textwidth]{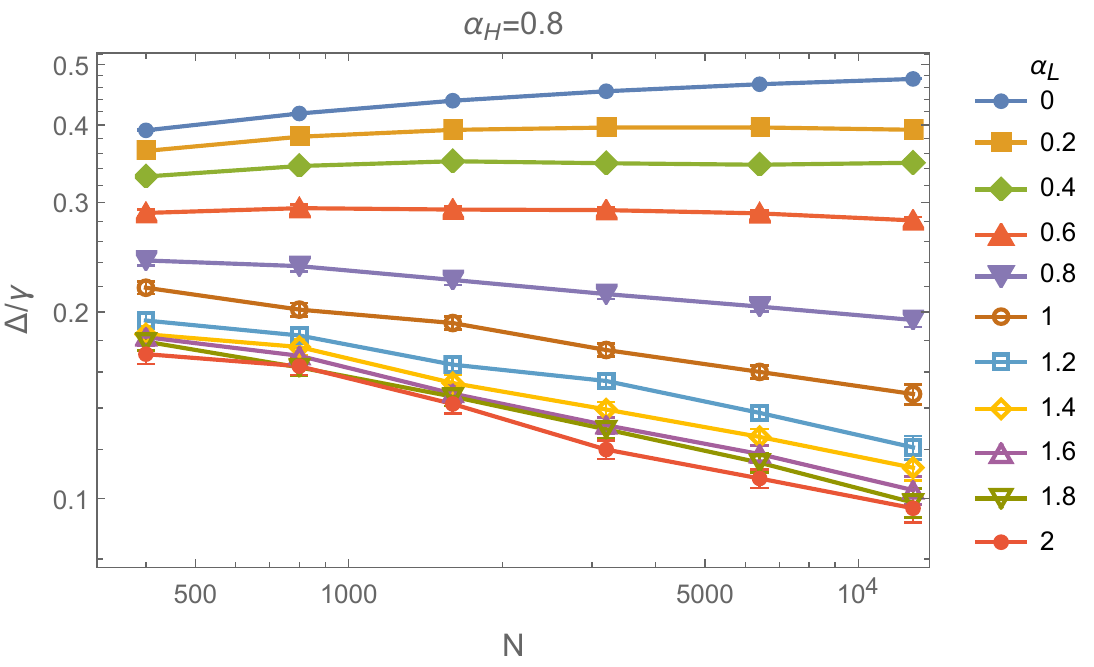}
\newline
\includegraphics[width = 0.48\textwidth]{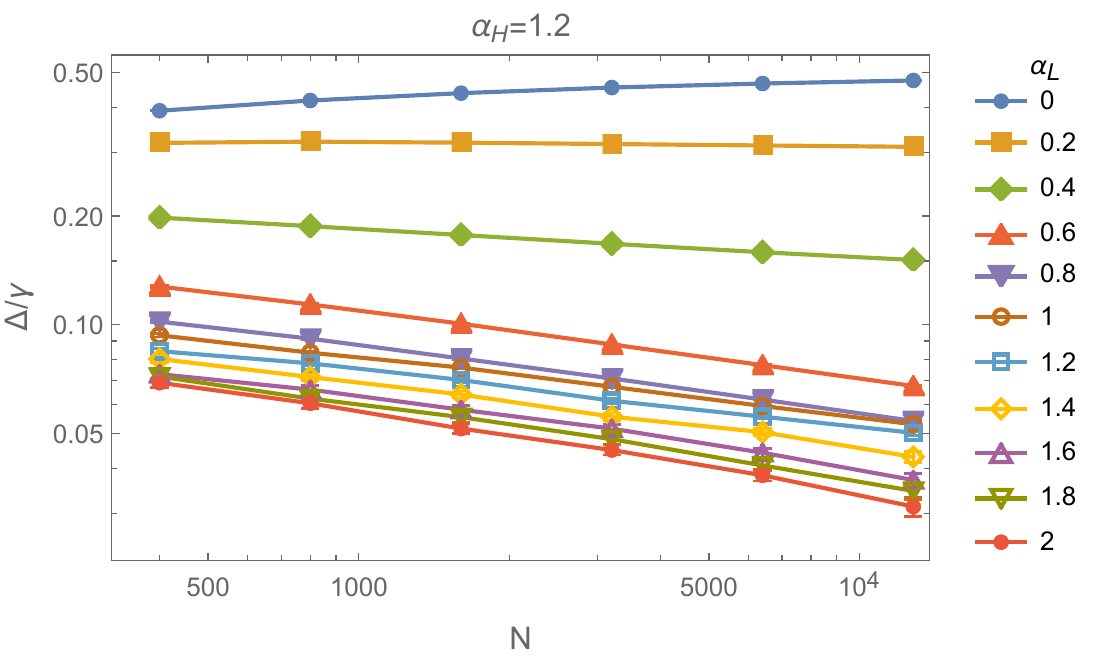}
\hspace{3mm}
\includegraphics[width = 0.482\textwidth]{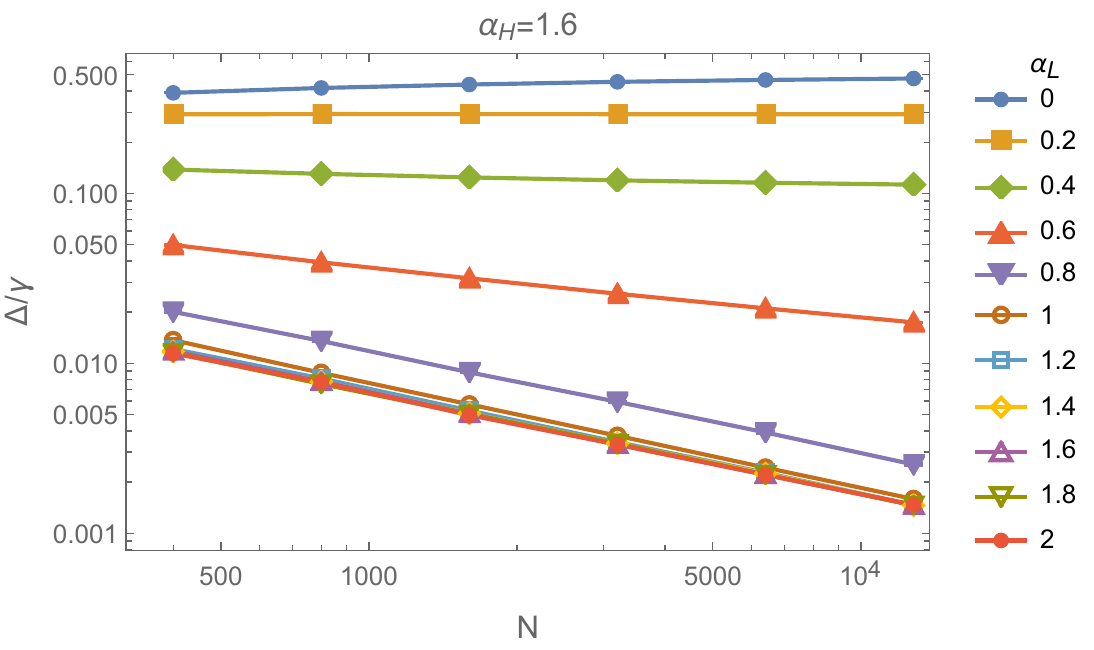}
\end{center}
\vspace{-0.5cm} \caption[]{\small $N$-dependence of the gap calculated from $A$ in the limit $\gamma\rightarrow 0$.}
\label{fig-smallg-Agap}
\end{figure}

\begin{figure}[!!!h]
\begin{center}
\vspace{2mm}
\includegraphics[width = 0.48\textwidth]{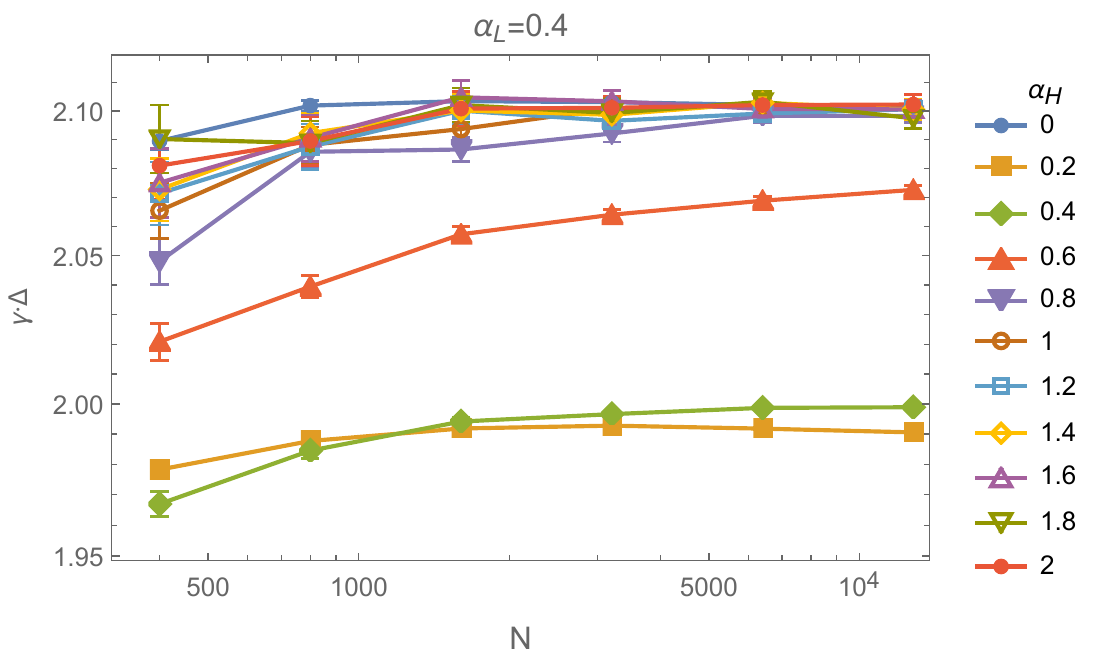}
\hspace{3mm}
\includegraphics[width = 0.48\textwidth]{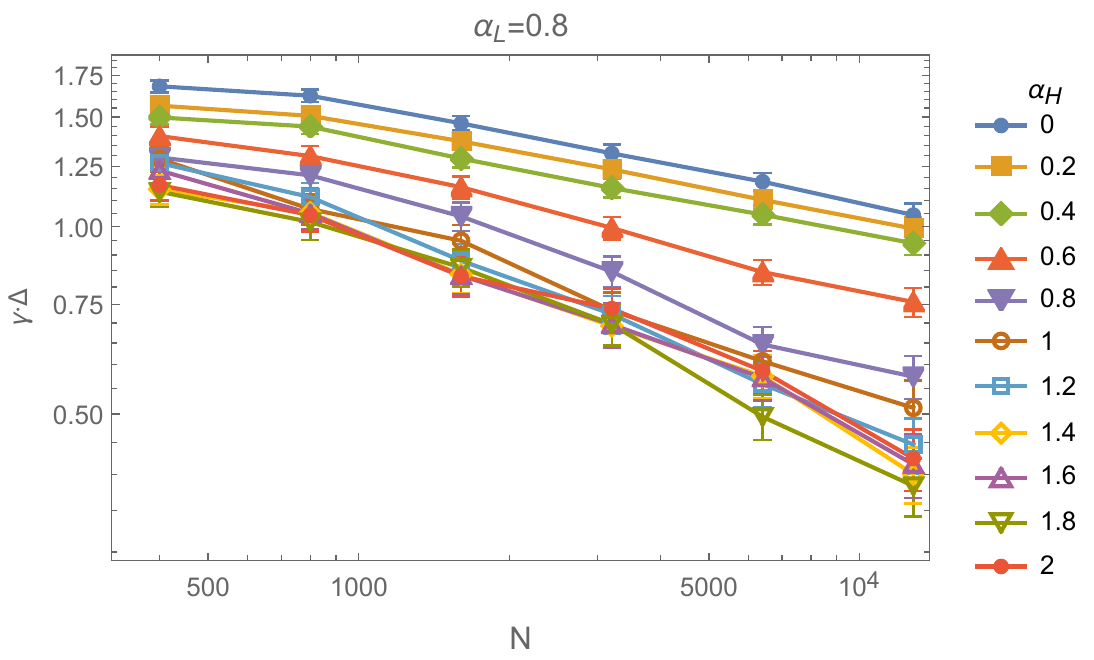}
\newline
\includegraphics[width = 0.475\textwidth]{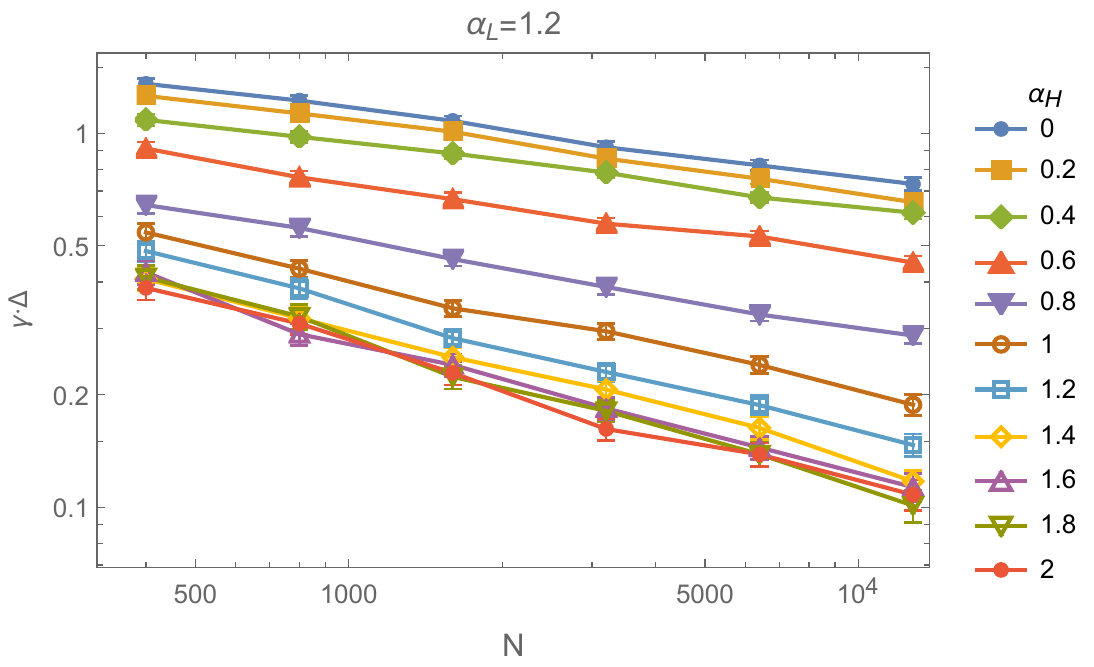}
\hspace{3mm}
\includegraphics[width = 0.482\textwidth]{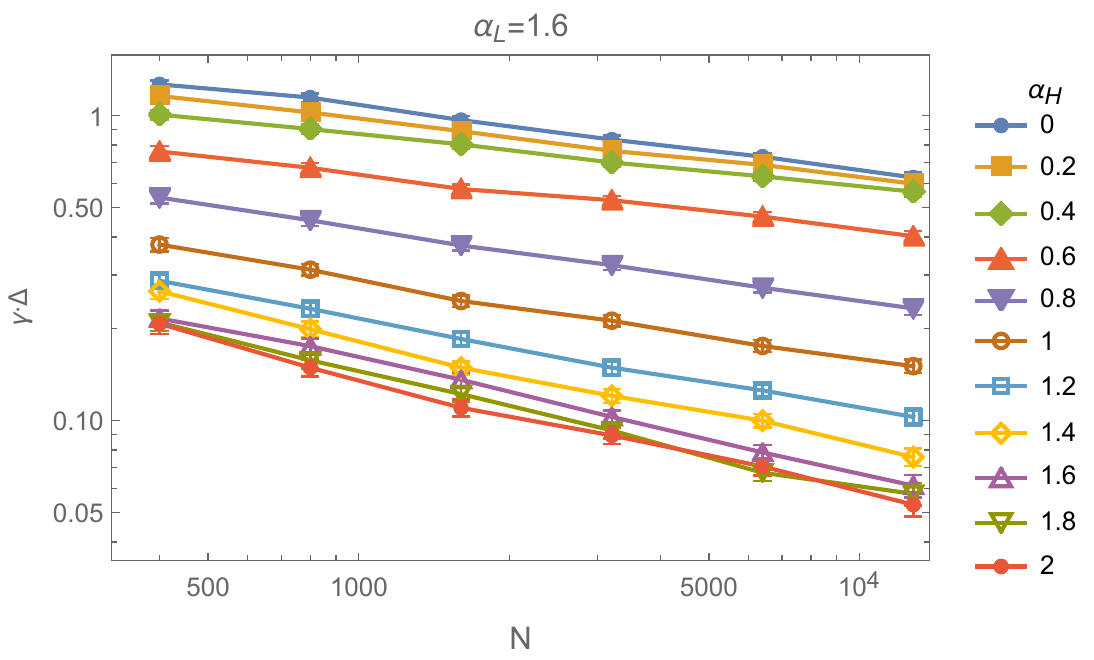}
\newline
\includegraphics[width = 0.482\textwidth]{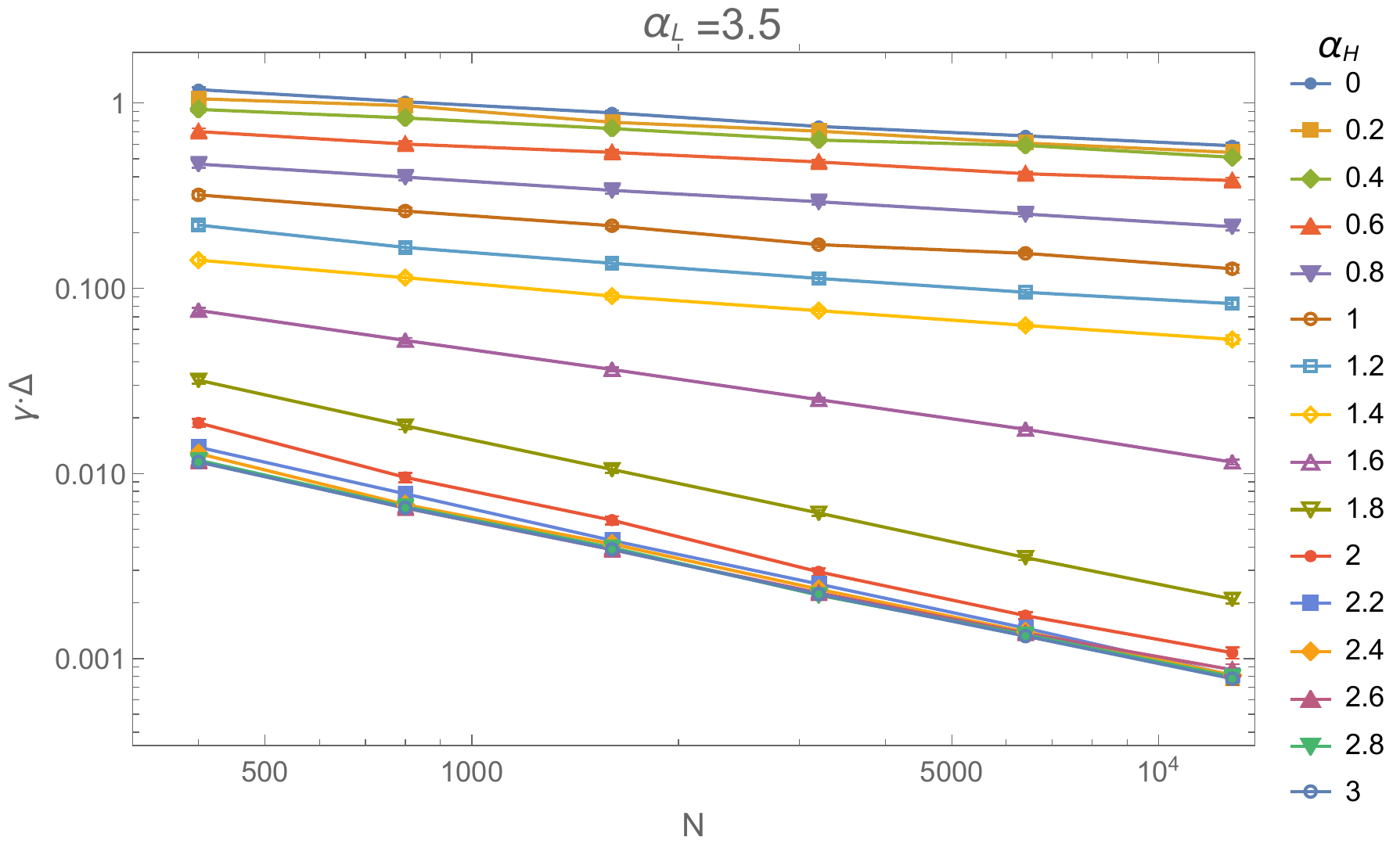}
\hspace{3mm}
\includegraphics[width = 0.475\textwidth]{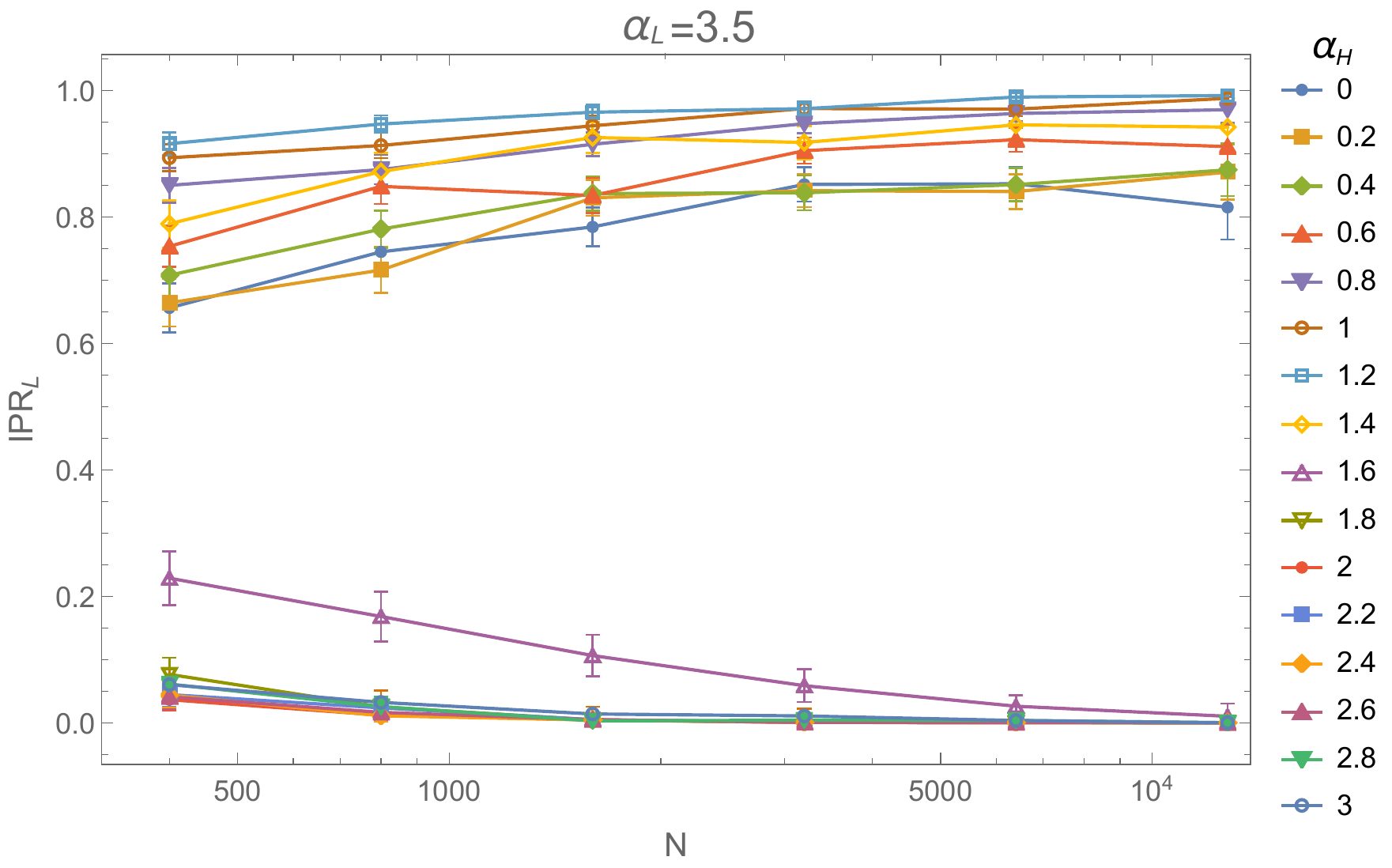}
\end{center}
\vspace{-0.5cm} \caption[]{\small $N$-dependence of the gap calculated from $A$ in the limit $\gamma\rightarrow\infty$.
The bottom right panel demonstrates that the slowest decaying mode in this limit becomes extended for $\alpha_L>3$ and $\alpha_H>3/2$.
Concomitantly, the gap decays more rapidly with $N$.}
\label{fig-largeg-Agap}
\end{figure}

Figs. \ref{fig-smallg-Agap} and \ref{fig-largeg-Agap} contain the finite size scaling of the spectral gap
calculated from $A$ for system sizes of up to $N=12800$. The results were obtained using Eq. (\ref{eq:Asmall}) and Eq. (\ref{eq:Alarge})
corresponding to the limits $\gamma\rightarrow 0$ and $\gamma\rightarrow \infty$, respectively.
Each point represents an average over 100-300 realizations.


\subsection{Comparing $\L$ and $A$}

\begin{figure}[!!!t]
\begin{center}
\includegraphics[width = 0.63\textwidth]{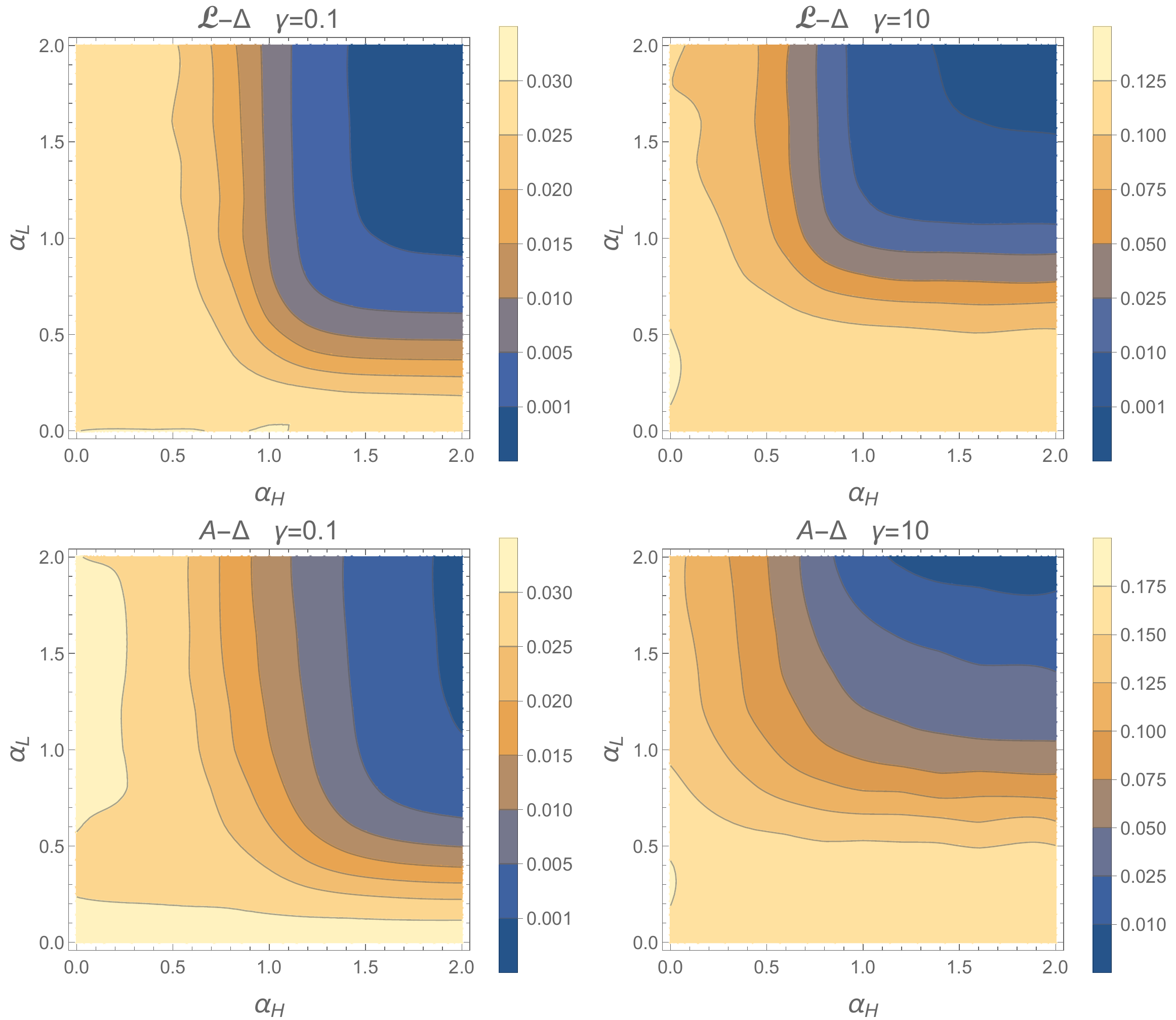}
\end{center}
\vspace{-0.5cm} \caption[]{\small Left: The spectral gap as calculated from the full Lindbladian and from the effective population
dynamics, Eq.(\ref{eq:Asmall}) for $\gamma=0.1$ and Eq. (\ref{eq:Alarge}) for $\gamma=10$, of a system with $N=100$.}
\label{fig-LAgap}
\end{figure}

\begin{figure}[!!!h]
\begin{center}

\includegraphics[width = 0.7\textwidth]{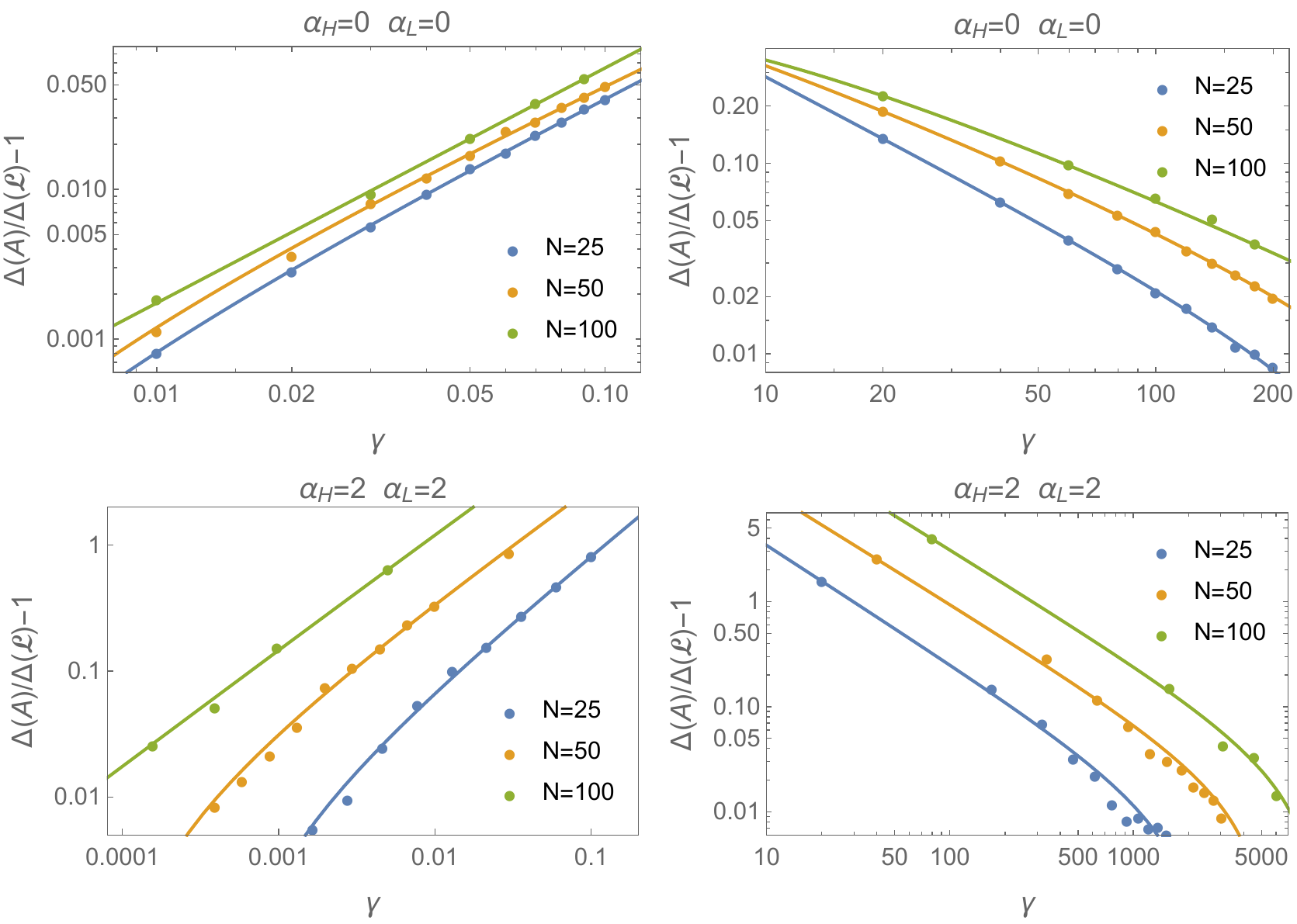}
\end{center}
\vspace{-0.5cm} \caption[]{\small Left: The relative error between the gap calculated from $A$ and the exact gap calculated from $\L$
for weak and strong decoherence. Results are shown for systems with $N=25,50$ and 100 in the cases where both $H$ and $L$ are GOE random matrices
($\alpha_H=\alpha_L=0$) and where they are both PRBMs in the localized regime ($\alpha_H=\alpha_L=2$).
The lines are guides to the eye.}
\label{fig-gaperror}
\end{figure}

The effective populations dynamics generated by Eqs. (\ref{eq:Asmall},\ref{eq:Alarge}) provides an asymptotically exact
description of the low-lying Lindbladian states for fixed $N$ and $\gamma\rightarrow 0,\infty$. However, we would like
to assess its validity away from the extreme $\gamma$ limits, and in particular when $N\rightarrow\infty$ while $\gamma$
is held fixed. To this end, we have calculated the properties of the slowest decaying Lindbladian mode, averaged over 100
realizations per $(\alpha_H,\alpha_L)$ point, and compared them with the corresponding results obtained by diagonalizing $A$,
given by Eqs. (\ref{eq:Asmall}) and (\ref{eq:Alarge}).
Fig. \ref{fig-LAgap} shows that the general dependence of the spectral gap on $\alpha_H$ and $\alpha_L$ is captured reasonably
well by $A$, at least for the system sizes amenable to exact diagonalization of $\L$.
However, qualitative deviations are expected at weak decoherence for $\alpha_H<1/2$ in the very large $N$ limit, see Sec. \ref{sec:weaklargeN}.
We also note that the relative error between the gap calculated from $A$ and the one calculated from $\L$,
$[\Delta(A)-\Delta(\L)]/\Delta(\L)$, increases with $N$, %especially when the spectra of $H$ and $L$ are localized,
as shown by Fig. \ref{fig-gaperror}.
Nevertheless, note that we always find $\Delta(A)>\Delta(\L)$, thus strongly suggesting an upper bound on the exact gap, and convergence
of the two gaps in the extreme $\gamma$ limits for fixed $N$.

We have also calculated the inverse participation ratio of the slowest decaying Lindbladian mode projected on the $H$-populations
space (IPR$_H$) and on the $L$-populations space (IPR$_L$). Concretely, if the expansion of the mode in the energy basis reads
$\rho^{(1)}=\sum_{i,j}\rho_{ij}|i\rangle\langle j|$, with $\sum_{i,j}|\rho_{ij}|^2=1$ then
\begin{equation}
\label{eq:IPRH-def}
{\rm IPR}_H=\frac{\sum_i \rho_{ii}^4}{\left(\sum_i \rho_{ii}^2\right)^2}.
\end{equation}
IPR$_L$ is defined in a similar manner using the expansion of $\rho$ in the $L$ basis. In addition, we calculated the real space
IPR of the mode, which for small $\gamma$ we define using the above energy basis expansion of $\rho$ as
\begin{equation}
\label{eq:IPRx-def}
{\rm IPR}_X=\frac{\sum_i \rho_{ii}^4\sum_x |\langle x|i\rangle|^4}{\left(\sum_i \rho_{ii}^2\right)^2}.
\end{equation}
This quantifier approaches 1 if the projected mode is largely composed of a single population of a real-space localized energy eigenstate,
and else vanishes in the thermodynamic limit. For large $\gamma$ we define IPR$_x$ in a similar fashion using the $L$-basis expansion of $\rho^{(1)}$.

\begin{figure}[!!!b]
\vspace{-0.5cm}
\begin{center}
\includegraphics[width = 0.63\textwidth]{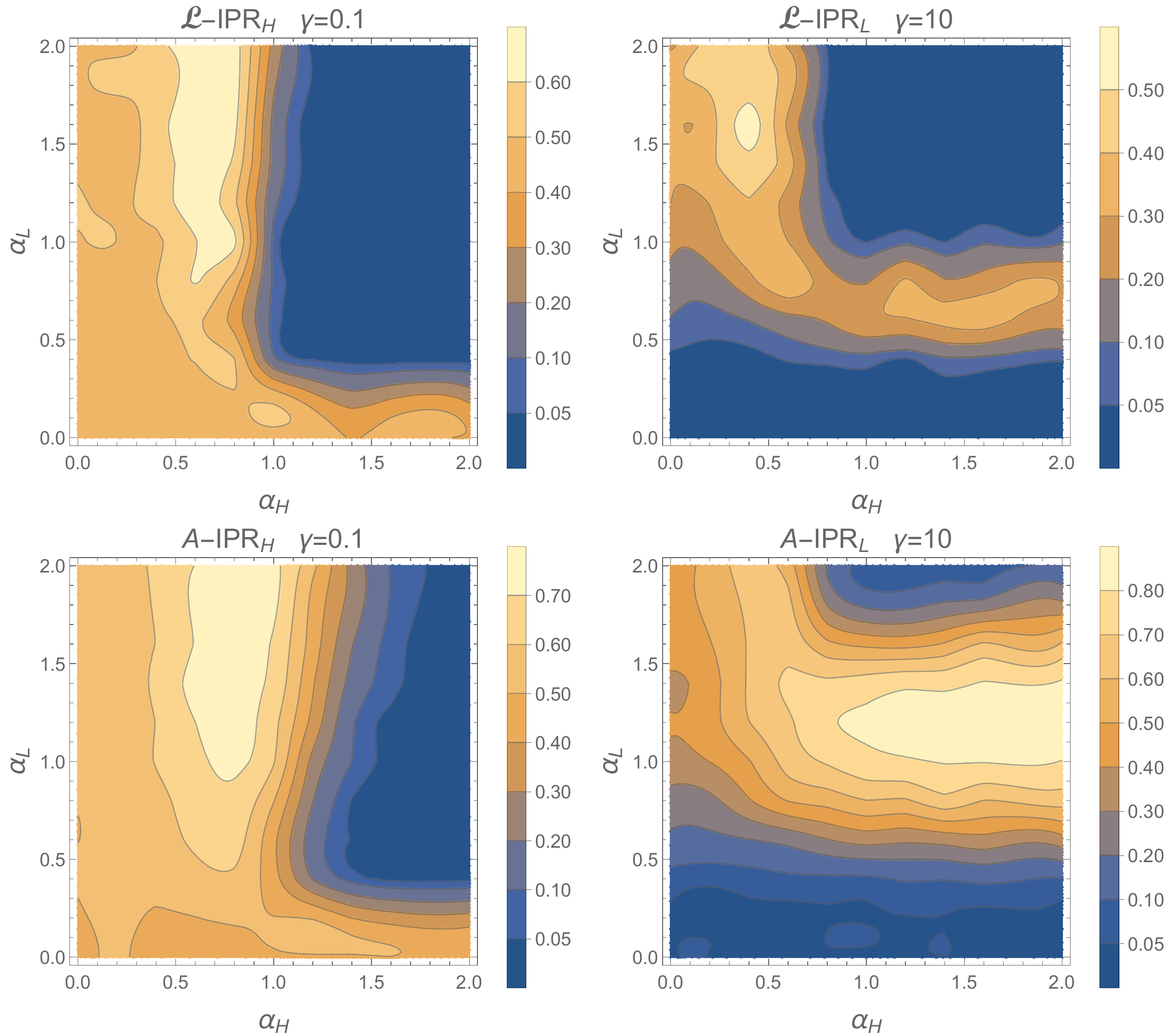}
\includegraphics[width = 0.25\textwidth, trim= 0 -5cm 0 0]{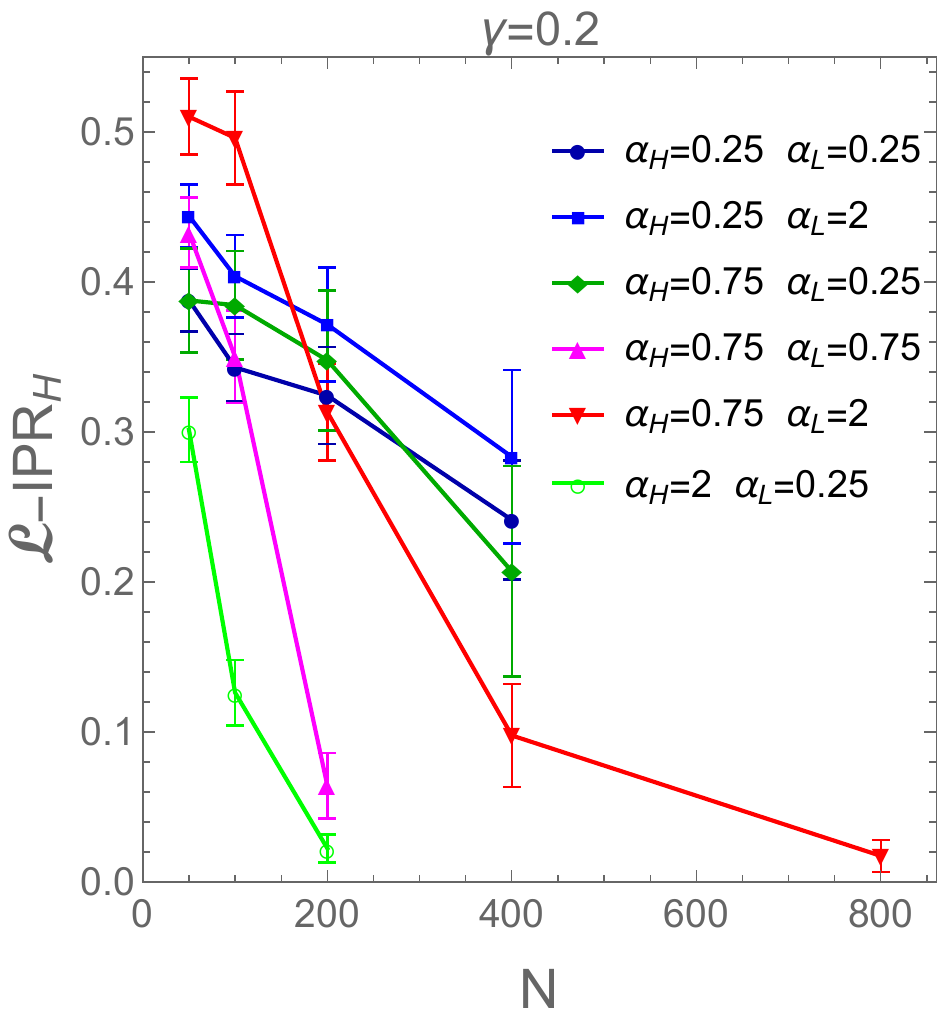}
\end{center}
\vspace{-0.5cm} \caption[]{\small Left: The average inverse participation ratio in energy space of the slowest decaying mode as calculated
from $\L$ and $A$ for a system with $\gamma=0.1$, $N=100$. Middle: The average participation ratio in the space of $L$-eigenvectors of
the slowest decaying mode for $\gamma=10$ and $N=100$. Right: System size dependence of the IPR$_H$ of the slowest decaying $\L$-mode for
a system with $\gamma=0.2$.}
\label{fig-LAIPRLH}
\end{figure}

\begin{figure}[!!!t]
\begin{center}
\includegraphics[width = 0.63\textwidth]{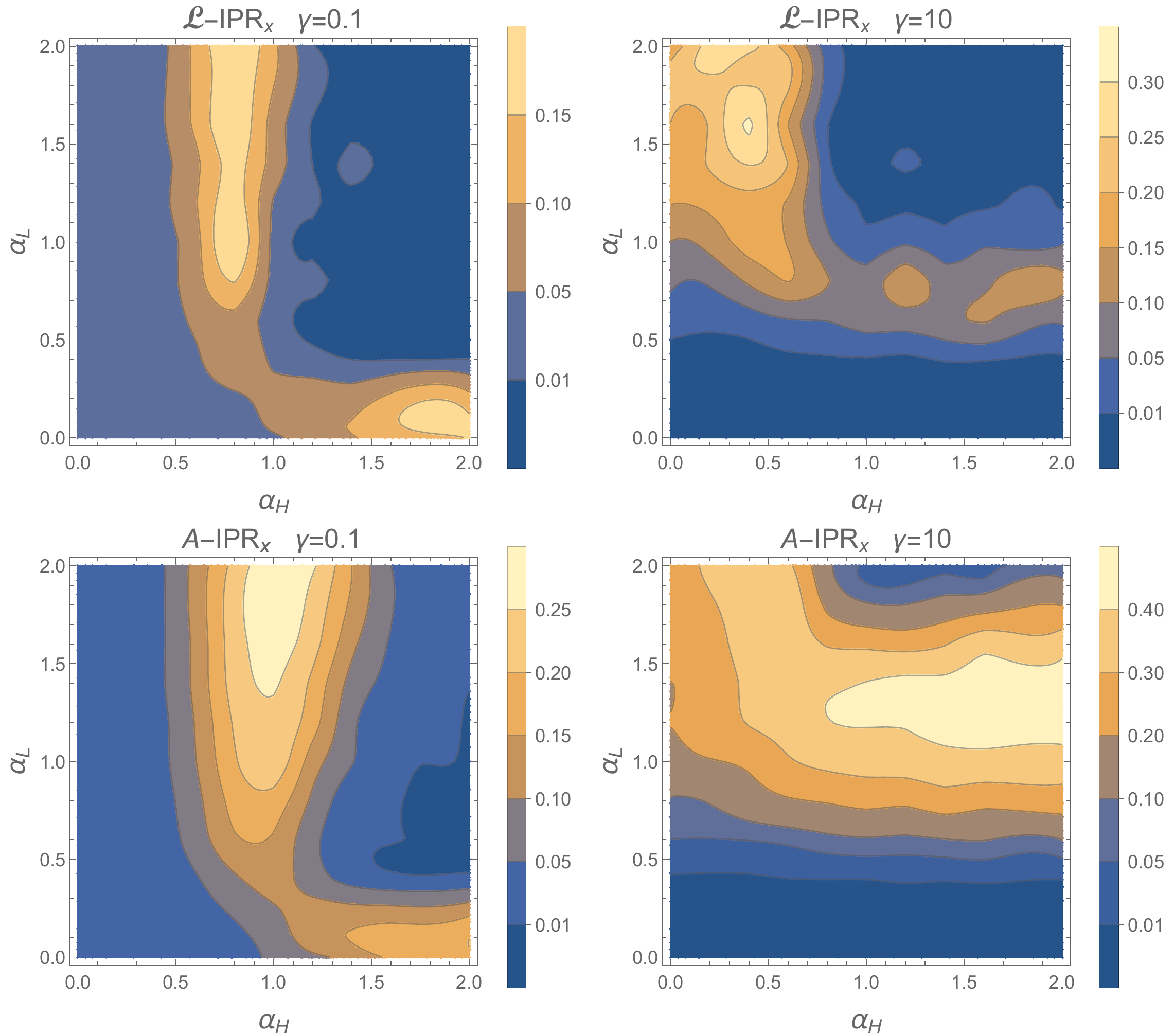}
\includegraphics[width = 0.25\textwidth, trim= 0 -5cm 0 0]{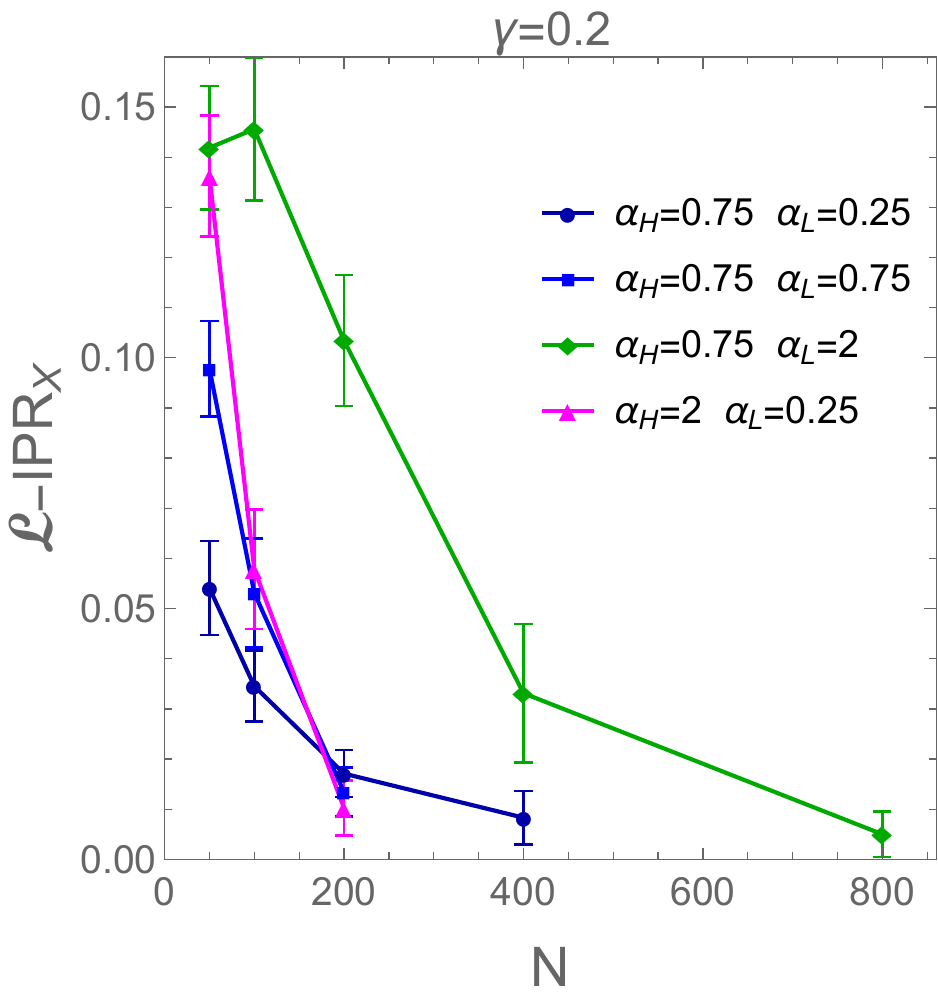}
\end{center}
\vspace{-0.5cm} \caption[]{\small Left: The average inverse participation ration in real space of the slowest decaying mode as calculated
from $\L$ and $A$ for an $N=100$ system. Right: System size dependence of the IPR$_X$ of the slowest decaying $\L$-mode for
a system with $\gamma=0.2$.}
\label{fig-LAIPRx}
\end{figure}

Figs. \ref{fig-LAIPRLH},\ref{fig-LAIPRx} show that the population dynamics broadly captures the
nature of the slowest mode within the $\alpha_H$-$\alpha_L$ plane, albeit with a tendency
to overestimate the range in which it is localized. This is particularly true  for $\gamma=10$
where the decrease of the IPR is recovered only at $\alpha_L>3/2$. We note that for this to
happen one must include the eigenvalue shifts in Eq. (\ref{eq:Alarge}).
Without it, the $A$-IPR is large for all $3>\alpha_L>1/2$. The figures also show that for small $\gamma$ the
$\L$-slowest mode becomes delocalized both in energy and real space. As shown in the main text, for large $\gamma$
this mode remains localized in the region $\alpha_H<1/2$, $\alpha_L>1/2$.
%We also note that the appearance of
%localized states for $\gamma=0.1$ at large $\alpha_H$ and small $\alpha_L$ is due to the relatively small size of the system.
%They become delocalized as $N$ increases, see Section \ref{ss-smallgloc}.

\begin{figure}[!!!h]
\begin{center}
\includegraphics[width = \textwidth]{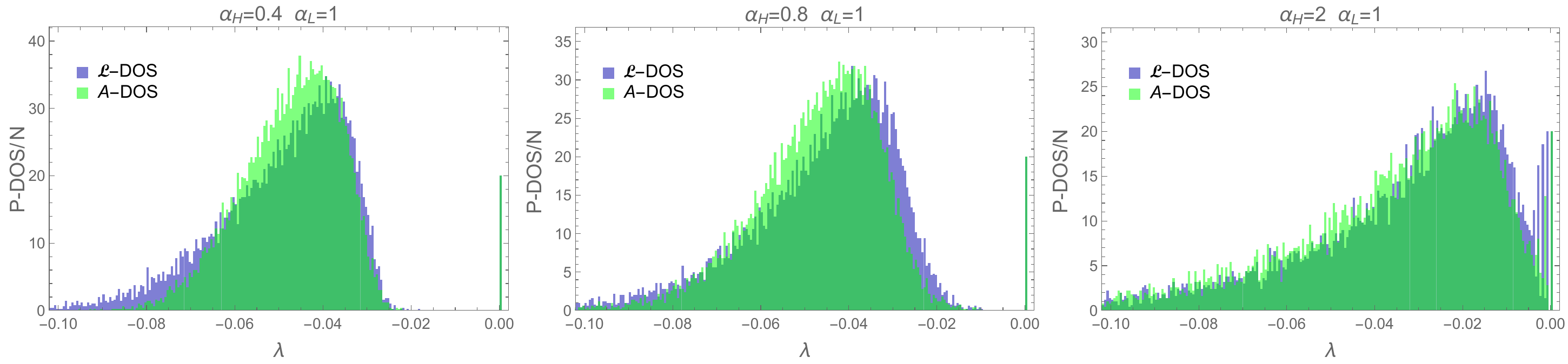}
\end{center}
\vspace{-0.5cm} \caption[]{\small The averaged population DOS calculated from $\L$ and $A$ for a system with $\gamma=0.1$, $N=100$ and
representative values of $(\alpha_H,\alpha_L)$.}
\label{fig-LADOS01}
\end{figure}

\begin{figure}[!!!h]
\begin{center}
\includegraphics[width = \textwidth]{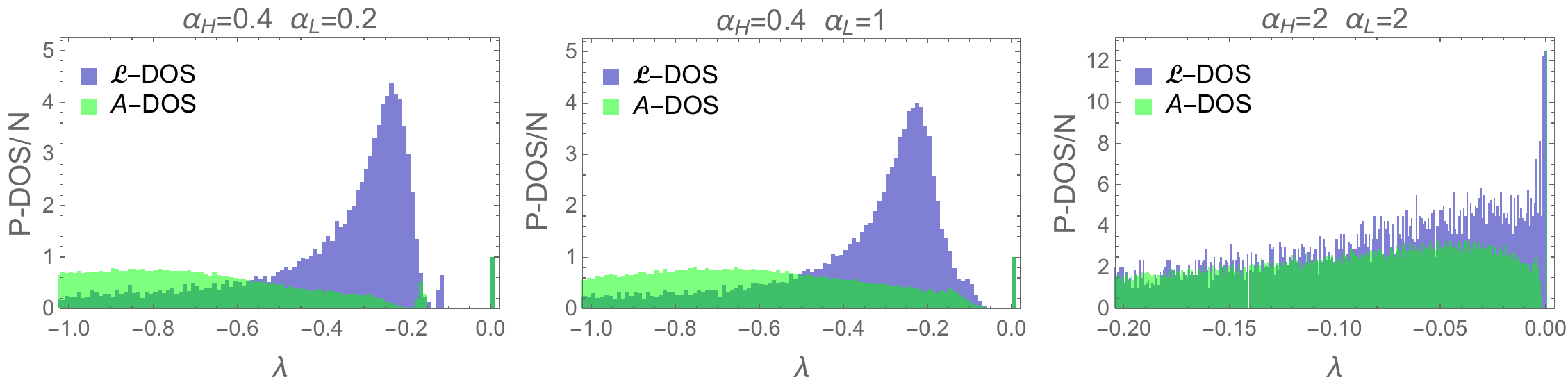}
\end{center}
\vspace{-0.5cm} \caption[]{\small The averaged populations DOS calculated from $\L$ and $A$ for a system with $\gamma=10$, $N=100$ and
representative values of $(\alpha_H,\alpha_L)$.}
\label{fig-LADOS10}
\end{figure}

Finally, we have compared the averaged populations density of states (P-DOS). For $\L$ the latter was obtained by picking
among the eigenstates with real eigenvalues the $N$ (there are at least as many for the case of an Hermitian $L$ \cite{ours})
whose overlaps with the $H$-populations (for small $\gamma$) or $L$-populations (large $\gamma$) are the largest.
Fig. \ref{fig-LADOS01} shows that $A$ essentially reproduces the exact DOS in the case of weak decoherence. For strong
decoherence it overestimates the width of the P-DOS, as seen in Fig. \ref{fig-LADOS10}. We note that this overestimation
is made much worst if one does not include the spectral shifts in the denominator of Eq. (\ref{eq:Alarge}).

\vspace{0.46cm}
\vspace{0.46cm}

\newpage

\subsection{Small $\gamma$ and $\alpha_H>1$ }
\label{ss-smallgloc}

Denote by $L_0$ the PRBM describing the
jump operator in the $x$-basis and by $U$ the unitary that diagonalizes the Hamiltonian: $H=U^\dagger \varepsilon U$,
with $\varepsilon$ a diagonal matrix containing the energy eigenvalues. Hence, $L=U L_0 U^\dagger$ and Eq. (\ref{eq:Asmall}) becomes
\begin{equation}
\label{eq:Atrans}
A_{ij}=-\gamma\sum_{mn}U_{im}{(L_0^2)}_{mn}U^\dagger_{nj}\delta_{ij} +\gamma \left|\sum_{mn}U_{im}{(L_0)}_{mn} U^\dagger_{nj}\right|^2.
\end{equation}

Consider the extreme case where the eigenstates of $H$ are localized on single sites. We expect that this limit is relevant for
$\alpha_H\gg 1$.  For such a model $U_{ij}=\delta_{j,P_i}$,
where $P_i=P(i)$, with $P$ some permutation of $\{1,\cdots ,N\}$. Consequently,
\begin{equation}
\label{eq:A0}
A_{ij}=-\gamma\left[{(L_0^2)}_{P_iP_j}\delta_{ij} - \left|{(L_0)}_{P_iP_j}\right|^2\right]
=-\gamma\sum_{mn}U_{im}\left[{(L_0^2)}_{mn}\delta_{mn}-\left|{(L_0)}_{mn}\right|^2 \right]U^\dagger_{nj}
\equiv\left(UA_0 U^\dagger\right)_{ij},
\end{equation}
and the spectrum of $A$ coincides with that of $A_0$.

Using Eq. \ref{eq:Hvar} we can decompose $A_0=\bar A_0+\delta A_0$, where the elements of the constant "mean-field" part are give by
\begin{equation}
\label{eq:barA0}
%{{}\bar{A}_0}_{ij}
{(\bar{A}_0)}_{ij}=\left\langle{(A_0)}_{ij}\right\rangle_L = \frac{\gamma}{2a(N,\alpha_L)} \left\{\begin{array}{lc}
-\sum_{m\neq i}f_{L,im}^2 & \;\;\;i=j \\
f_{L,ij}^2 & \;\;\; i\neq j
\end{array}\right. ,
\end{equation}
where $\langle\rangle_L$ denotes the mean with respect to the statistics of $L_0$, and $f_L$ is the power-law form factor associated with $L$.
The elements of the fluctuating part are random variables with zero mean. They are not independent. Clearly,
${(\delta A_0 )}_{ij}={(\delta A_0)}_{ji}$ and ${(\delta A_0 )}_{ii}$ contains ${(\delta A_0 )}_{ij}$, which also
means that it is not independent of ${(\delta A_0 )}_{jj}$. Thus
\begin{equation}
\label{eq:deltaA0}
\left\langle{(\delta A_0)}_{ij}{(\delta A_0)}_{ji}\right\rangle_L=
(-1)^{1-\delta_{ij}}\left\langle{(\delta A_0)}_{ii}{(\delta A_0)}_{ij}\right\rangle_L=\left\langle{(\delta A_0)}_{ii}{(\delta A_0)}_{jj}\right\rangle_L=
\frac{\gamma^2}{2 a^2(N,\alpha_L)} \left\{\begin{array}{lc}
\sum_{m\neq i}f_{L,im}^4 & \;\;i=j \\
f_{L,ij}^4 & \;\; i\neq j
\end{array}\right. \!\!\!.
\end{equation}

For a CPRBM jump operator the eigenvalues of $\bar A_0$ are given by (we assume here odd $N$)
\begin{equation}
\label{eq:eigsA0}
\lambda_n=-\frac{2\gamma}{a(N,\alpha_L)}\sum_{j=1}^{(N-1)/2}f_{L,j}^2\sin^2\left(\frac{j\pi n}{N}\right).
\;\;\;\;\;\;  n=0,\cdots ,\frac{N-1}{2},
\end{equation}
and apart from the zero mode ${(v_0)}_j=1/N$, every eigenvalue is doubly degenerate with the corresponding (normalized) eigenvectors
\begin{equation}
\label{eq:vecsA0}
{(v_n)}_j=\sqrt{\frac{2}{N}}\cos\left(\frac{2\pi n j}{N}\right),
\;\;\;\;\;\; {(u_n)}_j=\sqrt{\frac{2}{N}}\sin\left(\frac{2\pi n j}{N}\right),
\;\;\;\;\;\;  n=1,\cdots ,\frac{N-1}{2}.
\end{equation}
In order to evaluate the spectral gap we require the large-$N$ behavior of
\begin{eqnarray}
\label{eq:lambdaint}
\nonumber
\sum_{j=1}^{(N-1)/2}f_j^2\sin^2\left(\frac{j\pi}{N}\right)&\xrightarrow{N\rightarrow\infty}&
N^{1-2\alpha}\int_{1/N}^{1/2}dx x^{-2\alpha}\sin^2(\pi x)\\
&\xrightarrow{N\rightarrow\infty}&\frac{(N/2)^{1-2\alpha}}{2(1-2\alpha)}
\left[1-{}_1F_2\left(\frac{1}{2}-\alpha,\left\{\frac{1}{2},\frac{3}{2}-\alpha\right\}
,-\frac{\pi^2}{4}\right)\right]-\frac{\pi^2}{3-2\alpha}\frac{1}{N^2},
\end{eqnarray}
where ${}_1F_2$ is the generalized hypergeometric function and which leads, together with Eq. (\ref{eq:adef}), to the gap
\begin{equation}
\label{eq:lambda1}
\Delta=\left\{\begin{array}{cc}
-\frac{\gamma}{2}\left[1-{}_1F_2\left(\frac{1}{2}-\alpha_L,\left\{\frac{1}{2},\frac{3}{2}-\alpha_L\right\},-\frac{\pi^2}{4}\right)\right]
\left[1-(1-2\alpha_L)[1+\zeta(2\alpha_L)]\left(\frac{2}{N}\right)^{1-2\alpha_L}\right] & \;\;\;\alpha_L<1/2 \\
-\frac{\gamma}{2(1-2\alpha_L)[1+\zeta(2\alpha_L)]}\left[1-{}_1F_2\left(\frac{1}{2}-\alpha_L,\left\{\frac{1}{2},\frac{3}{2}-\alpha_L\right\}
\right)\right] \left(\frac{2}{N}\right)^{2\alpha_L-1} &\;\;\;3/2>\alpha_L>1/2 \\
-\frac{\gamma\pi^2}{(2\alpha_L-3)[1+\zeta(2\alpha_L)]}\frac{1}{N^2} & \;\;\;\alpha_L>3/2
\end{array}\right. .
\end{equation}

Consider now the effects of $\delta A_0$ on the gap. The zero-mode is guaranteed and to lowest order we need to diagonalize $\delta A_0$
within the degenerate subspace spanned by $v_1$ and $u_1$ (note that for $\alpha_L=0$ the $\bar A_0$ spectrum is $(N-1)$-fold degenerate
at $-\gamma/2$). To this end we note that $\left\langle\langle u|\delta A_0|v\rangle\right\rangle_L=0$ and use Eq. (\ref{eq:deltaA0}) to obtain
\begin{eqnarray}
\label{eq:varshift}
\nonumber
\sqrt{\left\langle\langle u|\delta A_0|v\rangle^2\right\rangle_L}&=&
\left\{\left\langle\left(\sum_{i\neq j}u_i v_j {(\delta A_0)}_{ij}+\sum_i u_i v_i {(\delta A_0)}_{ii}\right)
\left(\sum_{k\neq l}u_k v_l {(\delta A_0)}_{kl}+\sum_k u_k v_k {(\delta A_0)}_{kk}\right)\right\rangle_L\right\}^{1/2}\\
&=&\frac{\gamma}{\sqrt{2}a(N,\alpha_L)}\left[\sum_{i\neq j}\left[u_i^2(v_i^2+v_j^2)+2u_i u_j v_i v_j -2u_i^2 v_i v_j
-2u_i u_j v_j^2 \right]f_{L,ij}^4\right]^{1/2},
\end{eqnarray}
which scales as $1/N$ and $1/\sqrt{N}$ in the limits $\alpha_L\rightarrow 0$ and $\alpha_L\rightarrow\infty$, respectively.
Together with Eq. (\ref{eq:lambda1}) this signals the breakdown of perturbation theory for large $\alpha_L$. Since the
difference $\lambda_1-\lambda_2$ scales to a constant as $\alpha_L\rightarrow 0$ we expect that in this limit the effects
of the fluctuating part diminish for $N\rightarrow\infty$.

\begin{figure}[!!!h]
\begin{center}
\vspace{0.3cm}
\includegraphics[width = 0.5\textwidth]{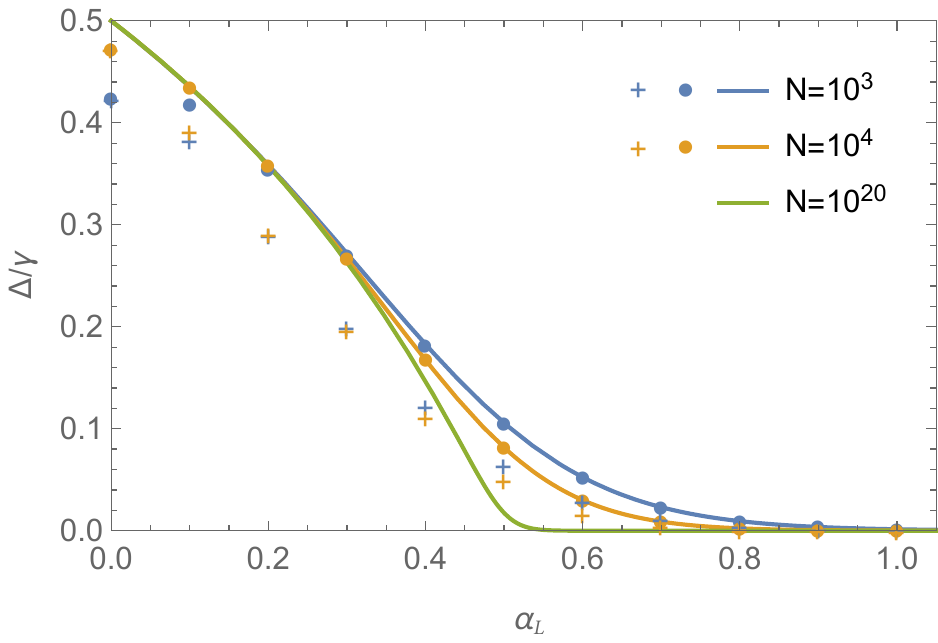}
\end{center}
\vspace{-0.5cm} \caption[]{\small The spectral gap of $A_0$. The solid lines are given by Eq. (\ref{eq:eigsA0}) while the
dots and crosses depict the averaged gaps (over 100 realizations) for the cases of a CPRBM and PRBM jump operator, respectively.}
\label{fig-smallgammagap}
\end{figure}

\begin{figure}[!!!h]
\begin{center}
\vspace{0.3cm}
\includegraphics[width=\textwidth]{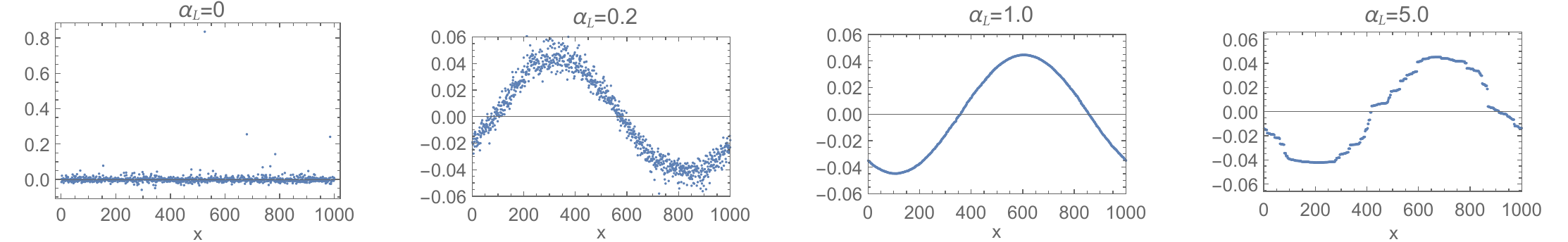}
\end{center}
\vspace{-0.5cm} \caption[]{\small Representative examples of the slowest decaying eigenvector of $A_0$ generated by an $N=1000$ CPRBM jump operator.}
\label{fig-v1-smallg-CPRBM}
\end{figure}

The above analysis is backed by our numerics. As demonstrated by Fig. \ref{fig-smallgammagap}, the gap largely follows
Eq. (\ref{eq:eigsA0}) with deviations at small $\alpha_L$ that diminish upon increasing $N$. An important lesson from the numerics is that
a decrease of the gap with $N$ does not necessarily imply that it vanishes in the thermodynamic limit. In particular, the gap shows
a flow-reversal as function of $N$ slightly above $\alpha_L=0.2$, while its $N\rightarrow\infty$ value is zero only for $\alpha_L>1/2$.
In addition, we find that the gap at large $\alpha_L$ decays as $1/N^3$ and not $1/N^2$, which is the prediction based on $\bar A_0$,
see Eq. (\ref{eq:lambda1}), thereby signaling the breakdown of perturbation theory.
The results for the lowest non-stationary eigenvector, Fig. \ref{fig-v1-smallg-CPRBM}, show that it is close
to a linear combination of the degenerate solutions, Eq. (\ref{eq:vecsA0}). In general, the mean-field eigenfunctions
provide a fair description of the low-lying states of $A$ within a window of eigenvalues whose
extent is largest at $\alpha_L=1$, and which decreases away from this point.
At $\alpha_L=0$ the lowest non-stationary eigenvector is almost localized on a single site.
%We find that a qualitatively similar picture also emerges for a
%system with a PRBM jump operator. The eigenvectors of $A_0$ are close to the eigenvectors of $\bar A_0$, whose structure, unlike in the CPRBM case,
%depends on $\alpha_L$.

\begin{figure}[!!!h]
\begin{center}
\includegraphics[width = 0.45\textwidth]{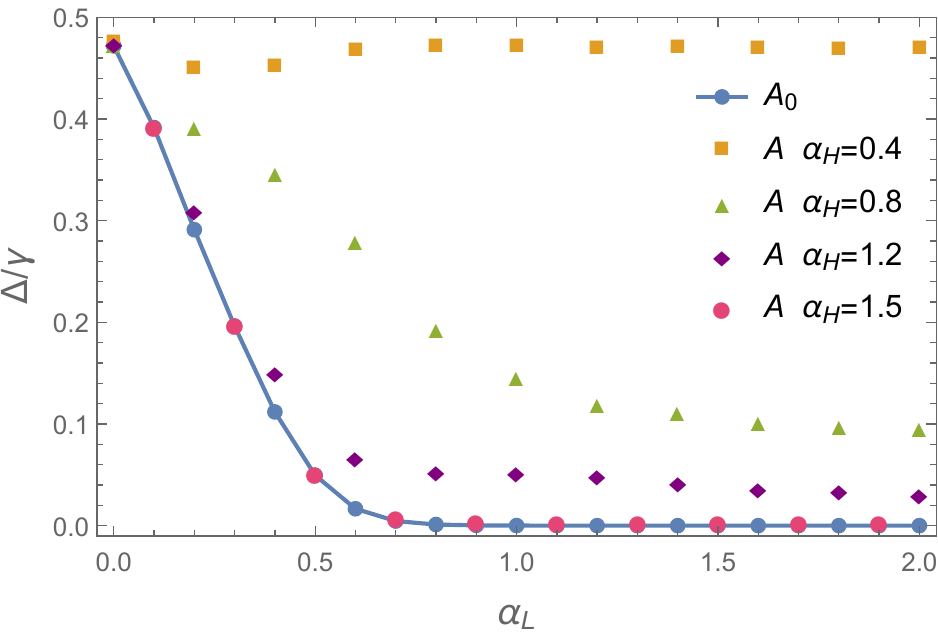}
\hspace{3mm}
\includegraphics[width = 0.44\textwidth]{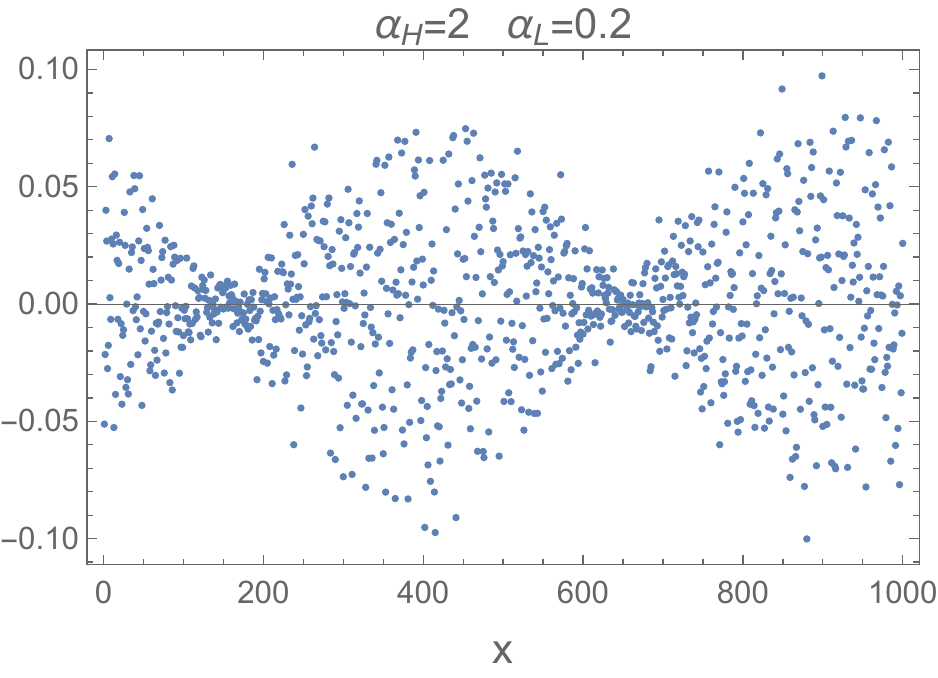}
\end{center}
\vspace{-0.5cm} \caption[]{\small Left: The spectral gap at small $\gamma$ calculated from averaging over 100 realization of $A$
with $N=12800$ PRBM $H$ and $L$, together with the gap calculated from $A_0$.
Right: A typical slowest decaying eigenvector of $A$ with $N=1000$ CPRBM, $\alpha_H=2$ and $\alpha_L=0.2$, shown in the
$x$-representation.}
\label{fig-smallg-A}
\end{figure}

\begin{figure}[!!!h]
\begin{center}
\includegraphics[width = \textwidth]{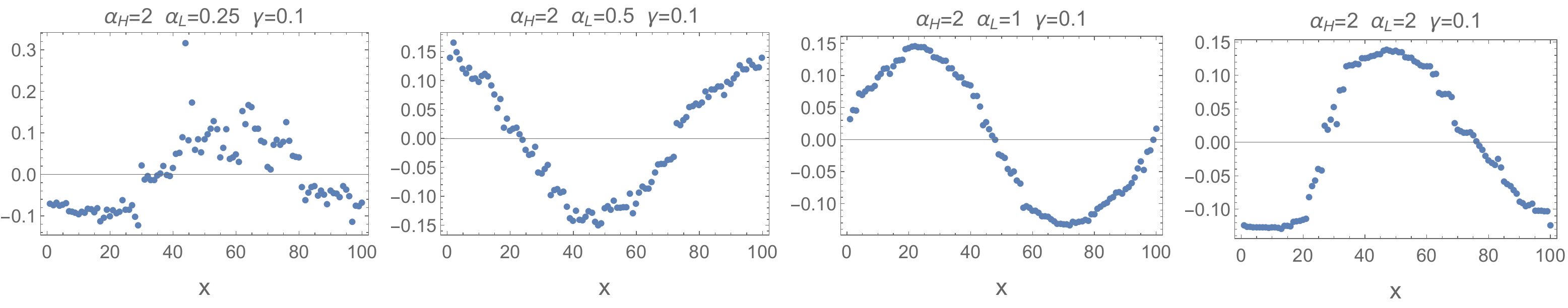}
\end{center}
\vspace{-0.5cm} \caption[]{\small Representative examples of the projection of the slowest decaying eigenvector of $\L$ onto the
subspace of position populations $|x\rangle\langle x|$. The data is for a system of size $N=100$ with CPRBM $H$ and $L$.}
\label{fig-vec1MaH2}
\end{figure}

In order to gauge the range at which the approximation of extremely localized energy-eigenstates provides a reasonable description
of a PRBM system we contrast, in Fig. \ref{fig-smallg-A}, the gap at weak decoherence calculated from $A$, as given by Eq. (\ref{eq:Asmall}),
with that calculated from $A_0$. The results indicate that for $\alpha_H>3/2$ the gap is very close to that of the extreme
localized system, although the gap calculated from $A_0$ diminishes faster with $N$ when $\alpha_L>1/2$. The same figure also shows that in the
range $\alpha_H>3/2$ the mode which sets the gap bears similarities to the one calculated from $A_0$ in terms of its envelope structure
(compare with Fig. \ref{fig-v1-smallg-CPRBM}), although it also exhibits short range fluctuations that are absent in the latter.
More importantly, we also find that in the same regime the slowest decaying mode of the full Lindbladian $\L$ is largely composed of
position populations and exhibits a structure that is close to that of the $A_0$ eignemodes, as can be seen by comparing
Figs. \ref{fig-vec1MaH2} and \ref{fig-v1-smallg-CPRBM}.

One can try to apply a similar "mean-field" analysis to understand the behavior in the range $\alpha_H>1$, but still away from the
extreme-localized regime of $\alpha_H\gg 1.$  For this purpose we model the energy eigenfunctions as having an envelope with a
single maximum from which it decays as $|x|^{-\alpha_H}$. This in turn implies that
$U_{ij}\sim u_{ij}|P_i-j+\delta_{ij}|^{-\alpha_H}/\sqrt{a(\alpha_H,N)-1}$, where $P_i$ is the position of the
maximum of the $i$th wavefunction and $u_{ij}$ is a fluctuating amplitude of order one.
Consequently, for a given $H$ realization the average over the $L_0$ ensemble of the off-diagonal elements $A_{ij}$ becomes

\begin{eqnarray}
\label{eq:modelA}
\nonumber
\left\langle A_{ij}\right\rangle_L &=&\left\langle\gamma \left|\sum_{mn}U_{im}{(L_0)}_{mn} U^\dagger_{nj}\right|^2\right\rangle_L \\
\nonumber
&=&\frac{\gamma}{2a(N,\alpha_L)}\sum_{mn}\left[(U_{im})^2(U_{jn})^2 + U_{im}U_{jm}U_{in}U_{jn}\right]f_{L,mn}^2 \\
\nonumber
&\sim & \frac{\gamma}{a(N,\alpha_L)\left[a(\alpha_H,N)-1\right]^2}{\Big\{}\sum_{mn}u_{im}^2 u_{jn}^2
|P_i-m+\delta_{mP_i}|^{-2\alpha_H} |P_j-n+\delta_{nP_j}|^{-2\alpha_H} |m-n+\delta_{mn}|^{-2\alpha_L} \\
\nonumber
&&\hspace{3.86cm}+\sum_{mn}u_{im}u_{jm}u_{in}u_{jn}
|P_i-m+\delta_{mP_i}|^{-\alpha_H} |P_j-m+\delta_{mP_j}|^{-\alpha_H}\\
\nonumber
&&\hspace{4.75cm}\times|P_i-n+\delta_{nP_i}|^{-\alpha_H} |P_j-n+\delta_{nP_j}|^{-\alpha_H}|m-n+\delta_{mn}|^{-2\alpha_L}{\Big\}} \\
&\sim & \frac{\gamma}{a(N,\alpha_L)\left[a(\alpha_H,N)-1\right]^2}\left\{ u_{iP_i}^2u_{jP_j}^2 |P_i-P_j|^{-2\alpha_L} +
\left[u_{iP_i}^2u_{jP_i}^2+u_{iP_j}^2u_{jP_j}^2\right]|P_i-P_j|^{-2\alpha_H} \right\},
%&\sim & \frac{\gamma}{a(N,\alpha_L)\left[a(\alpha_H,N)-1\right]^2}|P(i)-P(j)|^{-{\rm min}(2\alpha_H,2\alpha_L)},
\end{eqnarray}
where we have assumed real $U$ and neglected edge effects (as appropriate for CPRBM).
Since we treat $\alpha_H>1$ the sums are controlled by summands of order 1. In the first sum this occurs
when $m=P_i$ and $n=P_j$ or when $m=n=P_i$ or $m=n=P_j$. The latter case also dominates the second sum.
Thus, under the specified assumptions and after replacing the fluctuating $u_{ij}^2$ by their mean $\langle A\rangle_L$
is approximately similar to $A_0$, Eq. (\ref{eq:barA0}), with $\alpha={\rm min}(\alpha_H,\alpha_L)$. This in turn would imply
that the gap closes for $\alpha_L>1/2$, as we indeed find by numerically diagonalizing $A$.
Note, however, that the $N$-scaling of the numerical results, Fig. \ref{fig-smallg-Agap}, deviates from the
expected behavior of $A_0$ for the corresponding $\alpha$, Eq. (\ref{eq:lambda1}), and roughly follows
$N^{-(\alpha_H-1)}$ in the range of parameters that we studied. This is most likely due to our neglect of
the mutual orthogonality of the energy wavefunctions, their more complicated structure, which especially in the range $3/2>\alpha_H>1$
often exhibits more than one peak and finally, the effects of fluctuations.

\subsection{Small $\gamma$ and $\alpha_H<1/2$ }
\label{sec:weaklargeN}

In this range of parameters we first discuss the consequences of Eq. (\ref{eq:Asmall})
and then indicate when and how its predictions are expected to fail in the $N\rightarrow\infty$ limit.

\subsubsection{1.~~~$N<\exp(4/\gamma^2)$}
\label{sss-small}

For $\alpha_H<1/2$ the statistical properties of the energy eigenfunctions are close to those exhibited by the GOE.
Hence, we approximate the elements of $U$ as independent (neglecting the
othonormality conditions) normal random variables with zero mean and variance $1/N$. Under this approximation it is
easy to verify that $U_{im}{(L_0)}_{mn} U^\dagger_{nj}$ form a set of independent random variables that obey
the Lyapunov condition for the central limit theorem. Hence, as $N\rightarrow \infty$, the off-diagonal elements $A_{ij}$,
see Eq. (\ref{eq:modelA}), are chi-squared distributed in the ensemble of Hamiltonians and jump operators
(whose statistics we denote by the subscript $LH$) with mean and standard deviation
\begin{eqnarray}
\label{eq:Aoffmeanvar}
\nonumber
\left\langle A_{ij}\right\rangle_{LH}&=&\frac{\gamma}{2a(N,\alpha_L)N^2}\left[\sum_{mn}f_{L,mn}^2 +N\right] =\frac{\gamma}{2N},\\
\sigma_{LH}(A_{ij})&=&\sqrt{2}\left\langle A_{ij}\right\rangle_{LH}=\frac{\gamma}{\sqrt{2}N},
\end{eqnarray}
where we have assumed CPRBM and used $\sum_n f_{L,mn}^2 +1=a(N,\alpha_L)+(N/2)^{-2\alpha_L}$, which follows from Eq. (\ref{eq:adef}).
Consequently, owing to the central limit theorem the diagonal elements $A_{ii}=-\sum_{j\neq i}A_{ij}$ are normally distributed with
slight dependence between them ($L_{ij}$ appears both in $A_{ii}$ and $A_{jj}$) and
\begin{eqnarray}
\label{eq:Aiimeanvar}
\nonumber
\left\langle A_{ii}\right\rangle_{LH}&=&=-\frac{\gamma}{2},\\
\sigma_{LH}(A_{ii})&=&\frac{\gamma}{\sqrt{2N}}.
\end{eqnarray}

We have shown that such a behavior leads to a gap $\Delta=\gamma/2$ \cite{ours}. One can reach the same conclusion based on
the observation that for $\alpha_H<1/2$ the slowest decaying mode is dominated by a single $H$-population  (see Fig. \ref{fig-LAIPRLH})
for which $|A_{ii}|$ is minimal. To estimate the gap, which is close to this matrix element, consider the probability distribution of the largest
value, $X$, among $N$ iid random variables with Gaussian probability distribution $f_G(x)$ and cumulative distribution function $F_G(x)$
\begin{equation}
\label{eq:largeval}
f_X(X)=Nf_G(X)[F_G(X)]^{N-1}=\frac{N}{2^{N-1} \sqrt{2\pi\sigma^2}}e^{-\frac{(X-\mu)^2}{2\sigma^2}}
\left[1+{\rm erf}\left(\frac{X-\mu}{\sqrt{2}\sigma}\right)\right]^{N-1}.
\end{equation}
Its mean can be estimated using the saddle-point approximation which up to logarithmic accuracy leads to
\begin{equation}
\label{eq:meanlargeval}
\mu(X)\approx \mu+\sigma\sqrt{2\ln\left(\frac{N}{\sqrt{2\pi}}\right)}.
\end{equation}
Applying this result to the diagonal elements of $A$ implies that as $N\rightarrow\infty$ the smallest $|A_{ii}|$ resides at $-\gamma/2$.

To evaluate the perturbative correction to the gap consider the difference, $u$, between the two largest diagonal
elements of $A$ (the ones with smallest absolute values). Its distribution is given by
\begin{eqnarray}
\label{eq:udist}
\nonumber
f_u(u)&=&\int_{-\infty}^\infty dx N(N-1)f_G(x)f_G(x-u)[F_G(x-u)]^{N-2} \\
&=&\frac{N}{2^{N-1}\sqrt{2\pi}\sigma}
\int_{-\infty}^\infty dx \left(x+\frac{u}{\sigma}\right)e^{-\frac{1}{2}\left(x+\frac{u}{\sigma}\right)^2}
\left[1+{\rm erf}\left(\frac{x}{\sqrt{2}}\right)\right]^{N-1}
\underset{u\ll\sigma}{\approx}\frac{1}{\sigma}\sqrt{2\ln\left(\frac{N}{\sqrt{2\pi}}\right)}.
\end{eqnarray}
The gap may close when the correction to the extremal eigenvalue $\sum_{j\neq i}|A_{ij}|^2/(A_{ii}-A_{ij})<\sum_{j\neq i}|A_{ij}|^2/u$
becomes of order $|A_{ii}|$. Since both the distributions of $A_{ii}$, Eq. (\ref{eq:Aiimeanvar}), and of $\sum_{j\neq i}|A_{ij}|^2$
(approximately normal with mean $3\gamma^2/4N$ and standard deviation $\sqrt{6}\gamma^2/N^{3/2}$) become sharp in the large $N$ limit
the condition for gap closure becomes $u<\left\langle \sum_{j\neq i} |A_{ij}|^2\right\rangle_{LH}/\langle|A_{ii}|\rangle_{LH}\approx \gamma/N$.
Since $\gamma/N\ll \sigma_{LH}(A_{ii})$
we can use Eq. (\ref{eq:udist}) to find that the probability for this to happen scales as
\begin{equation}
\label{eq:pertbreak}
P\left(u<\gamma/N\right)\approx \frac{\gamma}{N\sigma_{LH}(A_{ii})}\sqrt{2\ln\left(\frac{N}{\sqrt{2\pi}}\right)}
\sim\sqrt{\frac{\ln N}{N}},
\end{equation}
and thus vanishes in the large $N$ limit.

Our numerical results, presented in Fig. \ref{fig-smallg-Agap}, agree with the conclusion that $\Delta=\gamma/2$.
The full Linbladian spectrum, as presented in the main text, also points towards a gapped phase for $\alpha_H<1/2$,
despite the fact that the $H$-populations fail to provide an accurate description of the slowest decaying mode
in the small-$\gamma$ large-$N$ limit. In particular, the mode is not dominated by a single $H$-population, see Fig. \ref{fig-LAIPRLH}.
As we argue next, this apparent gap is expected to close down in the limit $N\rightarrow\infty$, owing to rare events
that occur when $\alpha_L>1/2$.

\subsubsection{2.~~~$N>\exp(4/\gamma^2)$}
\label{sss-large}

The deduction of the Lindbladian gap from Eq. (\ref{eq:Asmall}) is based on the assumption that in the limit of
weak decoherence the lowest lying Lindbladian eigenstates originate from the $H$-populations, which are zero modes
in the strict $\gamma=0$ case. There is, however, another way by which a low-lying state may emerge.
For this to happen, the extremal $L$-eigenvalue, \ie, the one with the largest magnitude eigenvalue $\kappa_{\rm ext}$,
needs to obey $|\kappa_{\rm ext}|\gg\gamma^{-1}$. Such an eigenstate may emerge for $\alpha_L>1/2$, where the $L$-spectrum is unbounded.
The resulting low-lying state is then adiabatically connected to the corresponding $L$-population
$|\kappa_{\rm ext}\rangle\langle\kappa_{\rm ext}|$.

\begin{figure}[!!!h]
\begin{center}
\includegraphics[width = 0.428\textwidth]{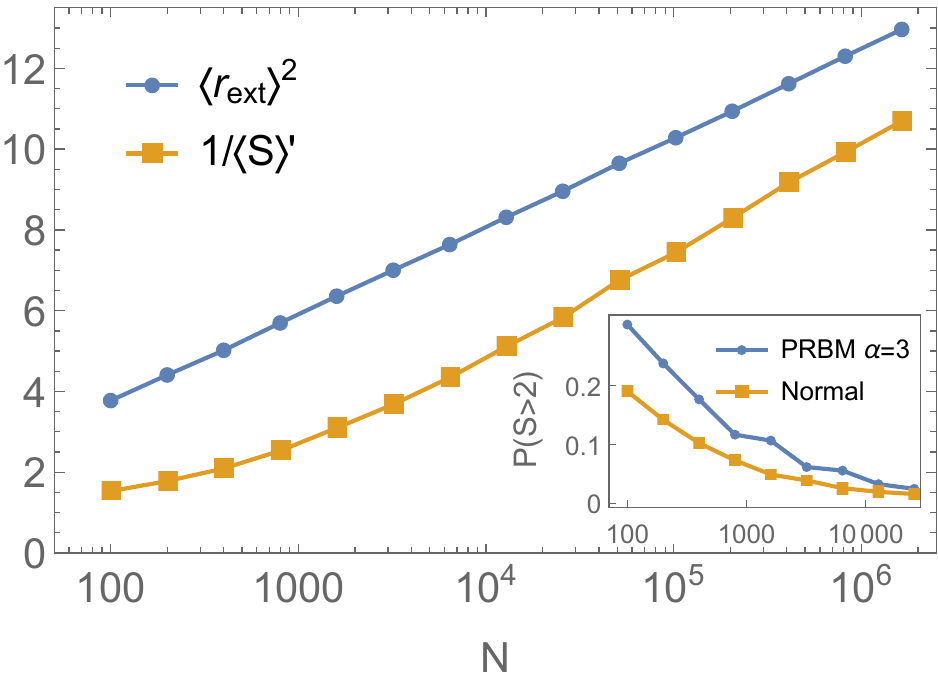}
\hspace{5mm}
\includegraphics[width = 0.42\textwidth]{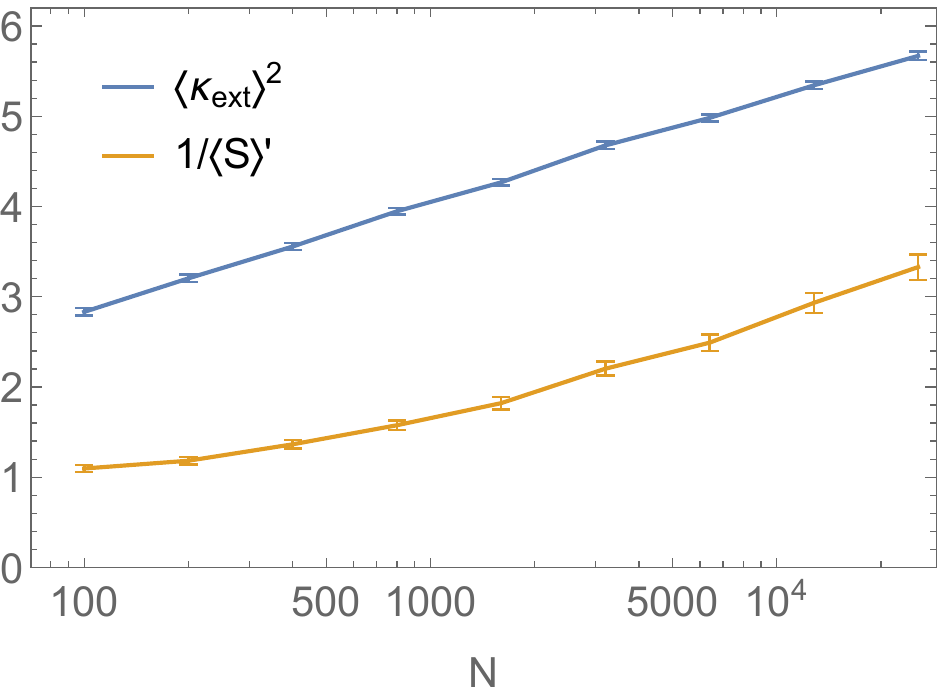}
\end{center}
\vspace{-0.5cm} \caption[]{\small Left: The (squared) mean of the extremal value $r_{\rm ext}$ among a sample of $N$ normal random variables $r_i$
with zero mean and variance $1/2$. Evidently, $\langle |r_{\rm ext}|\rangle\sim \sqrt{\log N}$. The average of
$S=(1/N)\sum_{r_i\neq r_{\rm ext}} (r_i-r_{\rm ext})^{-2}$ appears to diverge with $N$ due to events with small spacing between $r_{\rm ext}$
and its neighbor. However, if one filters out such events by eliminating from the sample the cases for which $S>2$ (they become rare with $N$,
as shown by the inset) one obtains an average of $S$ that scale approximately as $1/\log N$. Right: a similar behavior is shown by the
distribution of eigenvalues of a PRBM with $\alpha=3$.}
\label{fig-extremes}
\end{figure}

To see this, start from the $H=0$ limit where $|\kappa_{\rm ext}\rangle\langle\kappa_{\rm ext}|$ (as all the other $L$-populations)
is a zero mode. A non-zero Hamiltonian couples $|\kappa_{\rm ext}\rangle\langle\kappa_{\rm ext}|$ to $L$-coherences of the form
$|\kappa_{\rm ext}\rangle\langle\kappa_i|$ and $|\kappa_i\rangle\langle\kappa_{\rm ext}|$. The latter are eigenmodes
of the $H=0$ Lindbladian with eigenvalues $-(\gamma/2)(\kappa_{\rm ext}-\kappa_i)^2$. Consequently, second order perturbation
theory leads to a shift in the eigenvalue of $|\kappa_{\rm ext}\rangle\langle\kappa_{\rm ext}|$ of magnitude
$\Delta\lambda=-(4/\gamma)\sum_{\kappa_i \neq \kappa_{\rm ext}}|H_{\kappa_i\kappa_{\rm ext}}|^2/(\kappa_{\rm ext}-\kappa_i)^2$,
where $H_{\kappa_i\kappa_j}=\langle\kappa_i|H|\kappa_j\rangle$. To make progress with estimating this sum we model the
$\kappa_i$s as independent normal variables. Our numerical calculations, presented in Fig. \ref{fig-extremes}, provide evidence
that this approximation is reasonable for our purposes.
The numerics also indicates that the average $\langle\Delta\lambda\rangle_{LH}$ is dominated by rare events with small
spacing between $\kappa_{\rm ext}$ and its neighboring eigenvalue. Such events make the average diverge with $N$
but their proportion decrease with $N$, see Fig. \ref{fig-extremes}. Moreover, if they occur then one should apply degenerate
perturbation theory to the couplings between the pair of close eigenvalues. Doing so, which is equivalent to diagonalizing $A$
in the form given by Eq. (\ref{eq:Alarge}) (see Fig. \ref{fig-gap-largeg-aH0-aL3} in Sec. \ref{sec:largeg-aLgt1})
results in $\Delta\lambda$ that is similar to its value when the level spacing is not small.

Consequently, to estimate $\langle\Delta\lambda\rangle_{LH}$ we may assume that the
main contribution to the sum comes from states in the bulk of the $L$-spectrum $|\kappa|\lesssim 1$.
Furthermore, we break the sum into windows $1/N\ll\Delta\kappa\ll 1$ containing $n\gg1$ levels.
Each such window contributes approximately $-(4/\gamma)/(\kappa_{\rm ext}-\overline{\kappa_i})^2\sum_{\kappa_i \in\Delta\kappa}|H_{\kappa_i\kappa_{\rm ext}}|^2$.
By the central limit theorem $\sum_{\kappa_i \in\Delta\kappa}|H_{\kappa_i\kappa_{\rm ext}}|^2$ is normally distributed with mean $n\left\langle |H_{\kappa_i\kappa_{\rm ext}}|^2\right\rangle_{LH}$
and standard deviation $\sqrt{n}\sigma_{LH}(|H_{\kappa_i\kappa_{\rm ext}}|^2)$. One readily finds for $i\neq j$ that $\left\langle |H_{\kappa_i\kappa_j}|^2\right\rangle_{LH}\simeq 1/(2N)$, and
in the limit of well localized $L$-eigenstates (large $\alpha_L$)
\begin{equation}
\label{eq:sigmanumscaling}
\sigma_{LH}((|H_{\kappa_i\kappa_j}|^2)\sim\left\{
\begin{array}{cc}
N^{-1} & \;\;\; \alpha_H<1/4 \\
N^{-3/2+2\alpha_H} & \;\;\; 1/4<\alpha_H<1/2 \\
N^{-1/2} & \;\;\; 1/2<\alpha_H
\end{array}
\right. .
\end{equation}
(numerically, we find a somewhat slower decay away from the large $\alpha_L$ limit for $\alpha>1/4$). Hence, the fluctuations in
$\sum_{\kappa_i \in\Delta\kappa}|H_{\kappa_i\kappa_{\rm ext}}|^2$ become negligible compared to its mean for $n>N^{{\rm max}(0,4\alpha_H-1)}$, as
long as $\alpha_H<1/2$. Replacing the numerator by its average we finally obtain
\begin{equation}
\label{eq:estlam1}
\langle\Delta\lambda\rangle_{LH}\simeq-\frac{2}{\gamma N}\sum_{\kappa_i \neq \kappa_{\rm ext}}
\left\langle\frac{1}{(\kappa_{\rm ext}-\kappa_i)^2}\right\rangle_{L}' \sim -\frac{2}{\gamma}
\left\langle\frac{1}{\kappa_{\rm ext}^2}\right\rangle_{L} \simeq -\frac{2}{\gamma\log N},
\end{equation}
where $\langle\rangle_{L}'$ denotes the restricted mean calculated by filtering out the rare events for which the denominator is much smaller than $1/N$.
In evaluating Eq. (\ref{eq:estlam1}) we have applied Eq. (\ref{eq:meanlargeval}) to get $\left\langle\kappa_{\rm ext}\right\rangle_L\sim\sqrt{\log N}$
and checked that $\left\langle 1/\kappa_{\rm ext}^2\right\rangle_L=1/[\left\langle \kappa_{\rm ext}\right\rangle_L]^2$ for large $N$.

%We are interested in the case where the main contribution to this sum
%comes from states in the bulk of the $L$-spectrum $|\kappa|\lesssim 1$, and not from the tails. To this end, we require (assuming $\kappa_N>0$) that there is a vanishing probability to find $|H_{\kappa_{N-1}\kappa_N}|^2/u^2>1/\kappa_N^2$, where
%$u=\kappa_N-\kappa_{N-1}$. If this holds true we may estimate $\langle \lambda_1\rangle_{LH}$ by
%\begin{equation}
%\label{eq:estlam1}
%\langle\lambda_1\rangle_{LH}\sim-\frac{1}{\gamma}\sum_{i \neq N}\frac{\left\langle|H_{\kappa_i\kappa_N}|^2\right\rangle_{LH}}
%{\langle(\kappa_N-\kappa_i)^2\rangle_{L}}\sim-\left\langle\frac{1}{\gamma\kappa_N^2}\right\rangle_{L}\sim-\frac{1}{\gamma\log N},
%\end{equation}
%%Here we have assumed that the main contribution to the integral comes from $|\kappa|\lesssim 1$. To this end, one requires that
%%$\kappa_N\gg 1$ and that there is vanishing probability to find the distance $u=\kappa_N-\kappa_{N-1}$ obeying
%%$\left\langle(H_{\kappa_{N-1}\kappa_N})^2\right\rangle_H/u^2>1/\kappa_N^2$.
%where we used $\left\langle |H_{\kappa_i\kappa_N}|^2\right\rangle_{LH}\sim 1/N$, and applied Eq. (\ref{eq:meanlargeval}) to
%the spectrum of $L$ to find $\left\langle\kappa_N\right\rangle_L\sim\sqrt{\log N}$. We also checked that
%$\left\langle 1/\kappa_N^2\right\rangle_L=1/[\left\langle \kappa_N\right\rangle_L]^2$ for large $N$.
%Thus, for $\kappa_N>\gamma^{-1}$ one finds that $|\lambda_1|$ is smaller than the gap $\gamma/2$ that was derived in the preceding
%subsection using the $H$-populations.

\vspace{2mm}
\begin{figure}[!!!h]
\begin{center}
\includegraphics[width = 0.51\textwidth]{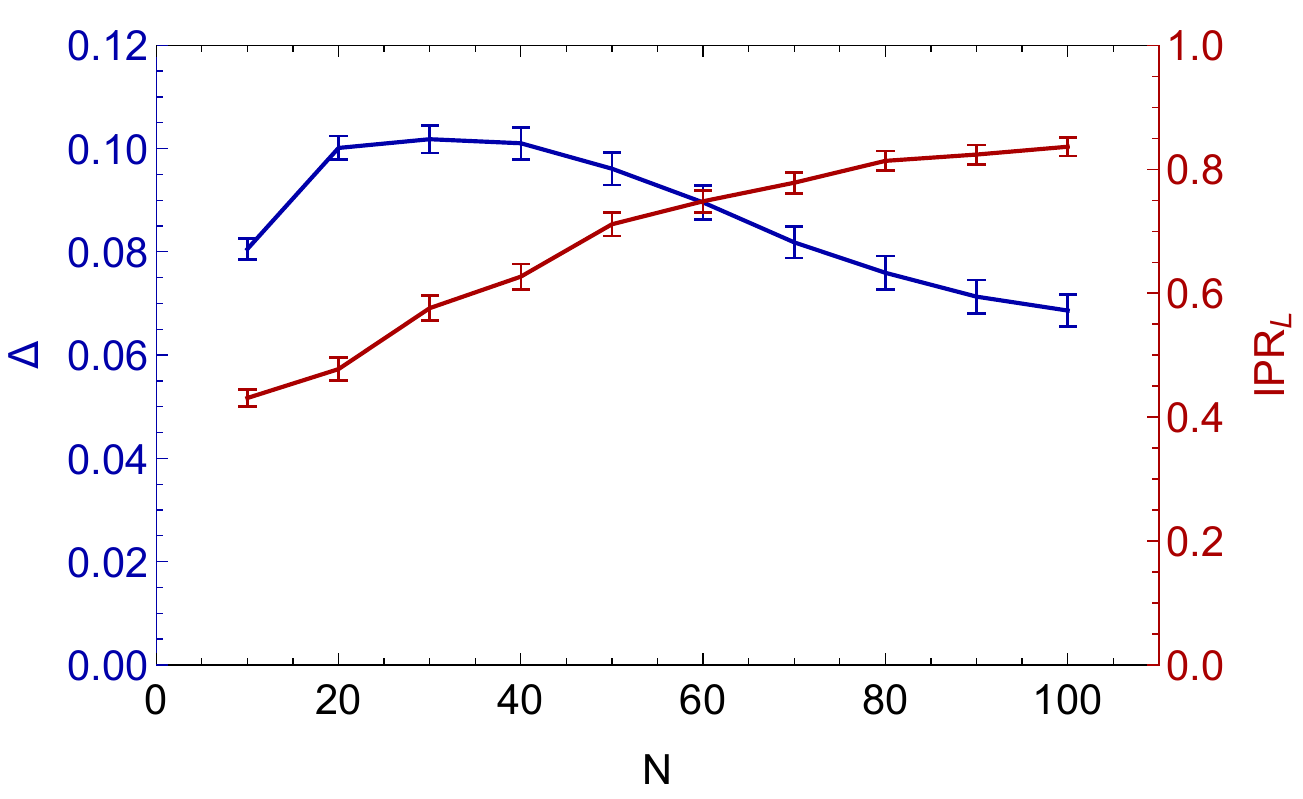}
\end{center}
\vspace{-0.5cm} \caption[]{\small The Lindbladian spectral gap and the IPR$_L$ of the slowest decaying mode of $\L$ in a system with $\gamma=0.2$,
$\alpha_H=0$ and a jump operator whose spectrum is distributed according to $2^{13/2}/\pi(4^4+\kappa^4)$.}
\label{fig-fattail}
\end{figure}

Therefore, we conclude that for $N> e^{4/\gamma^2}$, where $|\left\langle\Delta\lambda\right\rangle_{LH}|<\gamma/2$,
the gap is set by the population of the extremal $L$-eigenstate and follows
\begin{equation}
\label{eq:largeNgaprare}
\left\langle\Delta\right\rangle_{LH}\sim\frac{1}{\gamma\log N}.
\end{equation}

%Applying Eq. (\ref{eq:meanlargeval}) to $\nu(\kappa)$ yields $\left\langle\kappa_N\right\rangle_L\sim\sqrt{\log N}$.
%We have checked that
%$\left\langle 1/\kappa_N^2\right\rangle_L=1/[\left\langle \kappa_N\right\rangle_L]^2$ for large $N$.
%Furthermore, Eq. (\ref{eq:udist}) implies $P\left(u<\sqrt{\langle\kappa_N\rangle_L^2\langle |H_{\kappa_{N-1}\kappa_N}|^2\rangle_H}\right)\sim \log N/\sqrt{N}$,
%thus allowing us to use Eq. (\ref{eq:estlam1}) to estimate
%$\left\langle\lambda_1\right\rangle_{LH}\sim -1/(\gamma\log N)$. Therefore, we conclude that for $N> e^{1/\gamma^2}$, where
%$|\left\langle\lambda_1\right\rangle_{LH}|<\gamma/2$, the gap is set by $\lambda_1$ and follows
%\begin{equation}
%\label{eq:largeNgaprare}
%\left\langle\Delta\right\rangle_{LH}\sim\frac{1}{\gamma\log N}.
%\end{equation}

The large system sizes needed to observe the decrease of the gap with $N$ render a direct numerical confirmation of the effect
within the PRBM model impossible. Instead, we have calculated the gap in a model where the jump operator is diagonal in the
$x$-basis with a fat-tailed distribution of eigenvalues decaying as $\nu(\kappa)\sim \kappa^{-4}$, such that its spectrum still
possesses a finite variance but exhibits a more favorable scaling of the extreme eigenvalue, which approximately follows $N^{1/3}$.
Fig. \ref{fig-fattail} depicts the $N$ dependence of the gap and the IPR$_L$ of the slowest decaying mode, and shows that at large $N$
the gap is decreasing and set by the population of the largest $L$-eigenvalue.

\subsection{Small $\gamma$ and $1>\alpha_H>1/2$}
\label{ss-smallgaHinter}

As discussed in the main text, the $H$-populations fail to provide an effective description
of the slowest-decaying $\L$-eigenstate when one considers the thermodynamic limit for small but fixed $\gamma$ and $1>\alpha_H>1/2$.
Here, instead, we focus on the fixed $N$ and $\gamma\rightarrow 0$ limit where $A$ still provides a faithful description of the asymptotic
dynamics.

Fig. \ref{fig-LAIPRLH} shows that the slowest-decaying mode of $A$ for $1>\alpha_H>1/2$ continues to be dominated by a single
$H$-population - the one with the smallest $|A_{ii}|$, which approximately sets the gap. However, in contrast to the case $\alpha_H<1/2$,
where all diagonal elements of $A$ are sharply concentrated around $-\gamma/2$, the elements associated with
the localized states at the edges of the spectrum for $1>\alpha_H>1/2$ are much more broadly distributed. To see this,
consider the mean and variance of the diagonal elements over the ensemble of $L_0$. One finds,
\begin{eqnarray}
\left\langle A_{ii}\right\rangle_L&=& -\frac{\gamma}{2a(N,\alpha_L)}\left[a(N,\alpha_L)-2\sum_{mn}(U_{im})^2 (U_{in})^2 f_{L,mn}^2\right], \\
\nonumber
\sigma_L^2(A_{ii})&=& \frac{\gamma^2}{2a^2(N,\alpha_L)}\sum_{mn}(U_{im})^2(U_{in})^2 f_{L,mn}^2\left[a(N,\alpha_L)+1+f_{L,mn}^2
-8\sum_p (U_{ip})^2f_{L,np}^2+4\sum_{pq}(U_{ip})^2(U_{iq})^2f_{L,pq}^2\right].
\end{eqnarray}

If $U$ follows a Porter-Thomas statistics, as outlined at the beginning of Sec. \ref{sec:weaklargeN}, one recovers Eq. (\ref{eq:Aiimeanvar}).
The same behavior is obtained in the opposite limit of an extremely localized state with $U_{im}=\delta_{im}$, as long as $\alpha_L<1/2$.
On the other hand, for $\alpha_L>1/2$ the above expressions lead in the large-$N$ limit to
\begin{eqnarray}
\label{eq:extremeloc}
\nonumber
\left\langle A_{ii}\right\rangle_L&=&-\frac{\zeta(2\alpha_L)}{2[1+\zeta(2\alpha_L)]}\gamma, \\
\sigma_L(A_{ii})&=&\frac{\sqrt{\zeta(2\alpha_L)}}{2[1+\zeta(2\alpha_L)]}\gamma.
\end{eqnarray}

\begin{figure}[!!!t]
\begin{center}
\includegraphics[width = 0.405\textwidth]{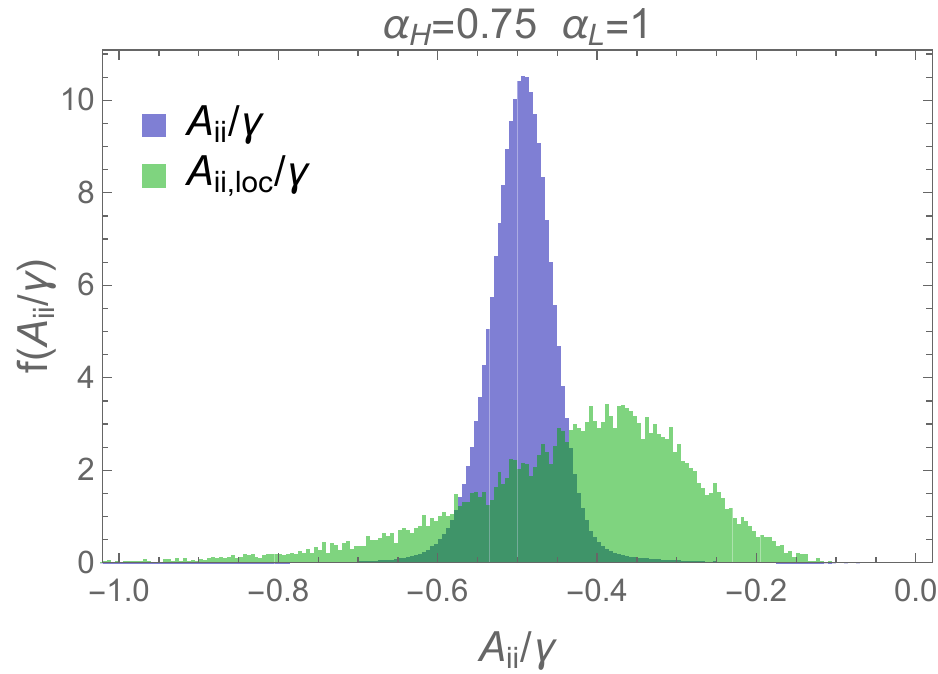}
\hspace{3mm}
\includegraphics[width = 0.423\textwidth]{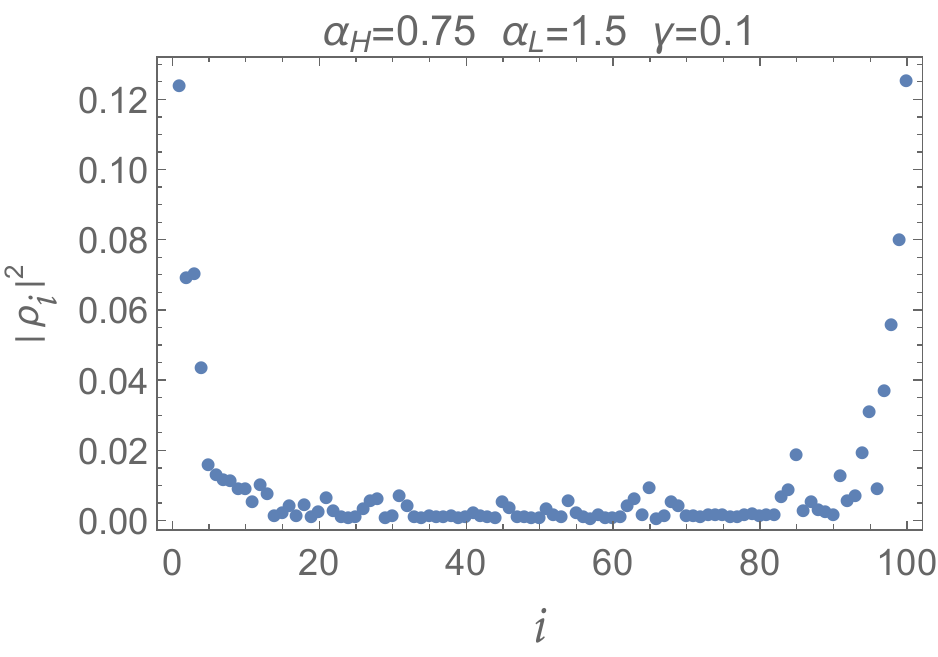}
\end{center}
\vspace{-0.5cm} \caption[]{\small Left:The distribution of all diagonal elements of $A$ (blue) for a system
with $\alpha_H=0.75$, $\alpha_L=1$, and $N=1000$. In green is the distribution of the subset of diagonal elements corresponding to
$H$-populations of energy eigenstates with IPR$>0.1$. Right: The squared components of the projected slowest decaying $\L$-mode onto
the $H$-population subspace $|i\rangle\langle i|$. Shown is an average over 180 realizations of a system with $N=100$, $\gamma=0.1$,
$\alpha_H=0.75$ and $\alpha_L=1.5$. The $H$-populations are indexed according to their energies. The two maxima at the extremes of
the spectrum are due to averaging. For a given realization the mode is typically concentrated near one of the edges.}
\label{fig-diagA}
\end{figure}

The off-diagonal elements $A_{ij}$ that couple a localized population to the rest of the population subspace do not obey the Lyapunov
condition and therefore the distribution of $A_{ii}=-\sum_{j\neq i}A_{ij}$ is not normal. Albeit, Eq. (\ref{eq:extremeloc}) indicates that
it is wide with comparable mean and standard deviation. Fig. \ref{fig-diagA} demonstrates that this conclusion holds also away from the
strongly localized limit and that the smallest $|A_{ii}|$ originates from populations of localized energy eigenstates. The same
conclusion is reached by examining the structure of the slowest decaying mode of $\L$ for a system with $N=100$, see
right panel of Fig. \ref{fig-diagA}.
%\begin{figure}[!!!h]
%\begin{center}
%\includegraphics[width = 0.4\textwidth]{vec1-g0p1-aH0p75-aL1p5}
%\end{center}
%\vspace{-0.5cm} \caption[]{\small The squared components of the projected slowest decaying $\L$-mode onto the $H$-population subspace
%$|i\rangle\langle i|$. Shown is an average over 180 realizations of a system with $N=100$, $\gamma=0.1$, $\alpha_H=0.75$ and $\alpha_L=1.5$.
%The $H$-populations are indexed according to their energies. The two maxima at the extremes of the spectrum are due to
%averaging. For a given realization the mode is typically concentrated near one of the edges.}
%\label{fig-vecsmallg}
%\end{figure}

Given the distribution $f_{\rm loc}(A_{ii})$ for localized populations and its cumulative distribution function $F_{\rm loc}(A_{ii})$
the distribution of $X={\rm max}(A_{ii})={\rm min}(|A_{ii}|)$ is given by (assuming independence between the diagonal terms)
$f_X(X)=n_{\rm loc}f_{\rm loc}(X)[F_{\rm loc}(X)]^{n_{\rm loc} -1}$, where $n_{\rm loc}$ is the number of localized energy eigenstates.
Despite the fact that we are not able to calculate $f_{\rm loc}(A_{ii})$, its breadth and the fact that $n_{\rm loc}$ increases with $N$,
see Fig. \ref{fig-loc}, imply a vanishing $\left\langle X\right\rangle_{LH}$ and thus a vanishing gap for large $N$.

Our numerical results (Fig. 1 in the main text) indicate that the $\L$-gap also vanishes as one fixes a small $\gamma$ and takes
$N\rightarrow\infty$. However, in this order of limits the slowest decaying $\L$-mode ceases to be dominated by a single $H$-population
and is delocalized both in energy and real spaces, see right panels of Figs. \ref{fig-LAIPRLH} and \ref{fig-LAIPRx}.

\vspace{-1mm}
\subsection{Large $\gamma$ and $\alpha_L<1/2$ }
\vspace{-0.5mm}

Consider the eigenvalue equation for $A$ in the $\gamma\rightarrow\infty$ limit, see Eq. (\ref{eq:Alarge}),
\begin{equation}
\label{eq:eigsA1}
\frac{4}{\gamma}\sum_{\substack{j=1\\ j\neq i}}^N\left[\sum_{mn}U_{im}(H_0)_{mn}U_{jn}\right]^2\frac{v_j-v_i}{(\kappa_j-\kappa_i)^2}=\lambda v_i,
\end{equation}
where $H_0$ is the $x$-representation of the Hamiltonian and $U$ is the unitary diagonalizing $L$.
For $\alpha_L<1/2$ the $L$-spectrum is largely delocalized. Approximating $U$ as a collection of random unit vectors
implies that $(H_{ij})^2=\left[\sum_{mn}U_{im}(H_0)_{mn}U_{jn}\right]^2$ are normally distributed
with $\langle|H_{ij}|^2\rangle_{LH}=1/2N$ and $\sigma_{LH}(\langle|H_{ij}|^2)=1/\sqrt{2}N$.
Indexing the eigenvector components not by the $L$-eigenstate index but by its corresponding eigenvalue $\kappa$,
and assuming that they change slowly with $\kappa$ such that the arguments of Sec. \ref{sss-small}2 allow to replace $(H_{ij})^2$
by its mean, turns Eq. (\ref{eq:eigsA1}) into
\begin{equation}
\label{eq:eigsA2}
\frac{2}{\gamma N}P\int_{-\sqrt{2}}^{\sqrt{2}}\!d\kappa\,\nu(\kappa)\frac{v(\kappa)-v(\tau)}{(\kappa-\tau)^2}=\lambda v(\tau),
\end{equation}
where $P$ stands for the principle value of the integral and the $\nu(\kappa)$ is the semicircle distribution of the $L$-states
\begin{equation}
\label{eq:semicircle}
\nu(\kappa)=\frac{N}{\pi}\sqrt{2-\kappa^2}\,\Theta(2-\kappa^2).
\end{equation}
The solutions to Eq. (\ref{eq:eigsA2}) take the form \cite{ours}
\begin{equation}
\label{eq:eigsAsol}
\lambda_n=-\frac{2}{\gamma}n,\;\;\;\;\;\; v_n(\kappa)=U_n\left(\frac{\kappa}{\sqrt{2}}\right),
\;\;\;\;\;\; n=0,1,2,\cdots
\end{equation}
where $U_n(x)$ are the Chebyshev polynomials of the second kind.

\begin{figure}[!!!h]
\begin{center}
\includegraphics[width = 0.43\textwidth]{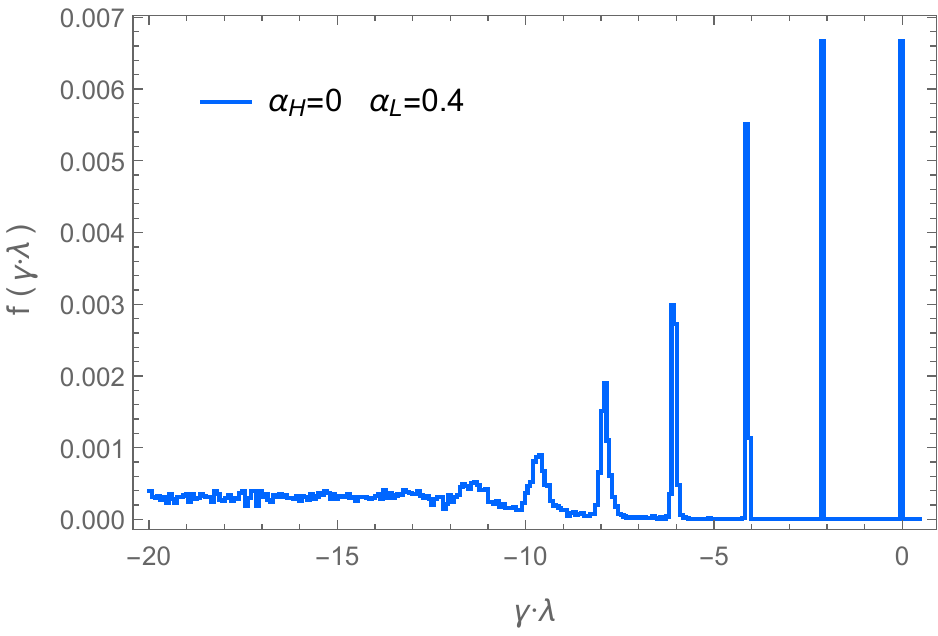}
\hspace{0.5cm}
\includegraphics[width = 0.43\textwidth]{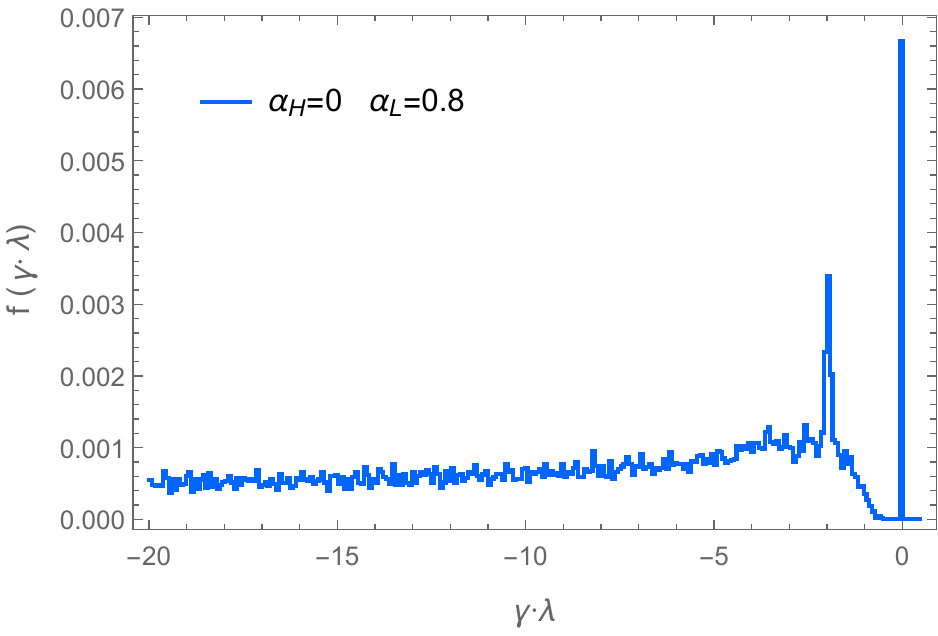}
\vspace{-0.3cm} \caption{Probability density of the eigenvalues of $A$ for $\gamma\rightarrow\infty$ with $N=2000$ averaged over 1000 realizations.}
\label{fig-largeg-DOS}
\end{center}
\end{figure}

\begin{figure}[!!!h]
\begin{center}
\includegraphics[width = 0.48\textwidth]{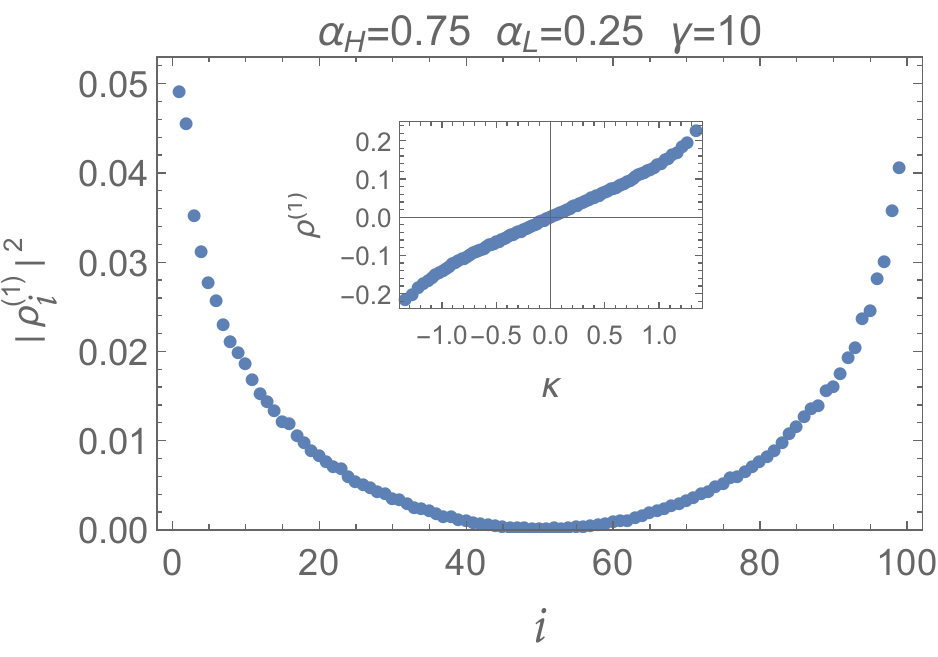}
\vspace{-0.3cm} \caption{The squared components of the projected slowest decaying $\L$-mode onto the $L$-population subspace
$|i\rangle\langle i|$. Shown is an average over 180 realizations of systems with $N=100$, $\gamma=10$.
The $L$-populations are indexed according to their $L$-eigenvalues.
The inset depicts the mode components as function of the $L$-eigenvalue $\kappa$ of the populations.}
\label{fig-vec1aH075aL025}
\end{center}
\end{figure}

The lowest-lying state near $2/\gamma$ is clearly visible in the $A$-DOS presented in Fig. \ref{fig-LADOS10}, where it also
shows up with a somewhat smaller eigenvalue in the DOS of the full Lindbladian. We find that more of the discrete states,
Eq. (\ref{eq:eigsAsol}), appear in the $A$-spectrum as $N$ is increased, see Fig. \ref{fig-largeg-DOS}. Our results suggest that
this also holds for the $\L$-spectrum, especially when $\gamma$ is increased as well. More directly, the inset of
Fig. \ref{fig-vec1aH075aL025} shows that the projected lowest-lying $\L$-mode for $\alpha_L<1/2$ is approximately linear in
the $L$-eigenvalue of the populations, as predicted by Eq. (\ref{eq:eigsAsol}).

\subsection{Large $\gamma$ and $1>\alpha_L>1/2$ }
\label{sec:largeg0p5-1}

\begin{figure}[!!!h]
\begin{center}
\includegraphics[width = 0.48\textwidth]{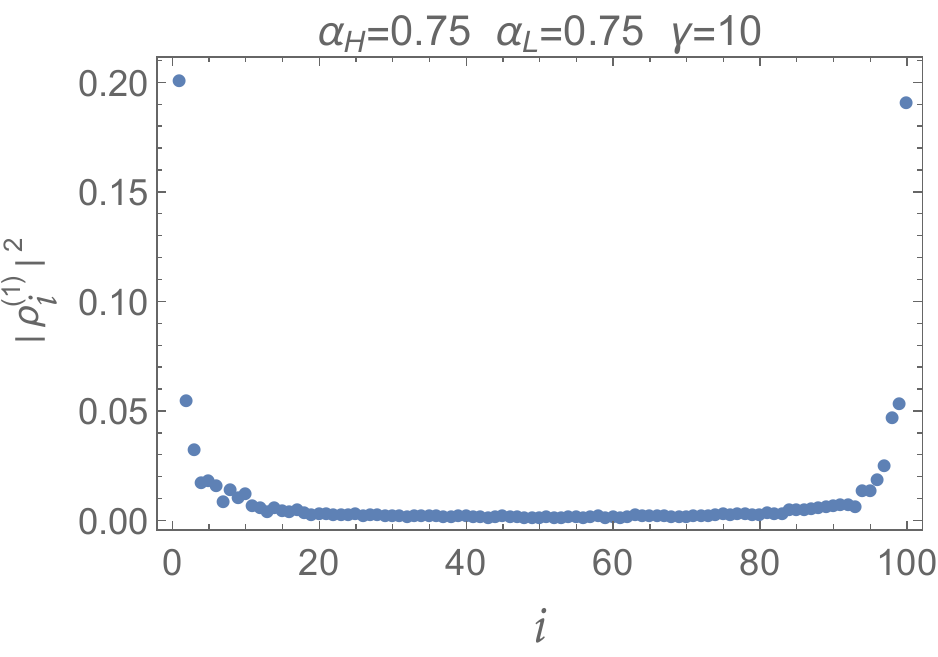}
\vspace{-0.3cm} \caption{The squared components of the projected slowest decaying $\L$-mode onto the $L$-population subspace
$|i\rangle\langle i|$. Shown is an average over 180 realizations of systems with $N=100$, $\gamma=10$.
The $L$-populations are indexed according to their $L$-eigenvalues. The two maxima at the extremes of
the spectrum are due to averaging. For a given realization the mode is typically concentrated near one of the edges.}
\label{fig-vec1aH075aL075}
\end{center}
\end{figure}

The numerical results, Fig. \ref{fig-largeg-Agap}, show that in this range the spectral gap of $A$ decreases very slowly with $N$.
Concomitantly, the discrete low-lying states that exist in the DOS at smaller values of $\alpha_L$ are largely smeared out and spectral
weight appears closer to the origin. This is evident in both the $A$-DOS and $\L$-DOS at relatively small values of $N$,
see Fig. \ref{fig-LADOS10}, but is particularly clear in the $A$-DOS for the larger values of $N$, as shown in
Fig. \ref{fig-largeg-DOS}. The mode which sets the gap becomes localized both in the $L$-populations subspace and in real space,
see Fig. \ref{fig-Alargeg}. It is dominated by the extremal eigenstates of $L$, as shown by Fig. \ref{fig-vec1aH075aL075}.

Since the bulk of the $L$-spectrum is delocalized both the mean and standard deviation of the Hamiltonian elements in the $L$-basis
scale approximately with $N^{-1}$, for all $\alpha_H$.
%have relatively weak fluctuations around $\left\langle |H_{\kappa\kappa'}|^2\right\rangle_{LH}=1/(2N)$, for all $\alpha_H$.
Hence, we may apply the analysis of Sec. \ref{sss-large}2
and conclude that the gap is expected to follow Eq. (\ref{eq:largeNgaprare}). The difference is
that now it suffices to have $\log N>1$ in order for the eigenvalue of the localized mode to reside below the one
produced by the extended mode, Eq. (\ref{eq:eigsAsol}).

The results that we present in the main text give evidence that the above discussion holds also for the spectrum of the full
Lindbladian in the limit of large but fixed $\gamma$ and $N\rightarrow\infty$, as long as $\alpha_H<1/2$. However, the
population dynamics fails to describe this limit in the range $\alpha_H>1/2$, where the slowest-decaying modes are hydrodynamical.

\subsection{Large $\gamma$ and $\alpha_L>1$ }
\label{sec:largeg-aLgt1}

When $\alpha_L>1$ the $L$-spectrum is unbounded and localized. If, in addition, $\alpha_H<1/2$, we may apply the analysis of Sec. \ref{sss-large}2.
Hence, one expects to find in this range of parameters a gap that decreases logaritmically with $N$ according to Eq. (\ref{eq:largeNgaprare})
and whose corresponding mode is comprised of the extremal $L$-population. These expectations are indeed borne out by the numerics.
\begin{figure}[!!!h]
\begin{center}
\includegraphics[width = 0.43\textwidth]{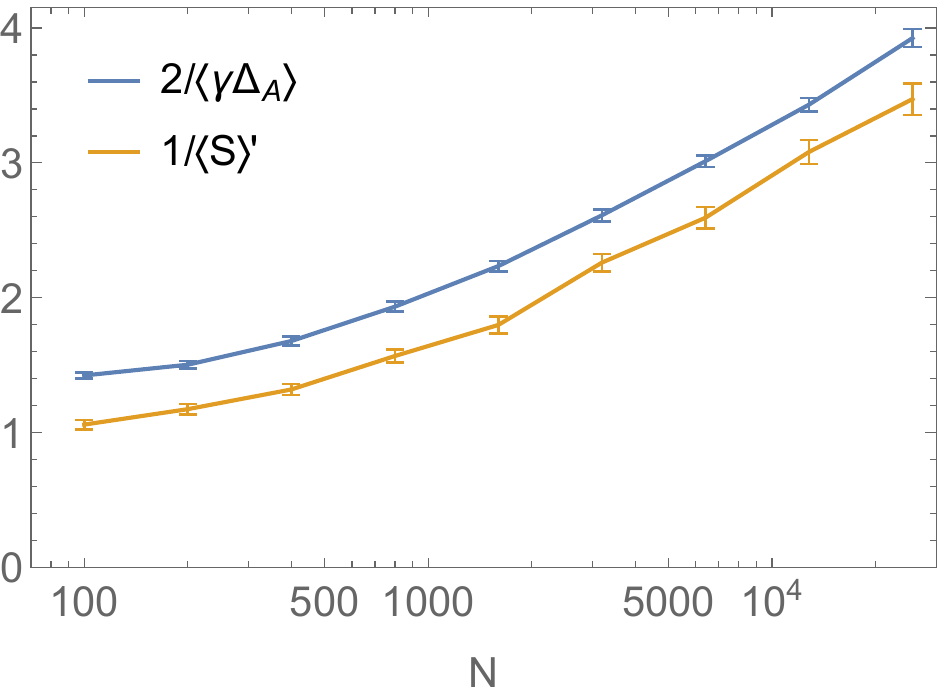}
\vspace{-0.3cm} \caption{The inverse of the $A$-gap for $\gamma\rightarrow\infty$ and $\alpha_H=0$, $\alpha_L=3$.
Also shown is the inverse mean of $S=\min[(1/N)\sum_{i\neq 1} (\kappa_i-\kappa_1)^{-2},(1/N)\sum_{i\neq N} (\kappa_i-\kappa_N)^{-2}]$,
for the same parameters. Here, $\kappa_1$ and $\kappa_N$ are the smallest (most negative) and largest $L$-eigenvalues. When calculating
the average of $S$ we filtered out events with $S>2$.}
\label{fig-gap-largeg-aH0-aL3}
\end{center}
\end{figure}

\begin{figure}[!!!t]
\begin{center}
\includegraphics[width = \textwidth]{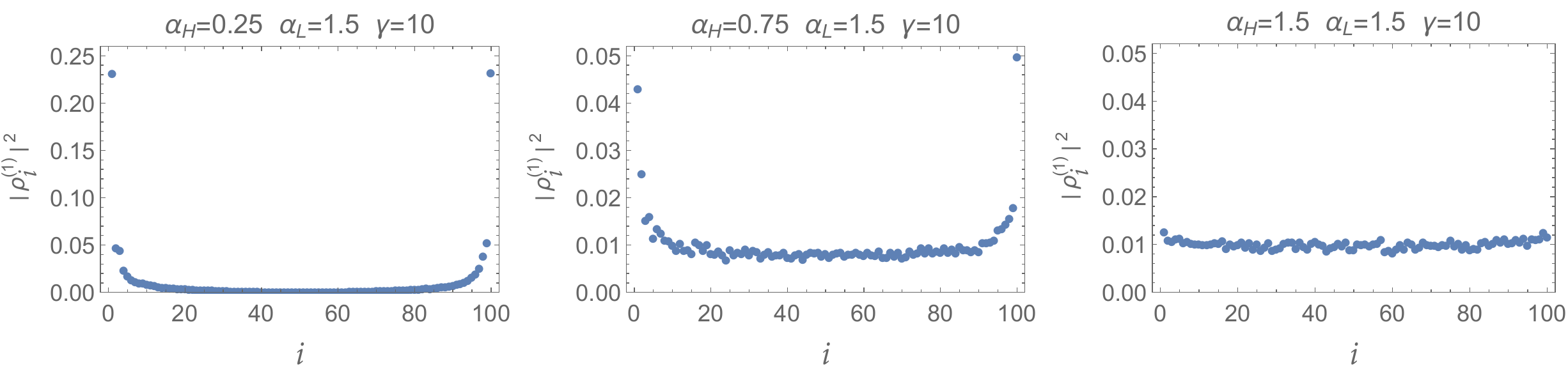}
\vspace{-0.3cm} \caption{
The squared components of the projected slowest decaying $\L$-mode onto the $L$-population subspace
$|i\rangle\langle i|$, averaged over 180 realizations of systems with $N=100$, $\gamma=10$.}
\label{fig-vec1aL1p5}
\end{center}
\end{figure}

Figure \ref{fig-gap-largeg-aH0-aL3} shows that the $A$-gap follows the sum of inverse square spacings
$\sum_{\kappa_i\neq \kappa_{\rm ext}} (\kappa_i-\kappa_{\rm ext})^{-2}$ (where here $\kappa_{\rm ext}$ is either the most
negative or the most positive $L$-eigenvalue), in accord with Eq. (\ref{eq:estlam1}). Concomitantly, Fig. \ref{fig-vec1aL1p5}
shows that for $\alpha_H<1/2$ the slowest decaying mode, not just of the effective population dynamics $A$, but of the full
Lindbladian is given approximately by the population of the extremal $L$-eigenstate. The same figure also demonstrates that
for $\alpha_H>1/2$ the nature of the slowest decaying $\L$-mode changes and becomes hydrodynamic. We note that this
stands in contrast to the slowest decaying eigenstate of $A$ in the $\gamma\rightarrow\infty$ limit,
which changes from being localized on the extremal $L$-population to being extended only when $\alpha_L>3$ and $\alpha_H>3/2$
(or when $\alpha_L>3/2$ and $\alpha_H>1/2$ for the case of the regularized $A$, Eq. (\ref{eq:Alarge}), as shown by Fig. \ref{fig-LAIPRLH}).

\section{Multiple jump operators}

Finally, we would like to consider the way by which our results for the case of a single jump operator generalize to
systems with multiple jump operators. To this end, consider the Lindblad master equation
\begin{equation}
\label{mastereq}
\partial_t \rho = \mathcal{L}\rho \equiv -i[H, \rho] + \sum_{k = 1}^{n_c} \gamma_k \left[ L_k \rho L_k
- \frac{1}{2} \{ L_k^2, \rho \} \right],
\end{equation}
where we have specialized to Hermitian $L_k$, each described by a PRBM governed by an exponent $\alpha_{L_k}$.
The number of noise channels $n_c$ can run up to $N^2-1$.

In the limit where $\sum_k\gamma_k\ll 1$ the effects of individual decoherence channels add up, and each channel can be analyzed
in a similar manner to the case of a single jump operator (Sec. \ref{ss-smallgloc} - \ref{ss-smallgaHinter}). Consequently,
if we denote $\tilde\alpha_L=\min(\alpha_{L_k})$, the perturbative and numerical analysis yields a gapless hydrodynamic phase
for $\alpha_H>1/2$ and $\tilde\alpha_L>1/2$. It predicts a gapped phase in the remaining part of the $\alpha_H$-$\tilde\alpha_L$ plane.
However, as we argue below, this conclusion fails in the range $\alpha_H<1/2$ and $\tilde\alpha_L>1/2$ where the system exhibits a
weakly gapless Lifshitz phase as $N\rightarrow\infty$.

The limit of strong decoherence is complicated by the fact that the jump operators generally do not commute.
Analytical progress can be made in the case where a single channel is much stronger than the others, i.e., $\gamma_1\gg 1$,
and $\gamma_1\gg \gamma_k$ for $k=2,\cdots ,n_c$. In this context, we concentrate on the case $n_c=2$ and consider the $L_1$-populations dynamics.
The generalization to larger number of channels is straightforward. If one turns off the unitary dynamics, namely sets $H=0$, the analysis
proceeds exactly along the lines of Sec. \ref{ss-smallgloc} - \ref{ss-smallgaHinter}, with the only difference being that the $A$ matrix,
Eq. (\ref{eq:Asmall}), is expressed in terms of matrix elements of $L_2$ in the eigenbasis of $L_1$ instead of $H$.
Borrowing the corresponding conclusions we expect to find a gapped phase when $\tilde\alpha_L<1/2$ and a gapless hydrodynamic
phase for $\tilde\alpha_L>1/2$

\begin{figure}[!!!t]
\begin{center}
\includegraphics[width = 0.48\textwidth]{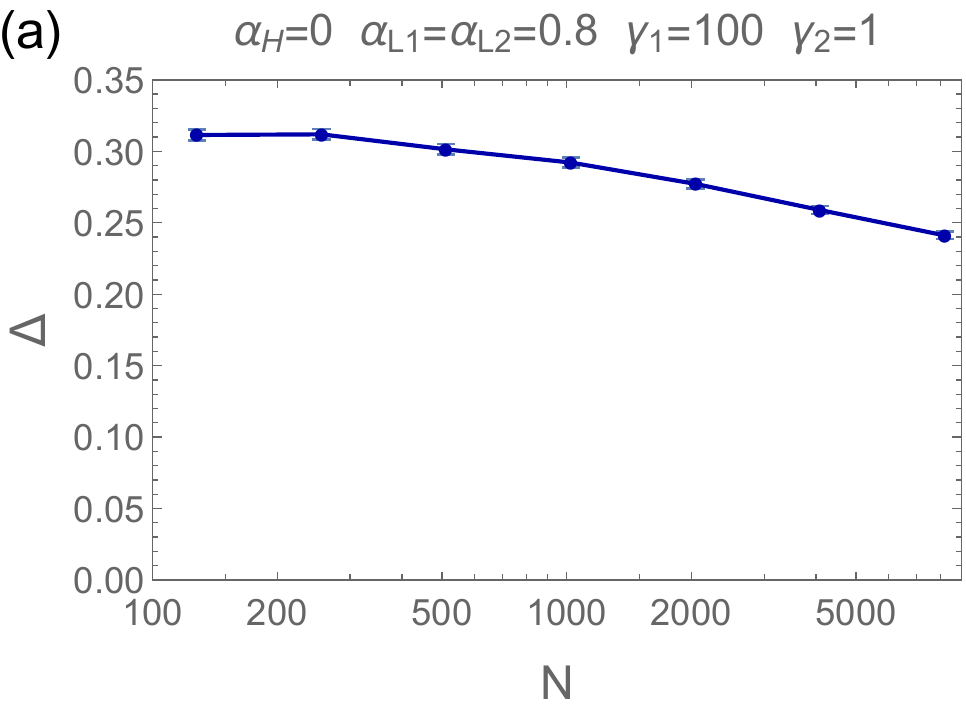}
\hspace{3mm}
\includegraphics[width = 0.471\textwidth]{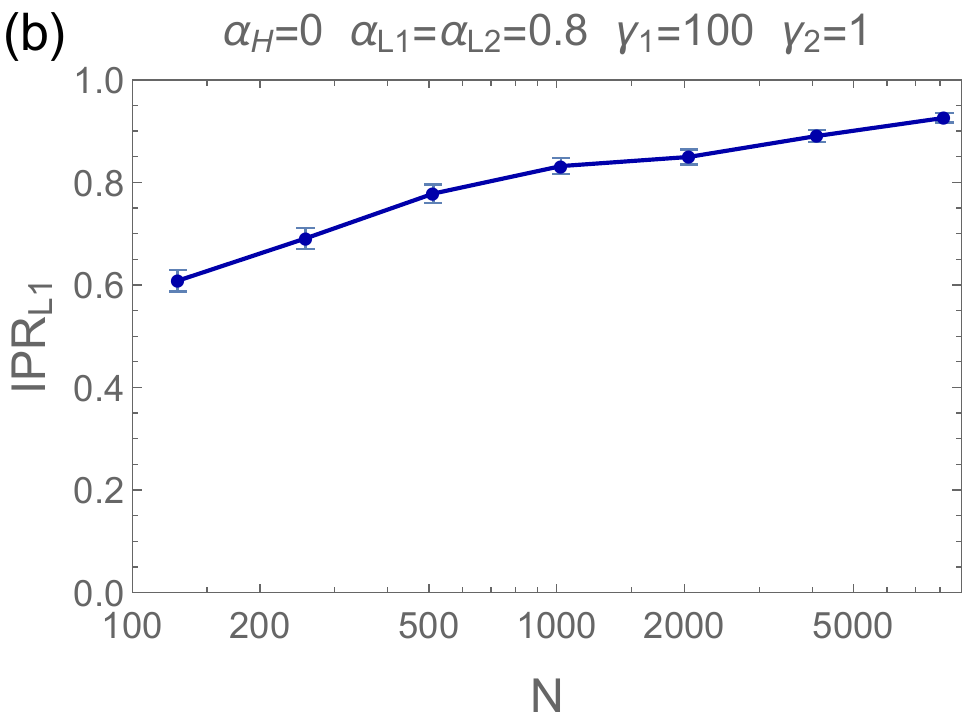}
\end{center}
\vspace{-0.5cm} \caption[]{\small (a) The spectral gap calculated from the effective dynamics of the $L_1$-populations (the $A$ matrix)
for a system with a Hamiltonian and two jump operators. (b) The IPR in the $L_1$ basis of the slowest decaying mode. }
\label{fig-A-HL2}
\end{figure}

We now reintroduce $H$ and take into account its effects on the $L_1$-populations dynamics, as expressed by Eq. (\ref{eq:Alarge}).
Hence, we end up with an effective $A$ which is the sum of Eqs. (\ref{eq:Asmall}) and (\ref{eq:Alarge}). Both parts lead to a
gapped phase for $\tilde\alpha_L<1/2$ and a gapless hydrodynamic phase when $\alpha_H>1/2$ and  $\tilde\alpha_L>1/2$.
However, following the discussion
of Sec. \ref{sec:largeg0p5-1} - \ref{sec:largeg-aLgt1}, we conclude that the $H$-induced part of $A$ leads to a Lifshitz phase
for $\alpha_H<1/2$ and $\alpha_{L_1}> 1/2$. We have checked numerically that this conclusion remains valid when the effects
of both $H$ and $L_2$ are included in $A$, as seen in Fig. \ref{fig-A-HL2}. Consequently, the arguments presented in Sec. \ref{sec:weaklargeN}
imply that a system with $\alpha_H<1/2$ and $\tilde\alpha_L>1/2$ is expected to exhibit a gapless Lifshitz phase also for small $\gamma_1$
(and $\gamma_2$) when $N\rightarrow\infty$.

\begin{figure}[!!!h]
\begin{center}
\includegraphics[width = 0.48\textwidth]{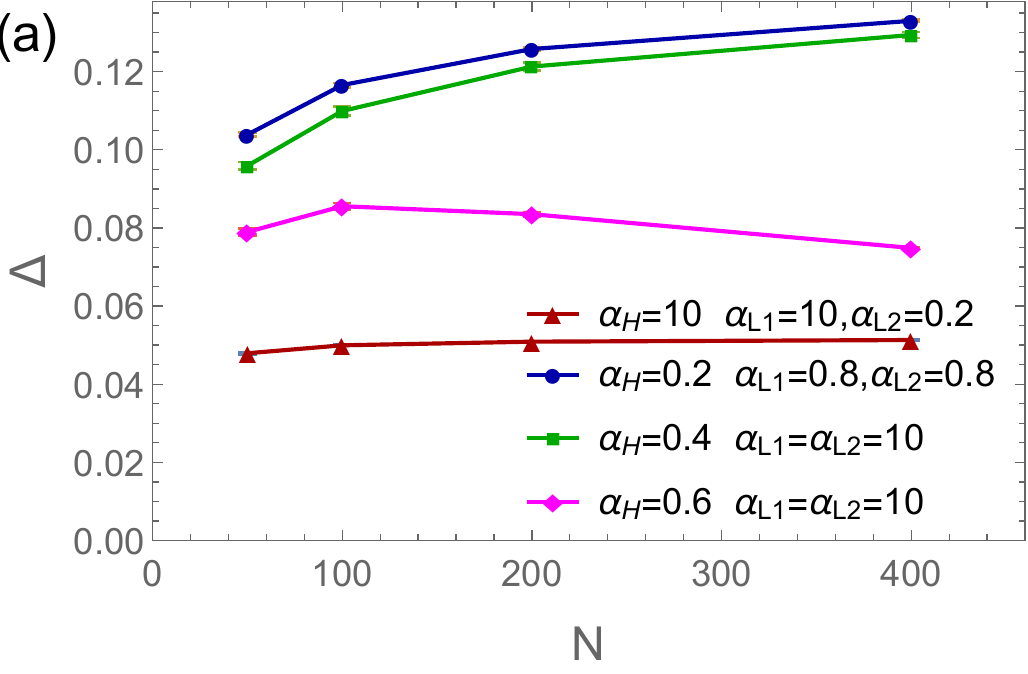}
\hspace{3mm}
\includegraphics[width = 0.48\textwidth]{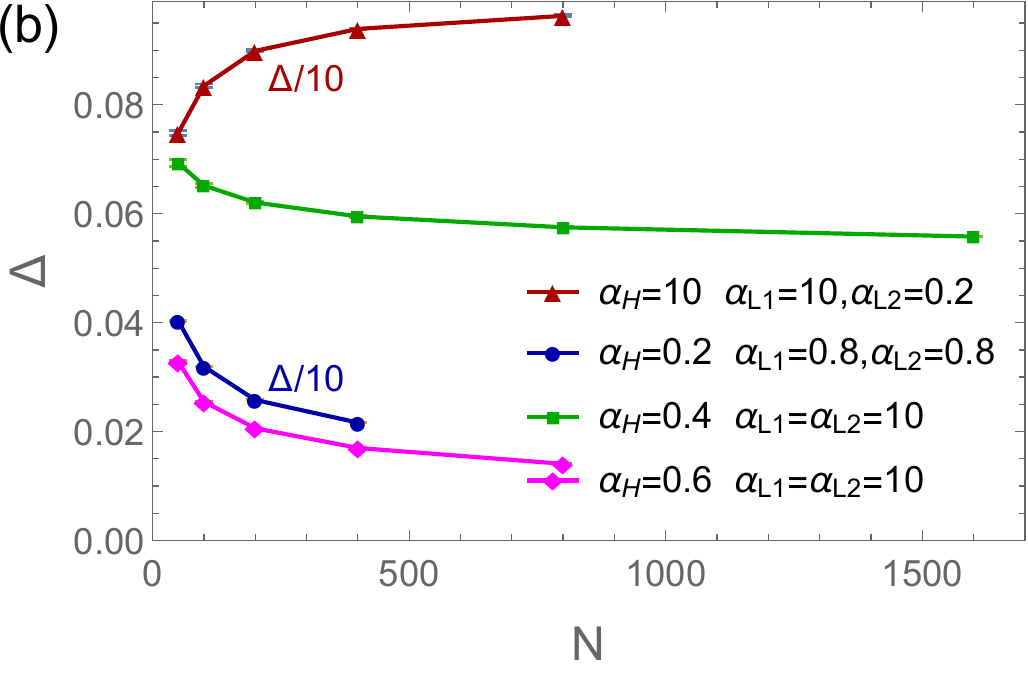}
\end{center}
\vspace{-0.5cm} \caption[]{\small (a) The Lindbladian spectral gap as a function of system size for the case of two
Hermitian PRBM jump operators at weak coupling $\gamma_1=\gamma_2=0.2$. Results are shown for four systems with the specified
values for the exponents $\alpha_H$ and $\alpha_{L_1}$, $\alpha_{L_2}$. (b) The spectral gap of the same systems as a function of
$N$ at strong coupling $\gamma_1=\gamma_2=10$. }
\label{fig-multi-L}
\end{figure}

Next, we contrast the above conclusions, based on perturbative analysis of the effective population dynamics expressed via the $A$ matrix,
with the spectrum of the full Lindbladian. We begin by considering the case of two jump operators with $\gamma_1=\gamma_2=\gamma$.  The results,
depicted in Fig. \ref{fig-multi-L}, exhibit the expected behavior, namely, for small $\gamma$ the spectrum appears gapless when $\alpha_H>1/2$
and $\tilde\alpha_L>1/2$ and gapped otherwise. For large $\gamma$ the system appears gapless when $\tilde\alpha_L>1/2$ and gapped otherwise.

\vspace{0.5cm}

\begin{figure}[!!!h]
\begin{center}
\includegraphics[width = 0.489\textwidth]{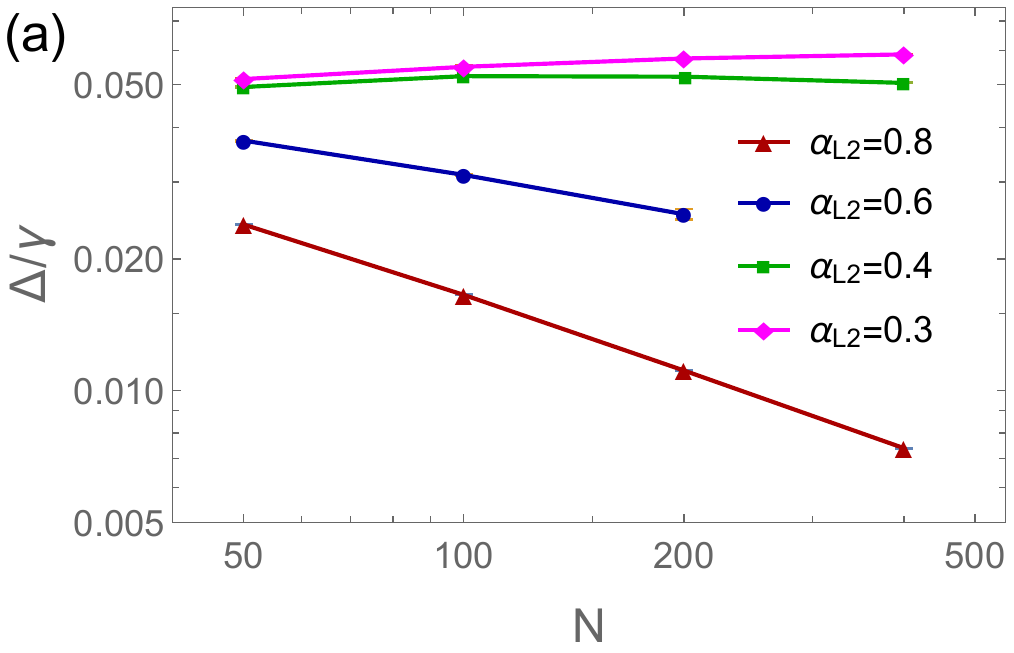}
\hspace{3mm}
\includegraphics[width = 0.471\textwidth]{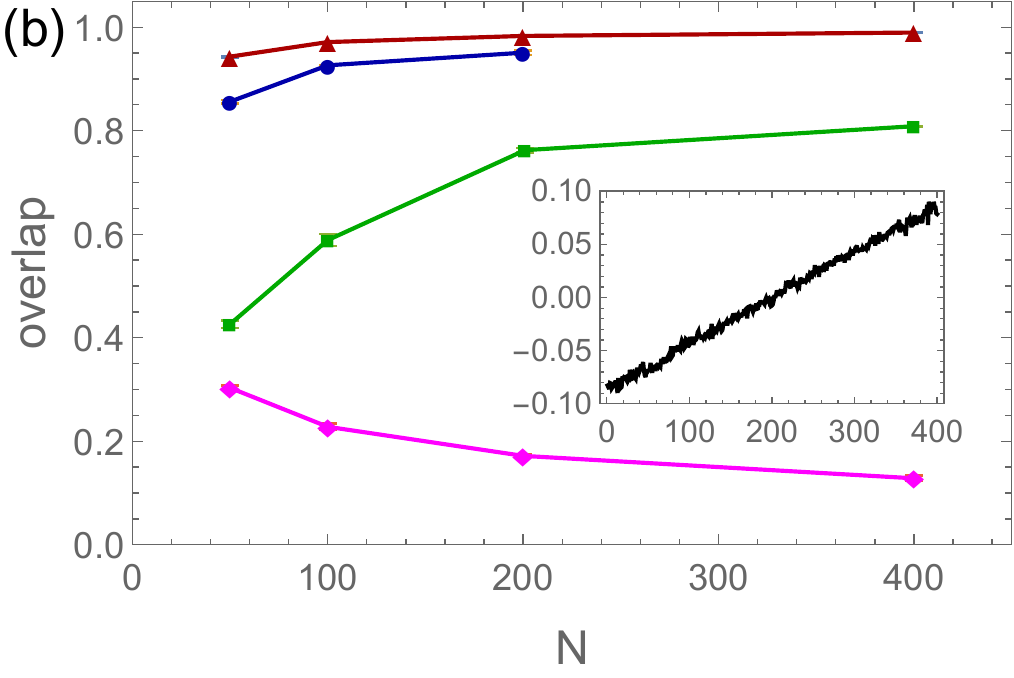}
\end{center}
\vspace{-0.5cm} \caption[]{\small (a) The Lindbladian spectral gap as a function of system size for the case of
pure noisy dynamics ($H=0$) with two Hermitian PRBM jump operators. The results are for systems with
$\alpha_{L_1}=0.8$, $\gamma_1=\gamma_2=\gamma$ and various values of $\alpha_{L_2}$.
(b) The overlap of the slowest decaying mode with the $x$-basis populations in the same systems. The inset depicts a representative
example of the slowest mode for $\alpha_{L_2}=0.4$. Shown are its components in the subspace of $x$-basis populations.}
\label{fig-pureL}
\end{figure}

In an effort to elucidate the nature of the gapless phase for strong decoherence, we follow the strategy outline above and treat
first the model of two jump operators and no Hamiltonian ($\gamma=\infty$). The results, depicted in Fig. \ref{fig-pureL} show
that: a) When $\gamma_1=\gamma_2$ (away from the perturbative limit $\gamma_1\gg\gamma_2$ considered above) the slowest decaying
Lindbladian eigenstate largely resides in $x$-basis populations subspace, at least for $\tilde\alpha_L>1/2$. b) The system is gapped
when $\tilde\alpha_L<1/2$ and gapless for $\tilde\alpha_L>1/2$, and c) The slowest decaying mode is hydrodynamic.

\begin{figure}[!!!h]
\begin{center}
\includegraphics[width = 0.48\textwidth]{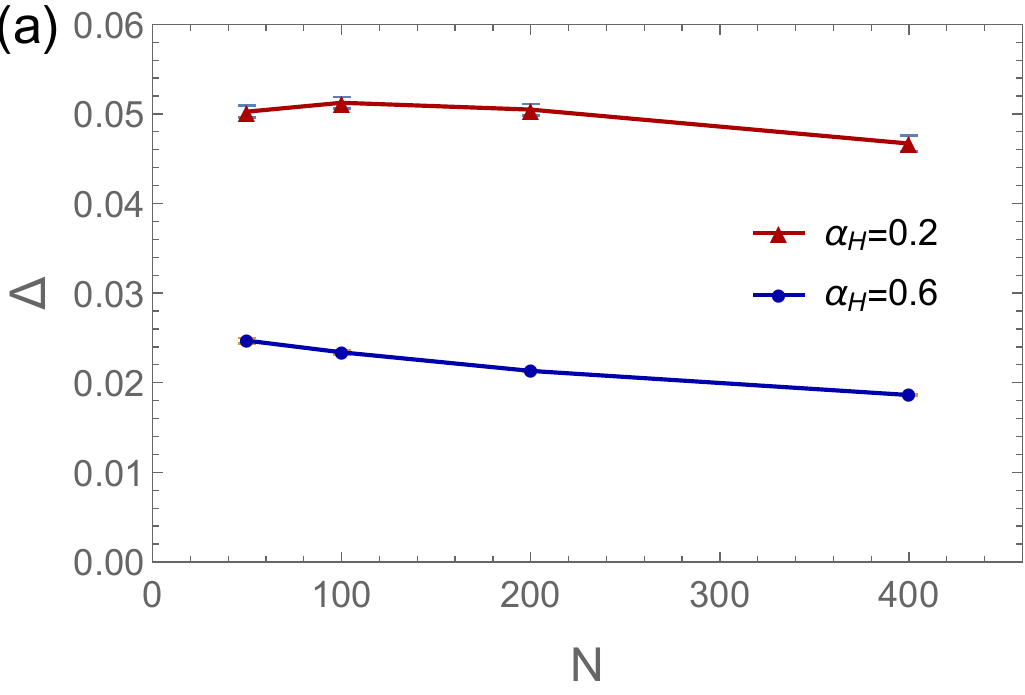}
\hspace{3mm}
\includegraphics[width = 0.48\textwidth]{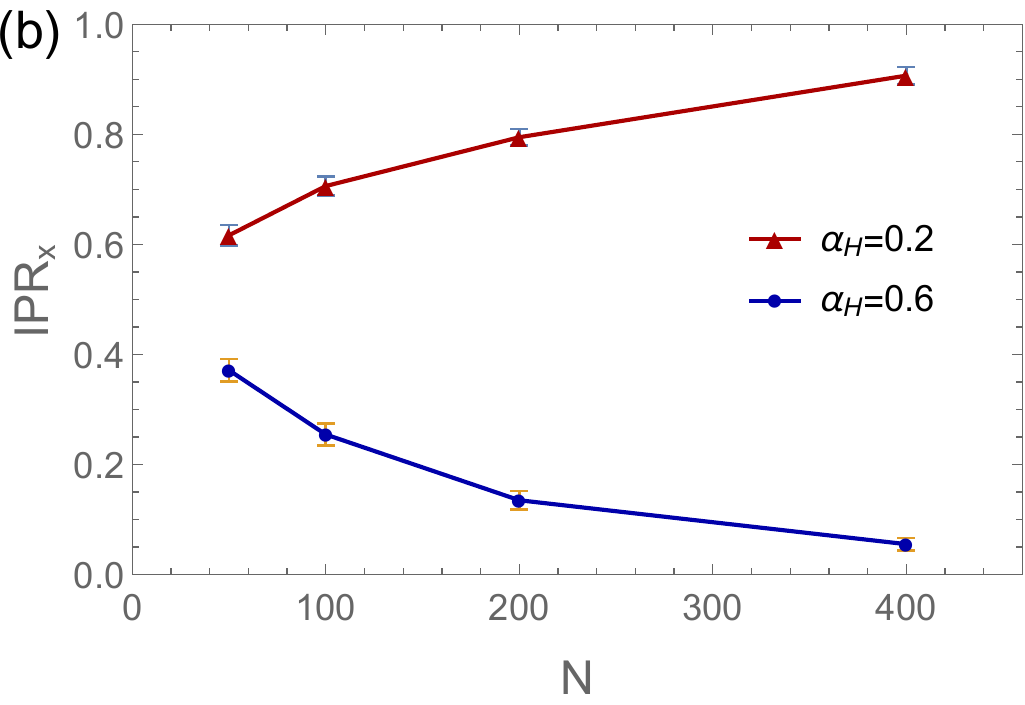}
\end{center}
\vspace{-0.5cm} \caption[]{\small (a) The Lindbladian spectral gap as a function of system size for the case of two
jump operators that are diagonal in the $x$-basis. Their diagonal elements are random permutations of the
eigenvalues of two $\alpha=10$ PRBMs. The system is at strong coupling $\gamma_1=\gamma_2=10$, and its Hamiltonian is a PRBM with
the specified values for $\alpha_H$. (b) The IPR in the $x$-basis of the slowest decaying eigenstate of the same systems. }
\label{fig-multidiag}
\end{figure}

Consequently, we add the Hamiltonian and concentrate on the character of the slowest decaying mode in the gapless phase $\tilde\alpha_L>1/2$.
We begin by considering the simpler case, where the two jump operators commute. Specifically, when they are diagonal in the $x$-basis with
matrix elements that are given by the eigenvalues of two $\alpha=10$ PRBMs (and are therefore unbound). Fig. \ref{fig-multidiag} shows
that while the spectral gap slowly decays with $N$ for both $\alpha_H<1/2$ and for $\alpha_H>1/2$, the corresponding mode is localized
in real space (Lifshitz phase) in the former case and delocalized (hydrodynamical phase) in the latter.

\begin{figure}[!!!h]
\begin{center}
\includegraphics[width = \textwidth]{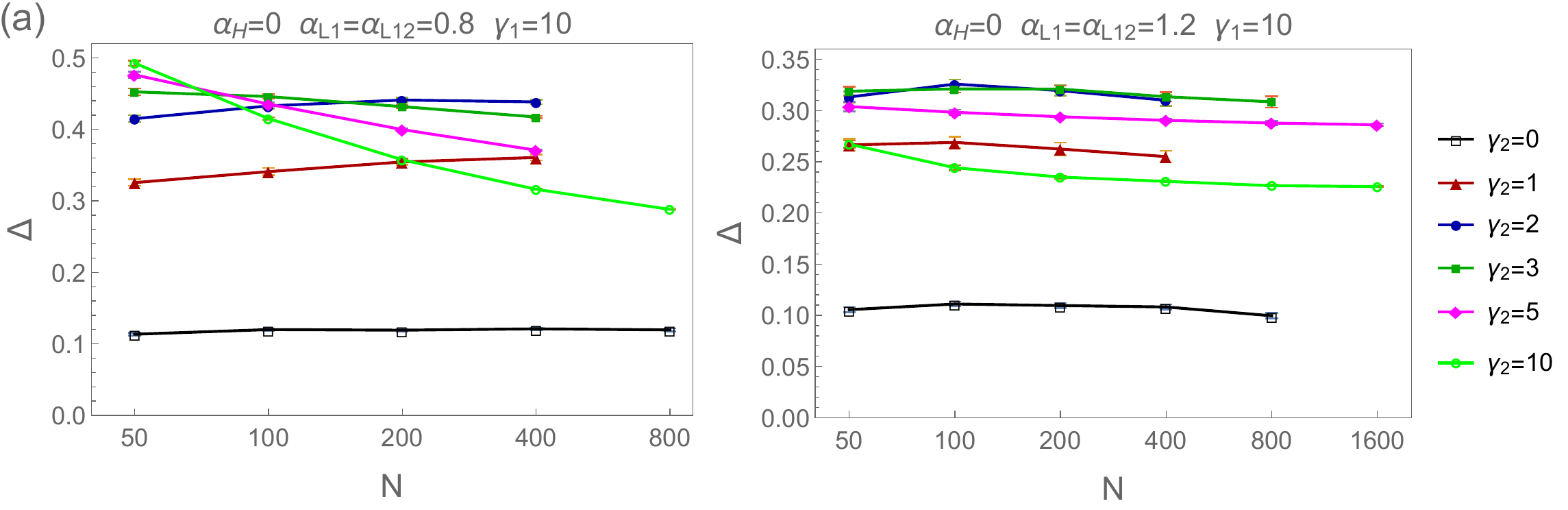}
\includegraphics[width = \textwidth]{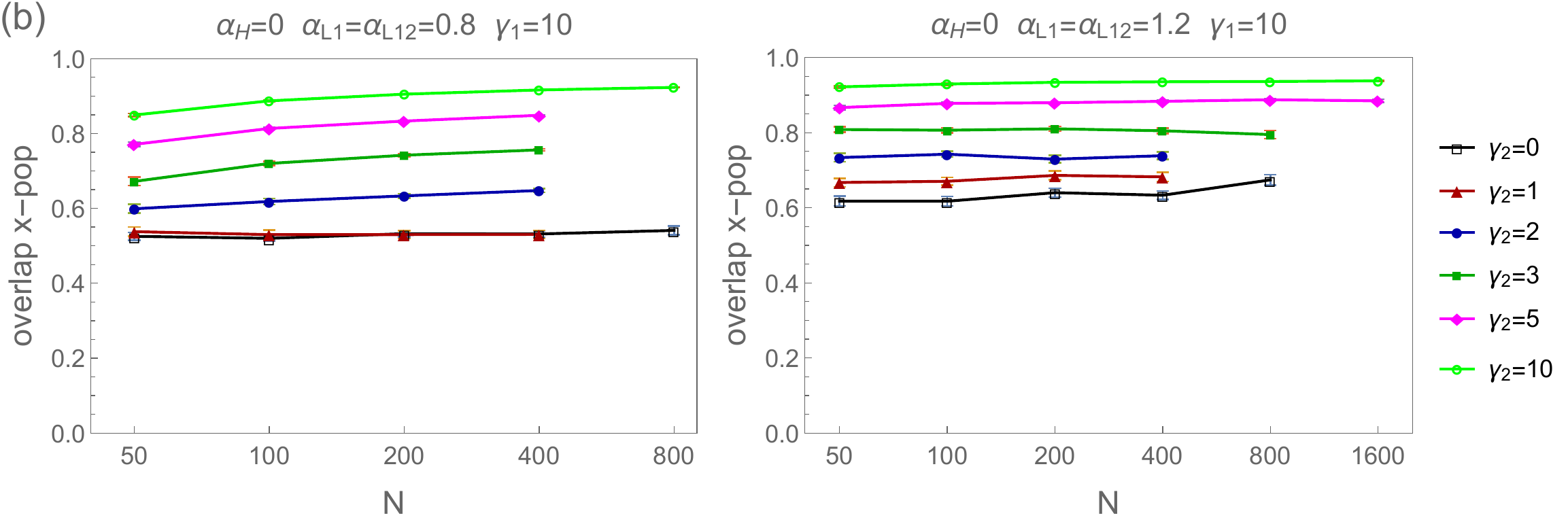}
\includegraphics[width = \textwidth]{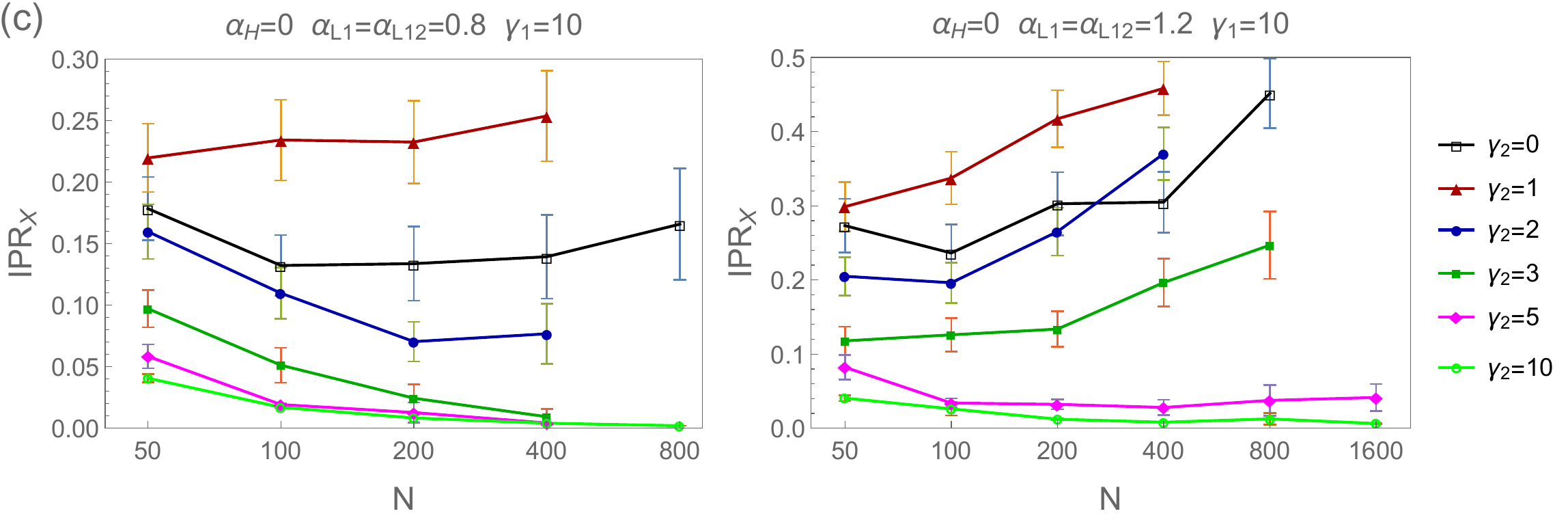}
\end{center}
\vspace{-0.5cm} \caption[]{\small The case of a Hamiltonian with $\alpha_H=0$ and two Hermitian PRBM jump operators
with $\alpha_{L_1}=\alpha_{L_2}=0.8$ (left), and $\alpha_{L_1}=\alpha_{L_2}=1.2$ (right). The results are for systems with $\gamma_1=10$
and various values of $\gamma_2$, as indicated. The figure depicts the dependence on system size for (a) The Lindbladian spectral gap,
(b) The overlap of the slowest decaying Lindbladian mode with the subspace of $x$-basis populations, and (c) The IPR$_X$
of the slowest decaying Lindbladian mode.}
\label{fig-twoL-aH0}
\end{figure}

\begin{figure}[!!!h]
\begin{center}
\includegraphics[width = 0.489\textwidth]{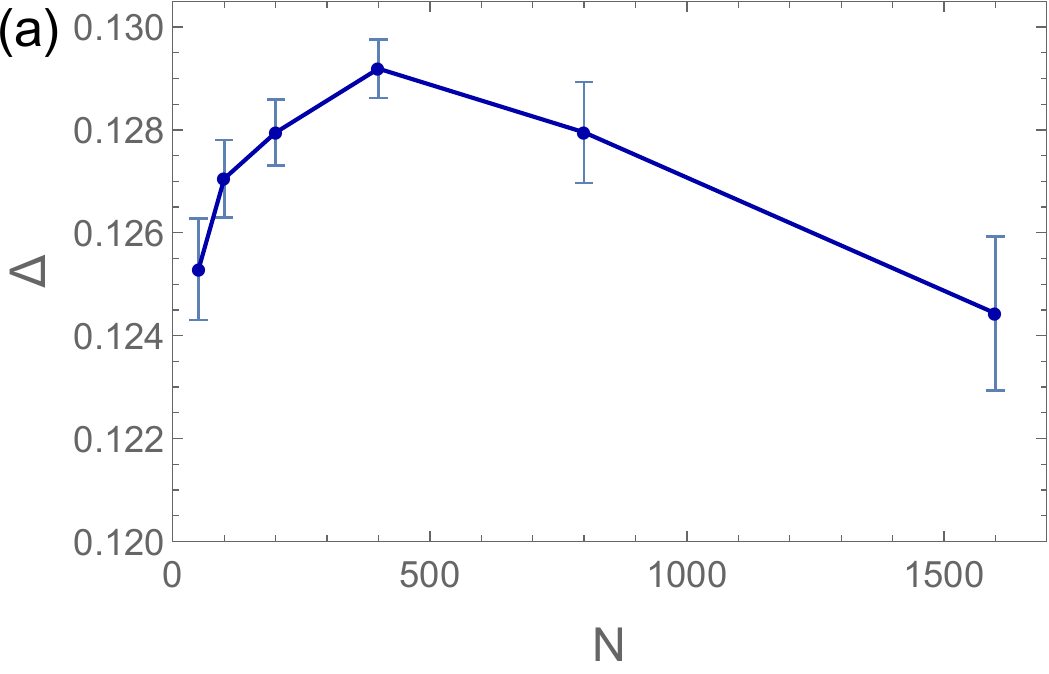}
\hspace{3mm}
\includegraphics[width = 0.471\textwidth]{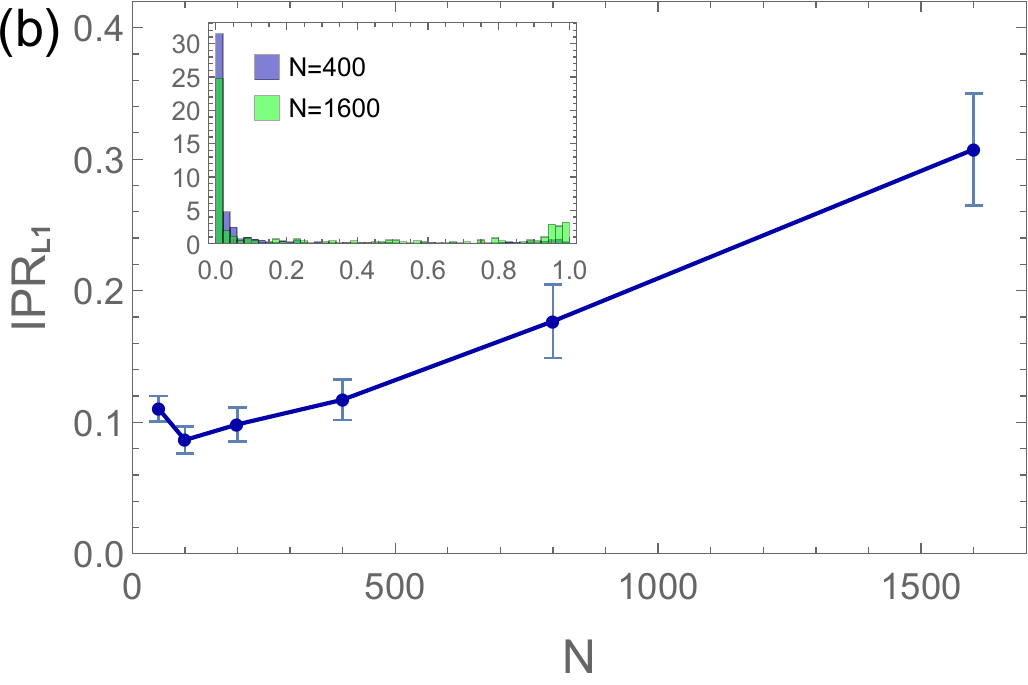}
%\newline\newline
%\includegraphics[width = 0.31\textwidth]{multi-L-largeg-aH02-wf1}
%\hspace{3mm}
%\includegraphics[width = 0.31\textwidth]{multi-L-largeg-aH02-wf2}
%\hspace{3mm}
%\includegraphics[width = 0.32\textwidth]{multi-L-largeg-aH02-wf3}
\end{center}
\vspace{-0.5cm} \caption[]{\small (a) The Lindbladian spectral gap as a function of system size for the case of two
Hermitian PRBM jump operators with $\alpha_{L_1}=\alpha_{L_2}=10$ and a Hamiltonian with $\alpha_H=0.2$,
at strong coupling $\gamma_1=\gamma_2=10$.
(b) The IPR of the slowest decaying mode in the basis of the $L_1$ eigenfunctions. The inset depicts the distribution function of the
IPR for $N=400$ and $N=1600$.}
%(c)-(e) Representative examples of the slowest
%decaying mode. Shown are the components along the $L_1$ populations. (c) The mode is dominated by an extreme population of $L_1$,
%(d) an extreme population of $L_2$, and (e) delocalized and spread over the entire populations subspace. }
\label{fig-multi-L-aH02}
\end{figure}

To further establish the existence of a Lifshitz phase in the regime $\alpha_H<1/2$ and $\tilde\alpha_L>1/2$ we turn to
the more general case, where the two jump operators do not commute. Fig. \ref{fig-twoL-aH0} provides evidence that
the Lifshitz phase survives in this case as well. Specifically, it follows the slowest decaying mode in a system with $\alpha_H=0$
and two Hermitian PRBM jump operators with $\alpha_{L_1}=\alpha_{L_2}=0.8$ or 1.2, as $\gamma_2$ is increased towards $\gamma_1=10$.
The findings for the limit $\gamma_2=0$ agree with our previous results. We observe a gap that begins to decay
slowly at large $N$ (at least for $\alpha_L=1.2$, although not yet for $\alpha_L=0.8$) with a concomitant increase of the real-space
IPR - both indicative of a Lifshitz mode. As $\gamma_2$ increases the rise in the IPR commences at larger values of $N$.
However, the fact that we observe the localization of the mode for $\gamma_2=3$ (and perhaps also for $\gamma_2=5$ in the $\alpha_L=1.2$
system) leads us to believe that the Lifshitz phase is established when $N\rightarrow\infty$ even for $\gamma_2=\gamma_1$.
This conclusion is supported by our observation, see Fig. \ref{fig-multi-L-aH02}, of Lifshitz behavior for $\gamma_1=\gamma_2$,
when $\alpha_{L_1}=\alpha_{L_2}=10$, i.e., when the non-commutativity of the jump operators is still present but weaker.

\newpage

\begin{figure}[!!!h]
\begin{center}
\includegraphics[width = 0.48\textwidth]{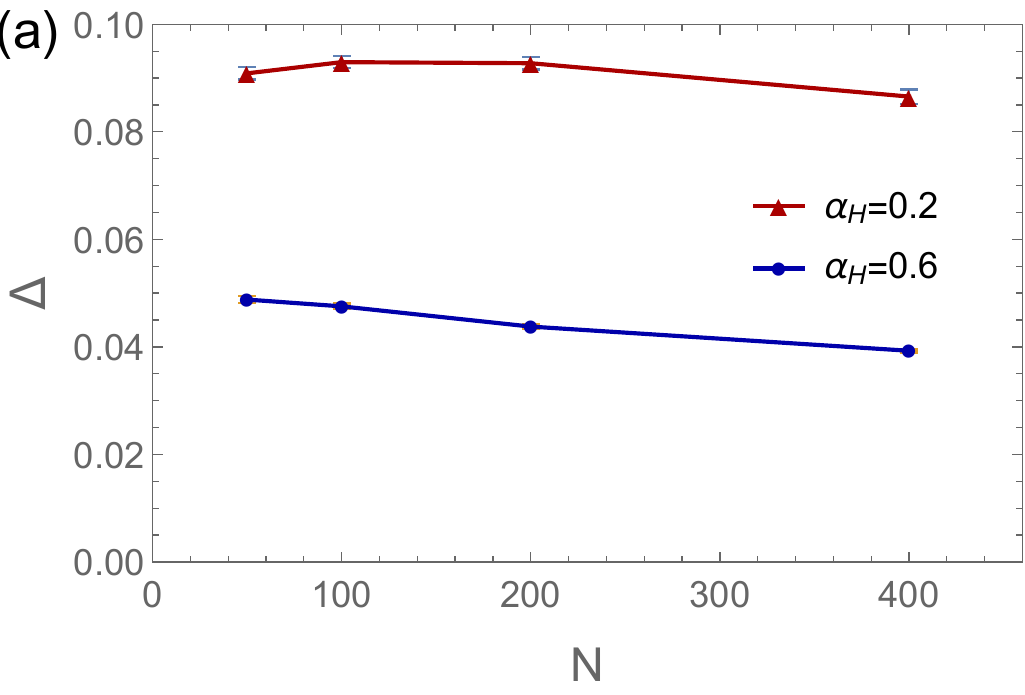}
\hspace{3mm}
\includegraphics[width = 0.48\textwidth]{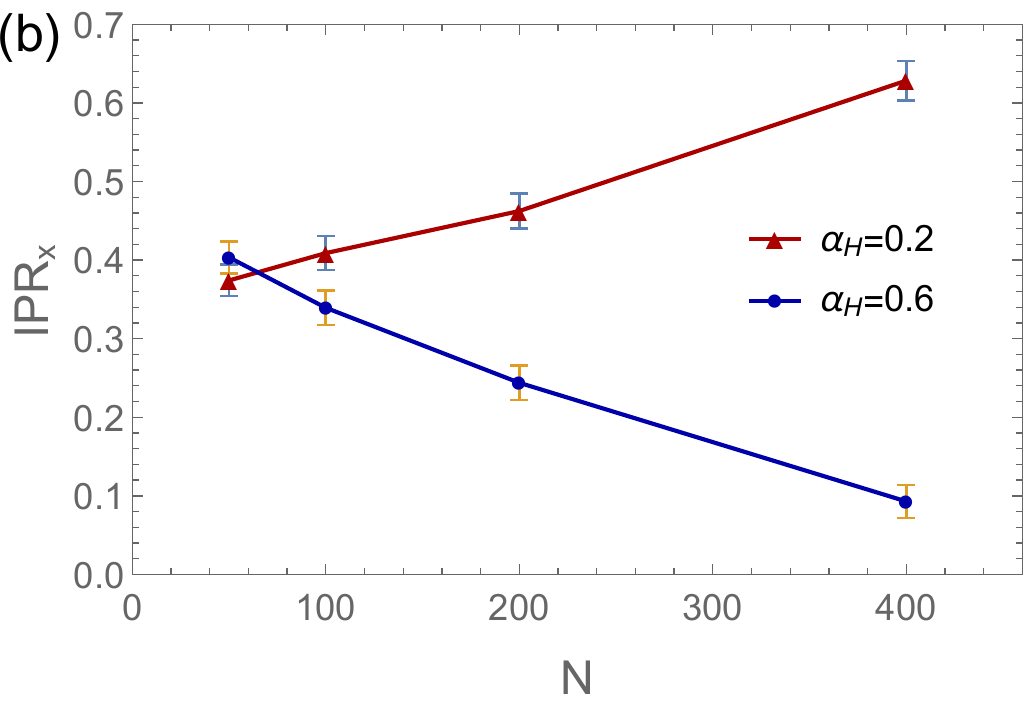}
\end{center}
\vspace{-0.5cm} \caption[]{\small (a) The Lindbladian spectral gap as a function of system size for the case of $N$ jump operators,
each localized on a different site. Their strengths are takes from a random permutation of the
eigenvalues of an $\alpha=10$ PRBM. The system is at strong coupling $\gamma_k=10$, and its Hamiltonian is a PRBM with
the specified values for $\alpha_H$. (b) The IPR in the $x$-basis of the slowest decaying eigenstate of the same systems. }
\label{fig-multilocal}
\end{figure}

\begin{figure}[!!!h]
\begin{center}
\includegraphics[width = 0.471\textwidth]{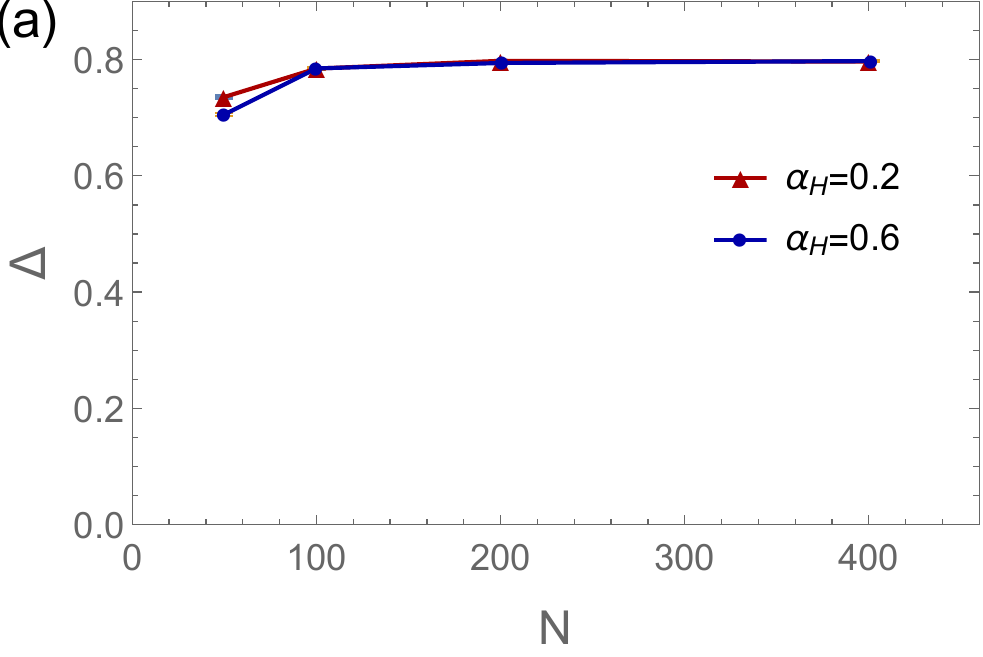}
\hspace{3mm}
\includegraphics[width = 0.48\textwidth]{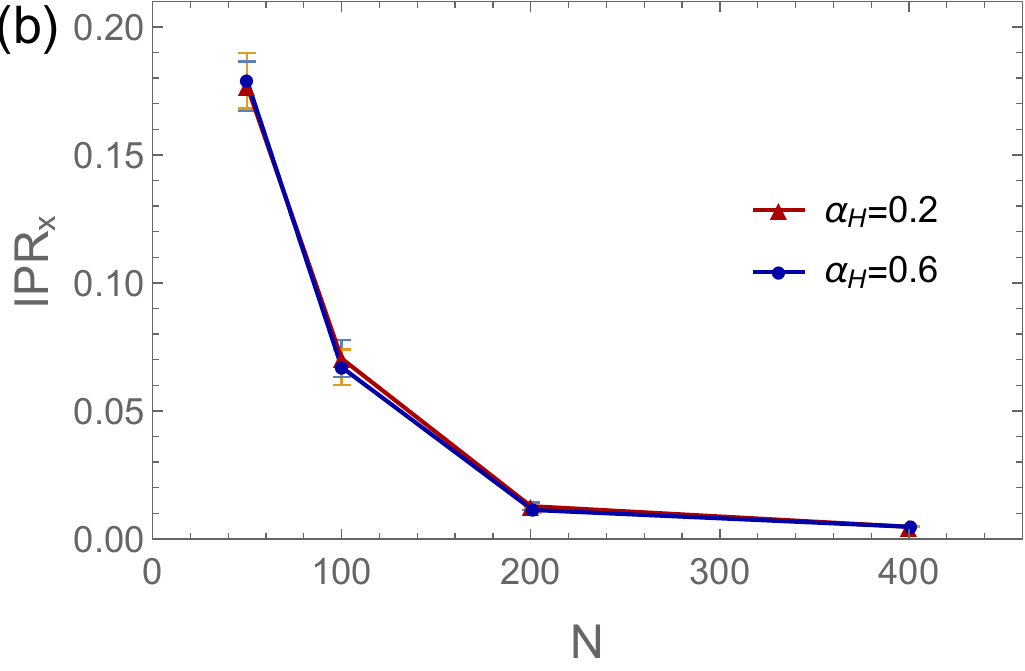}
\end{center}
\vspace{-0.5cm} \caption[]{\small Same as Fig. \ref{fig-multilocal} with an additional jump operator that is a PRBM in the
$x$-basis with $\alpha_L=0.3$.}
\label{fig-multilocalplus}
\end{figure}

Finally, we briefly turn to the case of many jump operators. Fig. \ref{fig-multilocal} considers the case where a system
of size $N$ is coupled to the environment via $N$ jump operators, each localized on a different site. When the couplings
are taken from a distribution of an $\alpha=10$ PRBM (which is therefore unbound), we again find a transition from a
localized Lifshitz phase to a delocalized hydrodynamic phase as $\alpha_H$ increases beyond $1/2$.
Fig. \ref{fig-multilocalplus} demonstrates that it suffices to add to the system a single jump operator with
$\alpha_L<1/2$ in order to establish a spectral gap. This provides further support to our claim for a gapped phase
whenever $\tilde\alpha_L<1/2$.

\section{Dynamics of local observables}

\begin{figure}[!!!h]
\begin{center}
\includegraphics[height = 0.3\textwidth]{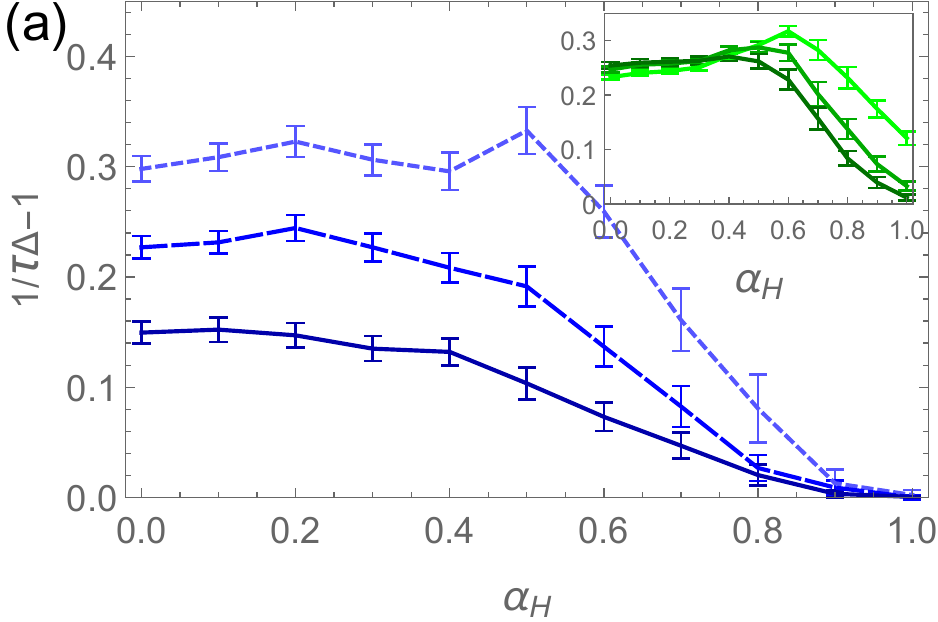}
\hspace{3mm}
\includegraphics[height = 0.3\textwidth]{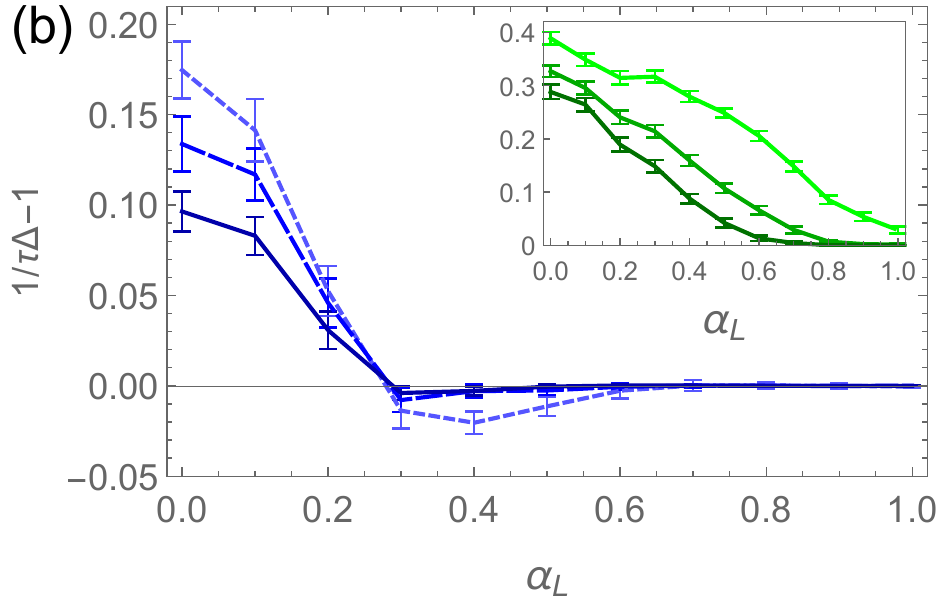}
\end{center}
\vspace{-0.5cm} \caption[]{\small(a) The relative difference between the spatial average of the relaxation rate $\tau^{-1}$ of local observables and $\Delta$ for
$\gamma=0.2$, $\alpha_L=1.5$ and $N=100$. The dotted, dashed and solid lines are based on $\tau$ extracted by fitting the
relaxation over the range $t=3-6$, $6-9$ and $9-12\Delta^{-1}$, respectively. The inset shows the standard deviation of the relative
difference. (b) The same quantities as a function of $\alpha_L$ for $\alpha_H=1.5$.}
\label{fig-evolsmallg}
\end{figure}

\begin{figure}[!!!h]
\begin{center}
\includegraphics[height = 0.205\textwidth]{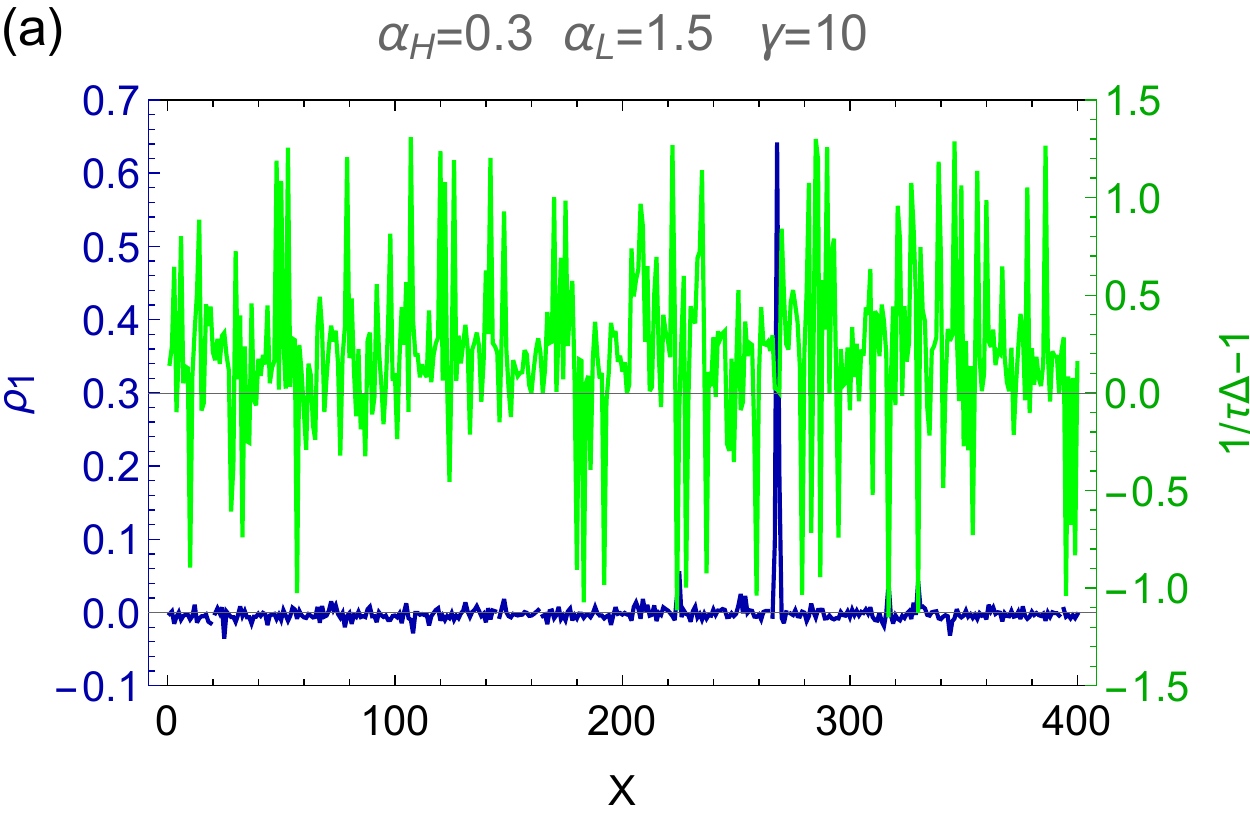}
\hspace{1mm}
\includegraphics[height = 0.205\textwidth]{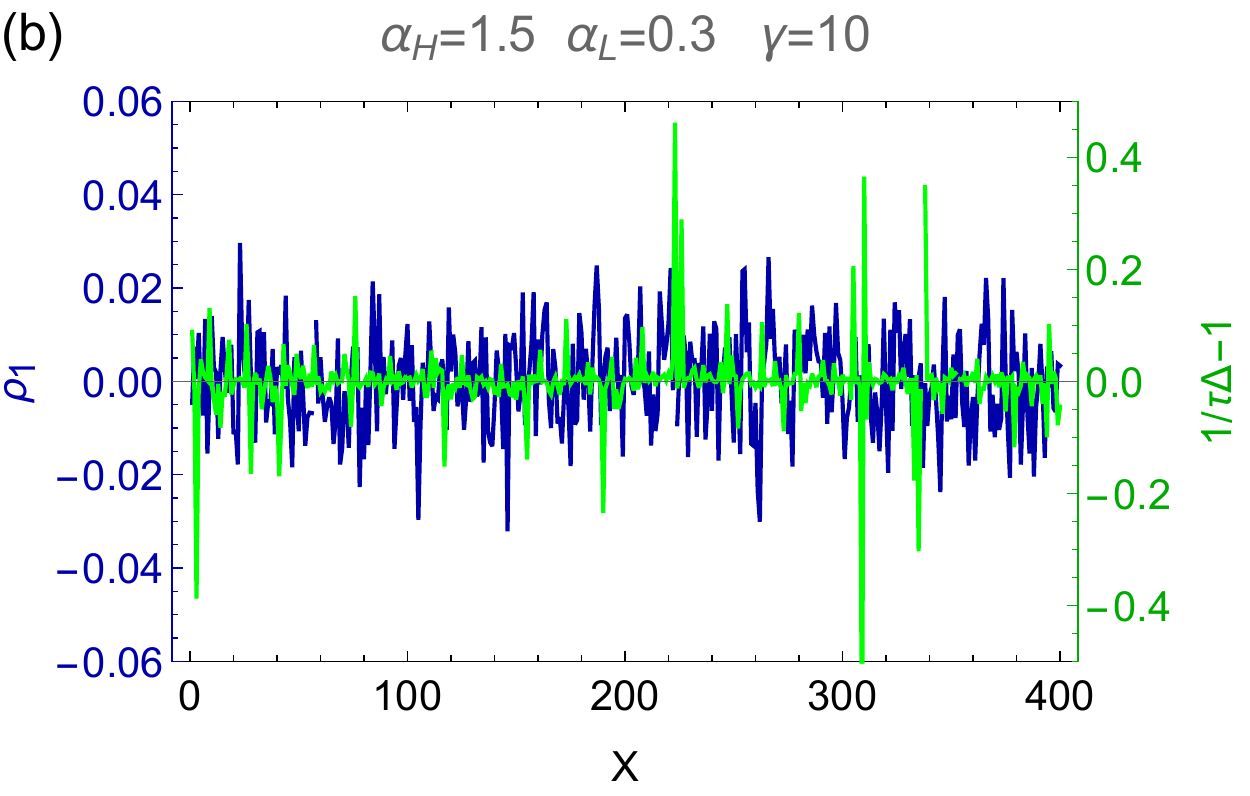}
\hspace{1mm}
\includegraphics[height = 0.205\textwidth]{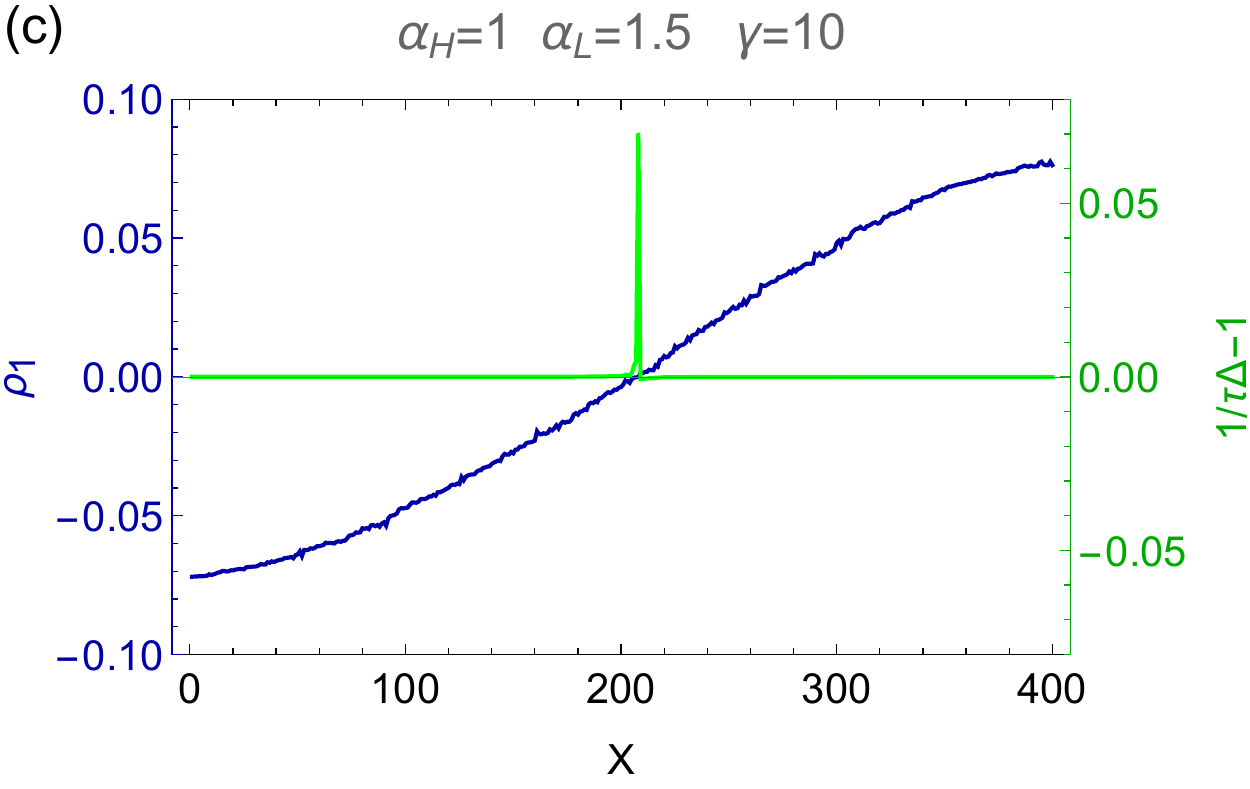}
\end{center}
\vspace{-0.5cm} \caption[]{\small Representative examples of the slowest relaxing mode $\rho_1$ (blue) and the relative difference between
the relaxation rate $\tau^{-1}$ of the local observables and $\Delta$ (green) for a system with $\gamma=10$, $N=400$. $\tau$
was extracted for each site by fitting the relaxation of $\rho_{ii}$ over the range $t=9-12\Delta^{-1}$. Results are shown for (a) the Lifshitz phase, (b) the gapped phase,
and (c) the hydrodynamic phase.}
\label{fig-rhofitlarge}
\end{figure}

\begin{figure}[!!!h]
\begin{center}
\includegraphics[height = 0.225\textwidth]{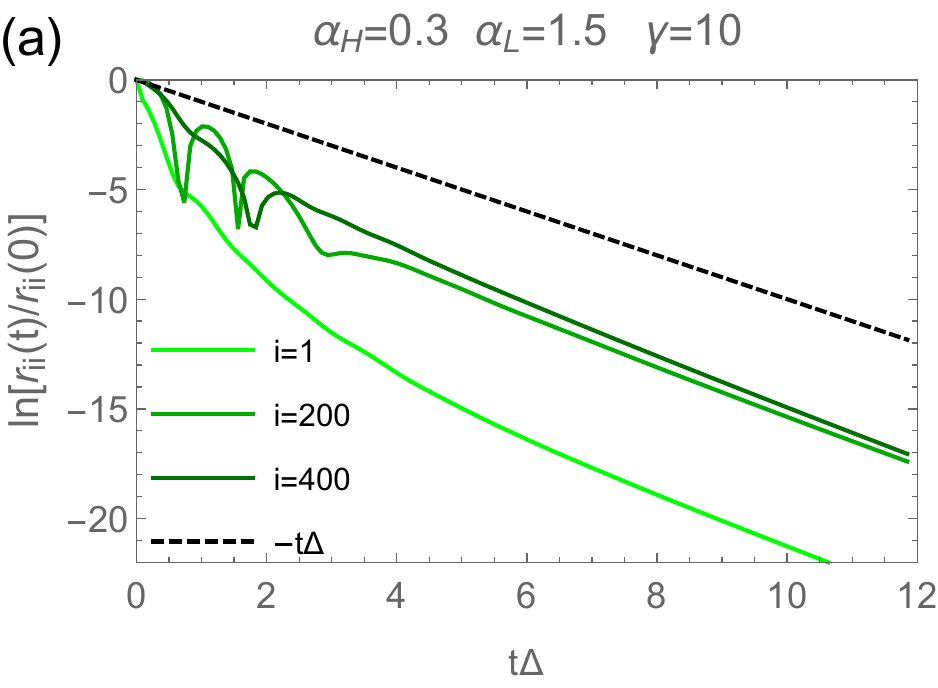}
\hspace{2mm}
\includegraphics[height = 0.2235\textwidth]{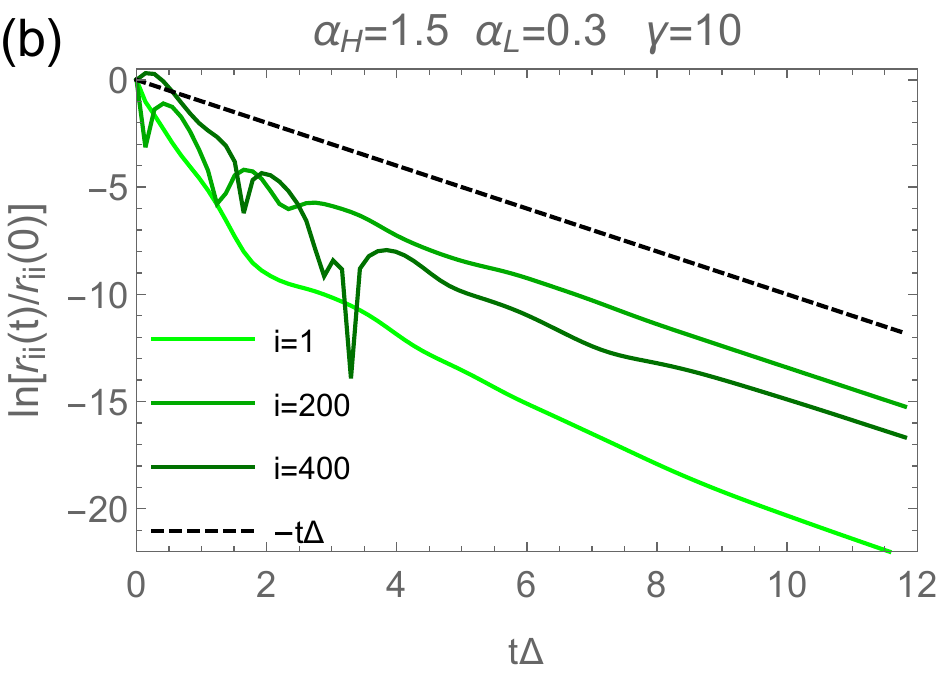}
\hspace{2mm}
\includegraphics[height = 0.225\textwidth]{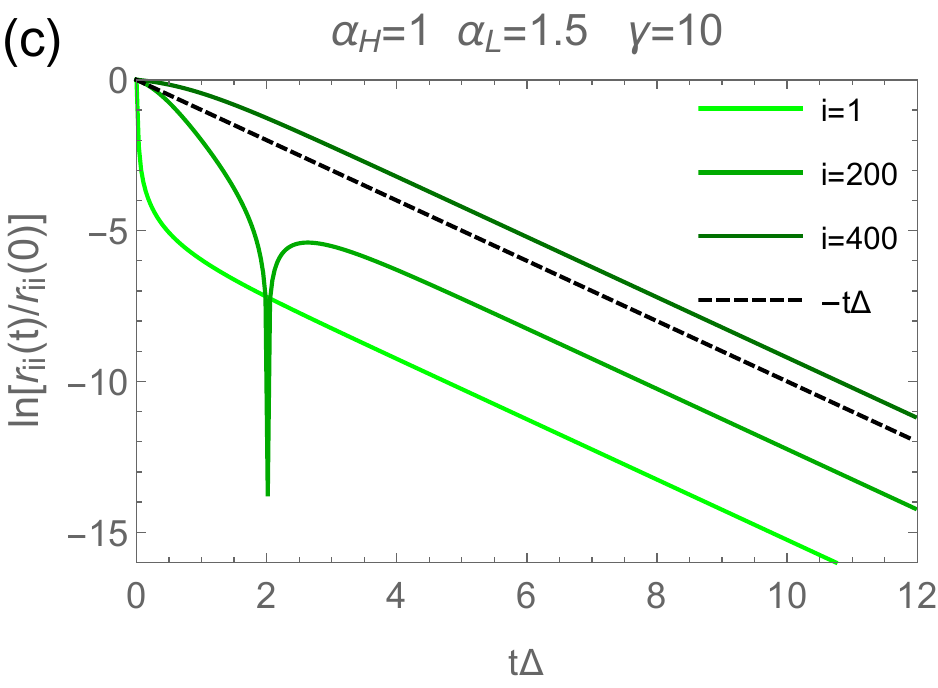}
\end{center}
\vspace{-0.5cm} \caption[]{\small The time evolution of $r_{ii}=|\rho_{ii}-1/N|$ on sites $i=1,200$ and $400$ in the samples presented in Fig. \ref{fig-rhofitlarge}.}
\label{fig-rhoevlarge}
\end{figure}

\begin{figure}[!!!h]
\begin{center}
\includegraphics[height = 0.204\textwidth]{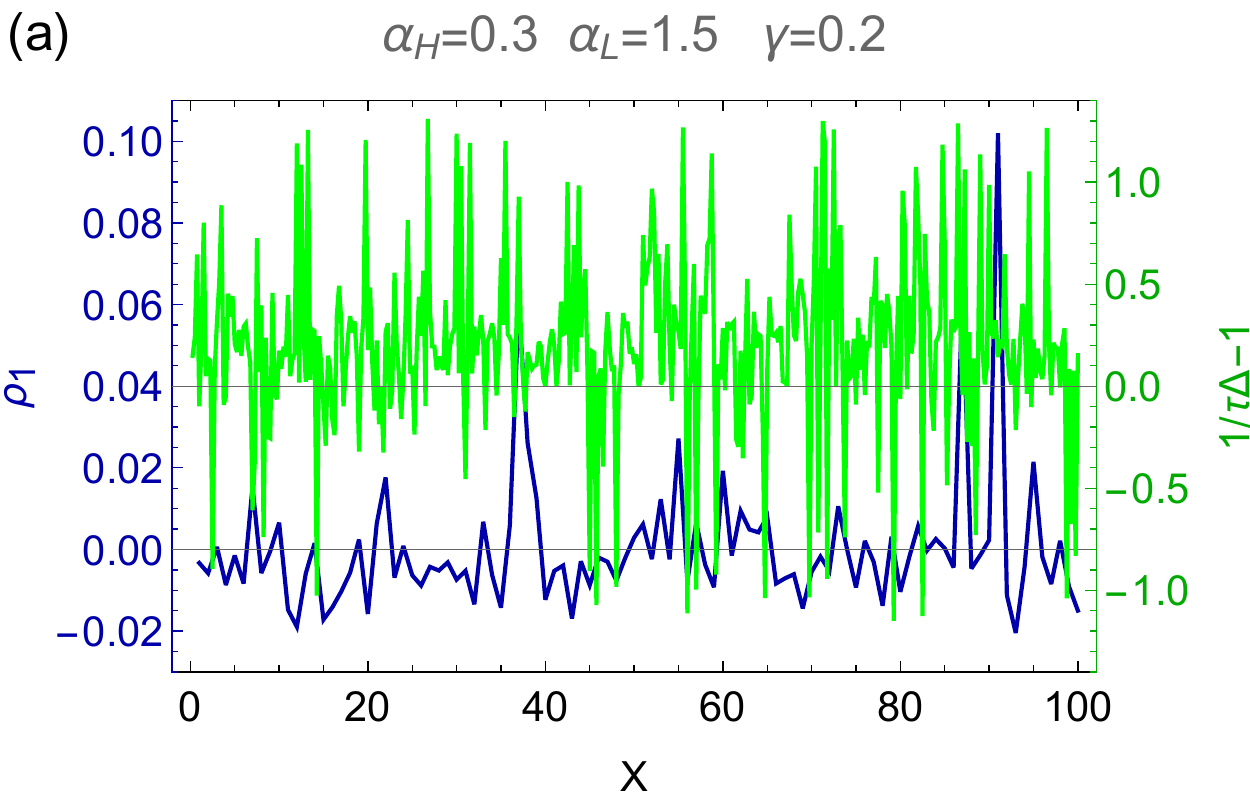}
\hspace{1mm}
\includegraphics[height = 0.204\textwidth]{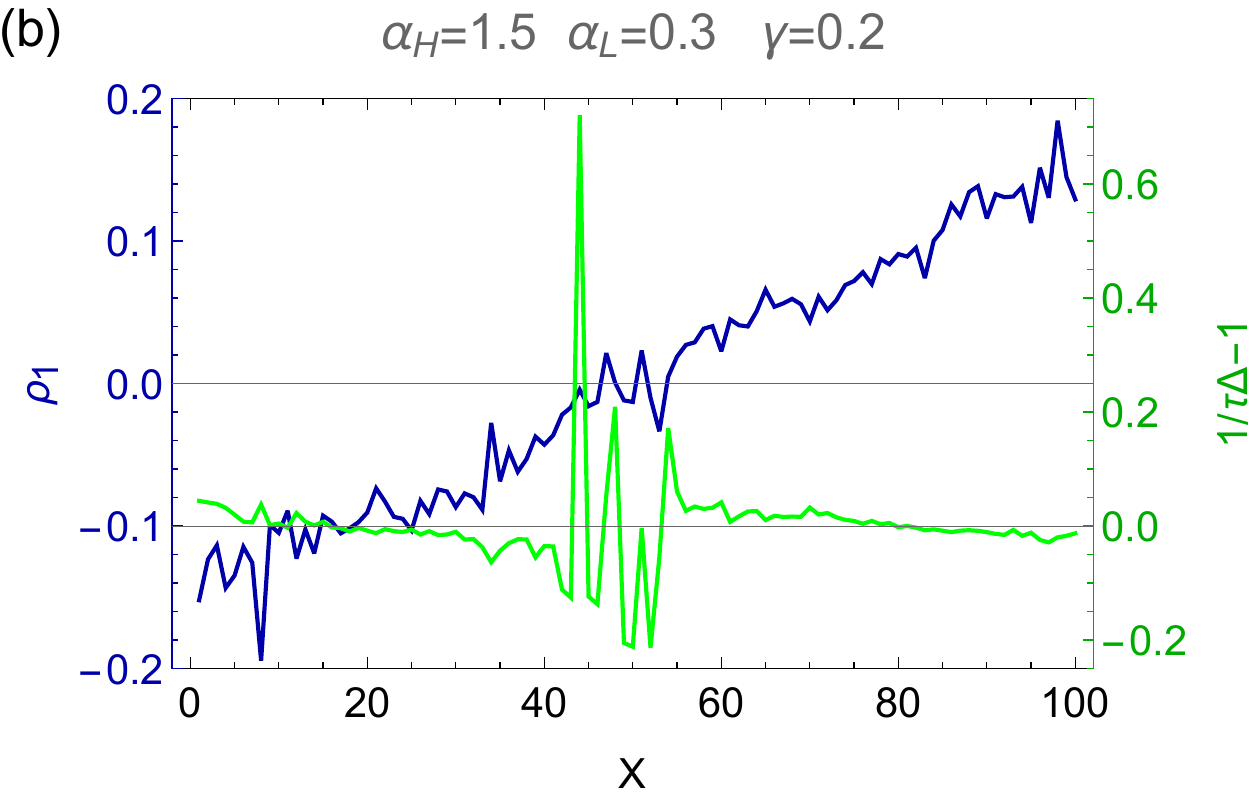}
\hspace{1mm}
\includegraphics[height = 0.204\textwidth]{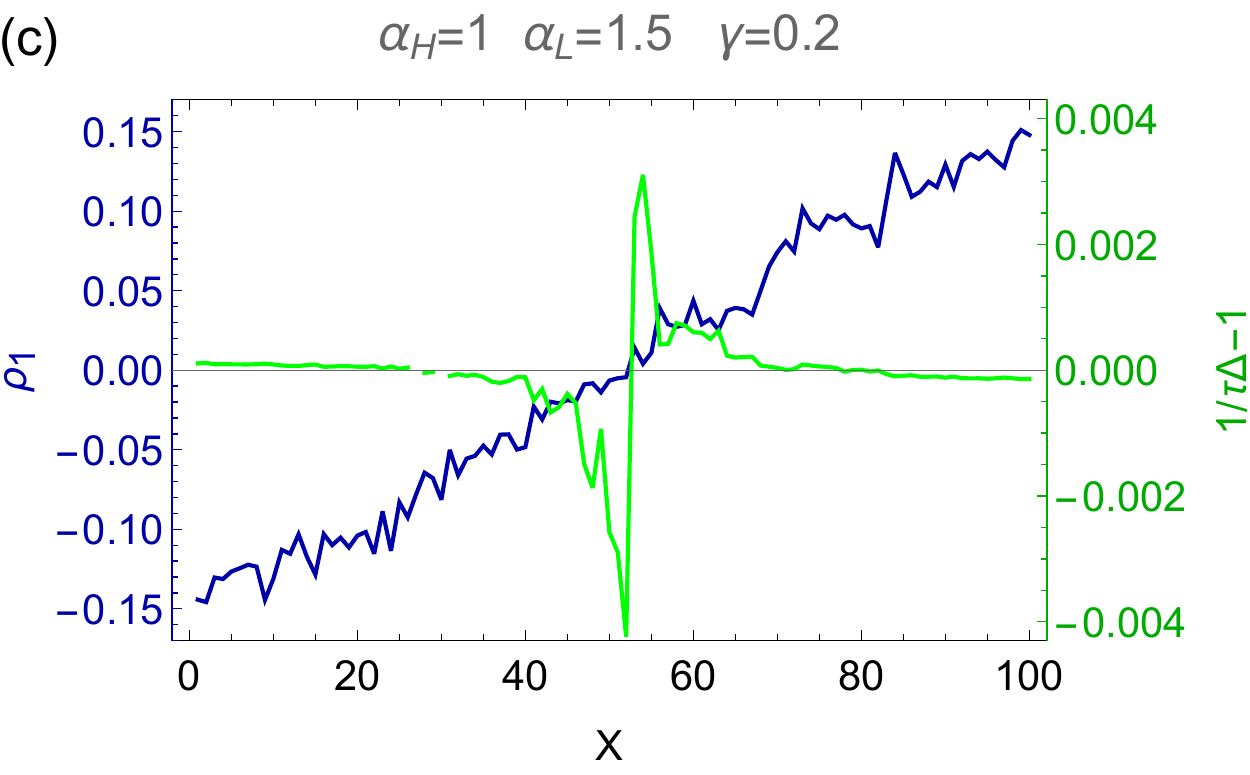}
\end{center}
\vspace{-0.5cm} \caption[]{\small Representative examples of the slowest relaxing mode $\rho_1$ (blue) and the relative difference between
the relaxation rate $\tau^{-1}$ of the local observables and $\Delta$ (green) for a system with $\gamma=0.2$, $N=100$. $\tau$
was extracted for each site by fitting the relaxation of $\rho_{ii}$ over the range $t=9-12\Delta^{-1}$. Results are shown
for (a) the Lifshitz phase (whose expected behavior is still not apparent for this system size), (b) the gapped phase, and (c) the hydrodynamic phase.}
\label{fig-rhofitsmall}
\end{figure}

\begin{figure}[!!!h]
\begin{center}
\includegraphics[height = 0.225\textwidth]{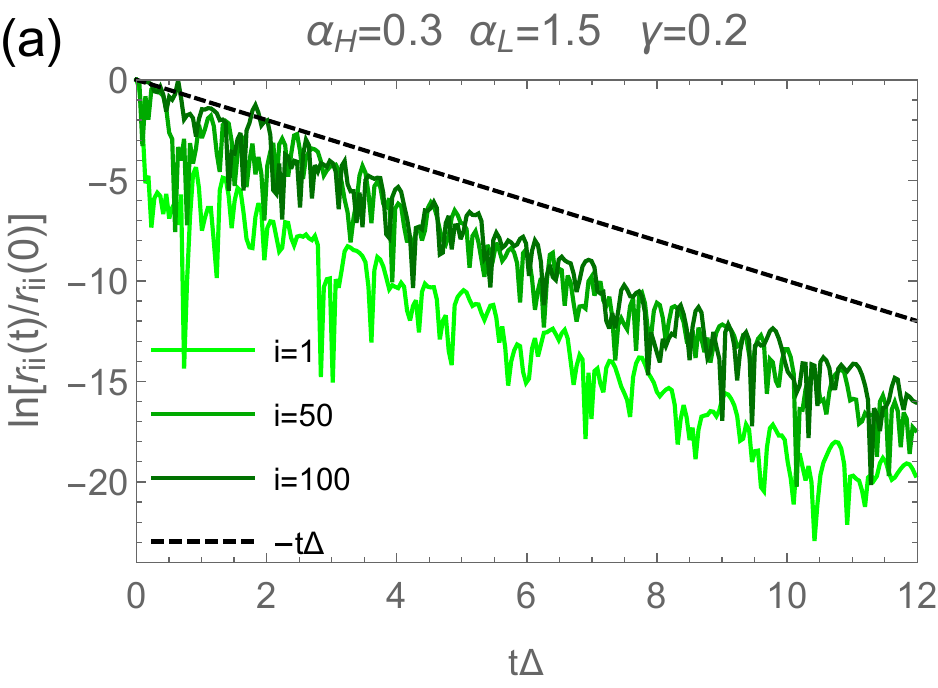}
\hspace{2mm}
\includegraphics[height = 0.225\textwidth]{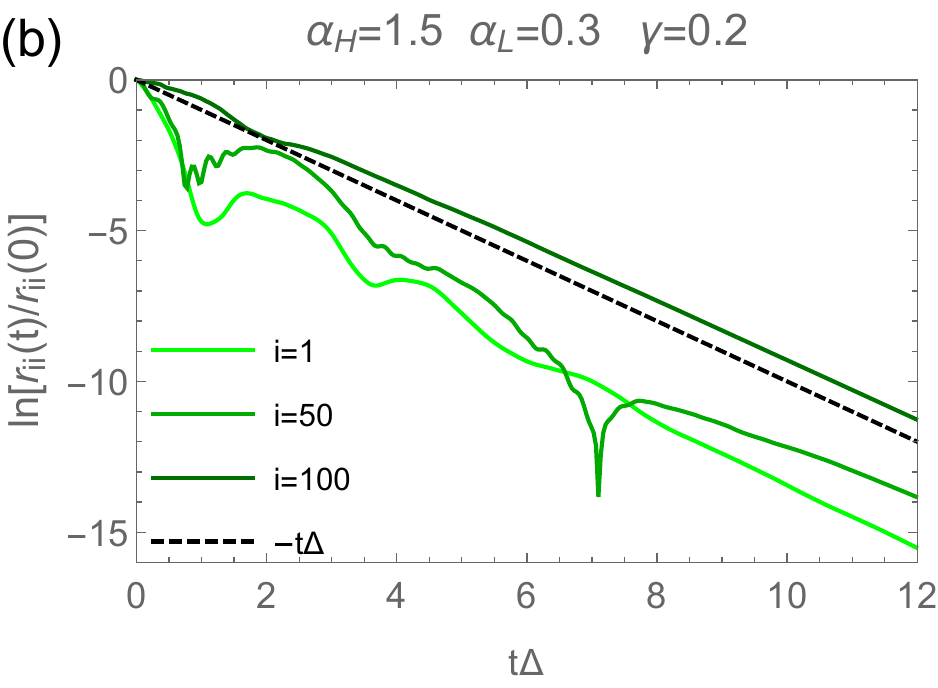}
\hspace{2mm}
\includegraphics[height = 0.225\textwidth]{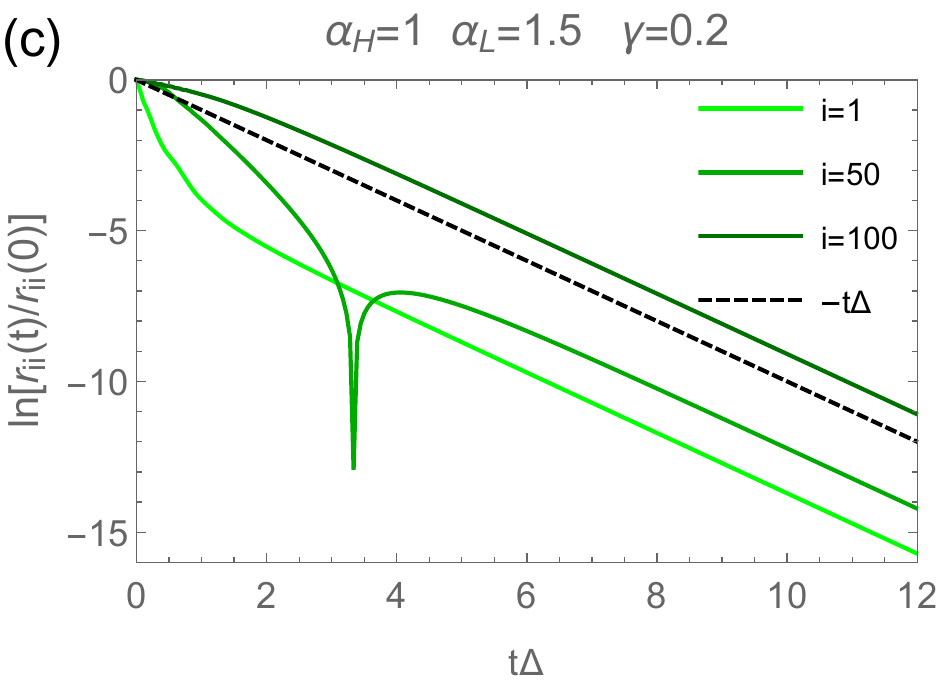}
\end{center}
\vspace{-0.5cm} \caption[]{\small The time evolution of $r_{ii}=|\rho_{ii}-1/N|$ on sites $i=1,50$ and $100$ in the samples presented in Fig. \ref{fig-rhofitsmall}.}
\label{fig-rhoevsmall}
\end{figure}

To tie the discussion of the Lindbladian gap $\Delta$ to the time dependence of local observables we have calculated the evolution
of an initial state of the form
\begin{equation}
\label{eq:rhoin}
\rho_{ij}(t=0)=\left\{
\begin{array}{cc}
0 & i=j=1 \\
\frac{1}{N-1} & i=j\neq 1\\
0 & i\neq j
\end{array}
\right. .
\end{equation}
We have fitted the diagonal elements $\rho_{ii}(t)$ (see examples in Figs. \ref{fig-rhoevlarge} and \ref{fig-rhoevsmall})
to an exponential relaxation over different time windows, thereby extracting the relaxation rate $1/\tau$. The results for
the limit of strong decoherence appear in Fig. 3 of the main text, and the results for weak decoherence are depicted here
in Fig. \ref{fig-evolsmallg}. In both cases, we find an asymptotic exponential relaxation of the local observable $\rho_{ii}$
with an asymptotic relaxation rate that equals the gap. However, for each $\rho_{ii}$ the asymptotic relaxation commences
at a different time, as demonstrated by Figs. \ref{fig-rhoevlarge} and \ref{fig-rhoevsmall}. This time correlates with the
overlap between $\rho_{ii}(t=0)$ and the Lindbladian slowest relaxing mode $\rho_1$. Smaller overlap implies a later onset of the
asymptotic relaxation. This is shown by Figs. \ref{fig-rhofitlarge} and \ref{fig-rhofitsmall} that present the
relative difference $(\tau^{-1}-\Delta)/\Delta$ between the relaxation rate extracted from the range $t=9-12\Delta^{-1}$
and the gap.

We find that the asymptotic relaxation starts early (in terms of the natural time scale $\Delta^{-1}$) in the hydrodynamic phase,
both in the limits of large and small $\gamma$. In addition, it exhibits small fluctuations in the onset time of the asymptotic
relaxation due to the smooth nature of the hydrodynamic modes. The final approach to the steady state starts later in the gapped
phase. In the strong decoherence limit it also exhibits larger fluctuations, due to the random nature of $\rho_1$, while the
more regular structure of $\rho_1$ in the limit of small $\gamma$ (at least for large $\alpha_H$, see Sec. \ref{ss-smallgloc})
implies smaller fluctuations in the onset time, see Figs. \ref{fig-rhofitlarge} and \ref{fig-rhofitsmall}. 
Finally, we find that in the Lifshitz phase the time where the dynamics starts
being controlled by the gap is the longest. This behavior is accompanied by large fluctuations in the onset time. Both of these
feature find their origin in the nature of $\rho_1$ that tends to be highly localized with only an algebraically-small overlap with the
majority of sites in the system.

In conclusion, let us comment about the dependence of the onset time on the system size. As $N$ grows, the number of
slowly-relaxing Lindbladian eigenstates near $\rho_1$ increases as well. This is particularly true for the gapped phase in
the limit of small $\gamma$, where the spectral width of the entire cloud of $\cal L$-eigenvalues along the real axis does not
scale with $N$ and is of order $\gamma$. Therefore the late-time dynamics involves many close spaced eignestates, and it is
difficult to isolate the truly asymptotic relaxation due to $\rho_1$. Nevertheless, the same reason also implies that the
relaxation necessarily takes place on the time scale $\Delta^{-1}$.

\vspace{1cm}

%\newpage

\begingroup
\renewcommand{\addcontentsline}[3]{}% Remove functionality of \addcontentsline
\renewcommand{\section}[2]{}% Remove functionality of \section